\documentclass[a4paper,11pt]{article}
\pdfoutput=1
\usepackage{jheppub}

\usepackage{amssymb,amsmath}
\usepackage[normalem]{ulem}
\usepackage[utf8x]{inputenc}
\usepackage{slashed}
\usepackage{graphicx}

\usepackage{csquotes} 
\usepackage{comment}
\usepackage{mathrsfs}
\usepackage{float}

\def\lsim{\mathrel{\rlap{\lower4pt\hbox{\hskip1pt$\sim$}}
    \raise1pt\hbox{$<$}}}         
\def\gsim{\mathrel{\rlap{\lower4pt\hbox{\hskip1pt$\sim$}}
    \raise1pt\hbox{$>$}}}         

\numberwithin{equation}{section}

\preprint{
\begin{minipage}{5cm}
\small
\flushright
KEK-TH-2430\\APCTP Pre2022 - 011\\KYUSHU-HET-243
\end{minipage}}

\title{Residual flavor symmetry breaking in the landscape of modular flavor models}

\author{Keiya Ishiguro$^{1}$,} 
\author{Hiroshi Okada$^{2,3}$,} 
\author{Hajime Otsuka$^{4}$} 
\affiliation{
$^1$ Graduate University for Advanced Studies (Sokendai), 1-1 Oho, Tsukuba, Ibaraki 305-0801, Japan\\
$^2$Asia Pacific Center for Theoretical Physics (APCTP), Pohang 37673, Republic of Korea \\
$^3$Department of Physics, Pohang University of Science and Technology, Pohang 37673, Republic of Korea \\
$^4$ Department of Physics, Kyushu University, 744 Motooka, Nishi-ku, Fukuoka 819-0395, Japan\\
}
\emailAdd{ishigu@post.kek.jp}
\emailAdd{hiroshi.okada@apctp.org}
\emailAdd{otsuka.hajime@phys.kyushu-u.ac.jp}

\abstract{
We study a symmetry breaking of residual flavor symmetries realized at fixed points of the moduli space. 
In the supersymmetric modular invariant theories, a small departure of the modulus from fixed points is required to realize fermion mass hierarchies and sizable CP-breaking effects. 
We investigate whether one can dynamically fix the moduli values in the vicinity of the fixed points in the context of Type IIB string theory. 
It is found that the string landscape prefers $|\delta \tau| \simeq 10^{-5}$ for the deviation of the complex structure modulus from all fixed points and the CP-breaking vacuum is statistically favored. 
To illustrate phenomenological implications of distributions of moduli values around fixed points, 
we analyze the lepton sector on a concrete $A_4$ modular flavor model. 
}
\makeatletter
\gdef\@fpheader{}
\makeatother

\begin{document}

\maketitle

\section{Introduction}

The flavor symmetry is a powerful approach to understand the 
flavor structure of quarks and leptons, and in addition, it provides the bridge between bottom-up and top-down approaches of model building. 
Indeed, when the flavor symmetry is embedded into a geometric 
symmetry of an extra-dimensional space, subgroups of the geometric 
symmetry would control the flavor structure of matter zero-modes. 
For instance, the $PSL(2,\mathbb{Z})$ modular symmetry incorporates the 
phenomenologically interesting non-Abelian discrete symmetries such as 
$S_3,S_4,A_4$ and $A_5$ in the principal subgroups \cite{deAdelhartToorop:2011re}. 
From the viewpoint of ultraviolet physics, it was known that the $SL(2,\mathbb{Z})$ modular group and its subgroups appearing in toroidal compactifications are connected to 
the flavor symmetries of matter zero-modes in heterotic orbifold models  \cite{Ferrara:1989qb,Lerche:1989cs,Lauer:1990tm,Baur:2019kwi,Baur:2019iai} and Type IIB superstring theories with magnetized D-branes \cite{Kobayashi:2018rad,Kobayashi:2018bff,Ohki:2020bpo,Kikuchi:2020frp,Kikuchi:2020nxn,Kikuchi:2021ogn,Almumin:2021fbk}. Such flavor symmetries are called modular flavor symmetries. 
The multi moduli cases such as $Sp(2h,\mathbb{Z})$ symplectic modular symmetry are also discussed in the context of heterotic string theory on toroidal orbifolds \cite{Baur:2020yjl} and Calabi-Yau manifolds \cite{Strominger:1990pd,Candelas:1990pi,Ishiguro:2020nuf,Ishiguro:2021ccl}.

From the phenomenological point of view, the modular flavor symmetries are attractive for predicting the masses and mixing angles of quarks and leptons 
under a certain value of the moduli fields \cite{Feruglio:2017spp,Kobayashi:2018vbk,Penedo:2018nmg,Novichkov:2018nkm,Ding:2019xna,Liu:2019khw,Chen:2020udk,Novichkov:2020eep,Liu:2020akv,Wang:2020lxk,Yao:2020zml,Ding:2020msi}. The higher-dimensional operators in the Standard Model effective field theory are also controlled by the modular symmetries \cite{Kobayashi:2021pav,Kobayashi:2022jvy}, taking into account the selection rule of the string theory \cite{Kobayashi:2021uam}. 
The flavor symmetry of quarks/leptons, as well as CP symmetry, is broken only by the modulus $\tau$ parametrizing the shape of the torus. 
Note that the CP transformation is regarded as an outer automorphism of the modular group for the single modulus \cite{Baur:2019kwi,Novichkov:2019sqv} and multi moduli cases \cite{Ishiguro:2021ccl}.
Once the modulus is fixed, there is no flavor symmetry in a generic moduli space. 
However, there exist so-called fixed points in the fundamental region of the $PSL(2,\mathbb{Z})$: $\tau= i, w, i\infty$ with $w=\frac{-1+i\sqrt{3}}{2}$, preserving $\mathbb{Z}_2$, $\mathbb{Z}_3$ and $\mathbb{Z}_2$ symmetries, respectively. 
Such fixed points play an important role for several phenomenological applications of the lepton sector \cite{Novichkov:2018ovf,Novichkov:2018yse,Novichkov:2018nkm,Ding:2019gof,Okada:2019uoy,King:2019vhv,Okada:2020rjb,Okada:2020ukr,Okada:2020brs,Feruglio:2021dte,Kobayashi:2021pav,Kobayashi:2022jvy} as well as controlling the effective action 
such as the dark matter (DM) stability \cite{Kobayashi:2021ajl}. 
To dynamically fix the moduli values gives a strong prediction on proposed modular flavor models. 
These attempts were performed in Refs. \cite{Kobayashi:2019xvz,Kobayashi:2019uyt,Kobayashi:2020uaj,Ishiguro:2020tmo,Novichkov:2022wvg}. 
However, in most of modular flavor models, one requires a slight difference in moduli values from fixed points to explain 
the observed masses and mixing angles of quarks/leptons as recently discussed in  Ref. \cite{Novichkov:2022wvg}.

In this paper, we adopt a top-down approach to dynamically fix the moduli values around the fixed points of the moduli space. 
In the string theory, background fluxes can stabilize the moduli fields such that subgroups of $SL(2,\mathbb{Z})$ are realized  \cite{Kobayashi:2020hoc} and also the CP symmetry is spontaneously broken \cite{Kobayashi:2020uaj,Ishiguro:2020nuf}. 
In addition, the flux landscape prefers the stabilization of moduli fields at fixed points with enhanced symmetries \cite{DeWolfe:2004ns,Ishiguro:2020tmo}. 
The purpose of this paper is to investigate the stabilization of moduli values at nearby fixed points and discuss the phenomenological implications. 
For concreteness, we focus on Type IIB string theory on toroidal orientifolds, where the complex structure moduli 
determine the flavor structure of quarks and leptons. 
Turning on background three-form fluxes, these complex structure moduli, as well as the dilaton, will be stabilized at statistically-favored symmetric points. 
To break enhanced symmetries in the complex structure moduli space, we incorporate non-perturbative effects whose existence is motivated by the stabilization of remaining volume moduli associated with the metric of extra-dimensional space. 
It is then expected that these non-perturbative effects can slightly shift the values of complex structure moduli from fixed points. 
Indeed, our systematic analysis of flux vacua with non-perturbative effects reveals that the complex structure moduli are 
stabilized at nearby fixed points whose magnitudes are controlled by non-perturbative effects.

Furthermore, we also incorporate the uplifting potential to obtain the de Sitter (dS) vacuum as discussed in the Kachru-Kallosh-Linde-Trivedi (KKLT) scenario \cite{Kachru:2003aw}. Such a supersymmetry (SUSY) breaking source also 
shift the value of the complex structure moduli from fixed points. \footnote{Soft SUSY breaking terms will keep the modular invariance in the moduli-mediated SUSY breaking scenario \cite{Kikuchi:2022pkd}, and their phenomenological aspects are discussed in Refs. \cite{Du:2020ylx,Kobayashi:2021jqu,Otsuka:2022rak}.}
It turns out that the string landscape prefers $|\delta \tau| \simeq 10^{-5}$ for the deviation of the complex structure modulus from fixed points $\langle \tau \rangle = i, w, 2i$, respectively.\footnote{Here, $\tau = i\infty$ is approximated as $\tau =2i$.} 
It corresponds to a specific SUSY breaking scale. 
In addition, the CP-breaking vacua are statistically favored due to the existence of non-perturbative effects as well as the uplifting source, 
although the number of CP-breaking vacua is statistically small in the finite number of flux vacua \cite{Ishiguro:2020tmo}. 
These moduli values are well fitted with observed masses and mixing angles in the lepton sector on a concrete $A_4$ modular flavor model. 
Furthermore, a quasi-stable dark matter (DM) would be realized due to the softly broken residual flavor symmetry at fixed points. 

This paper is organized as follows. 
After reviewing the structure of Type IIB flux vacua on $T^6/(\mathbb{Z}_2\times \mathbb{Z}_2^\prime)$ orientifolds in section \ref{sec:flux}, we incorporate the non-perturbative effects to stabilize the volume moduli in section \ref{sec:delta}. 
We numerically estimate the deviations of the complex structure modulus $\tau$ from fixed points in section \ref{sec:up}, taking into 
account SUSY breaking effects. 
These effects slightly break the enhanced symmetries in the moduli space of toroidal orientifolds. 
Given these moduli values, we study the concrete $A_4$ modular flavor model in section \ref{sec:A4} with an emphasis on the lepton sector. The distributions of 
$A_4$ model and the string landscape are compared. 
We summarize the paper in section \ref{sec:con}. In Appendix \ref{app}, we list the $A_4$ modular forms used in this paper.

\section{Moduli distributions in the string landscape}
\label{sec:moduli}

In section \ref{sec:flux}, we first review the flux vacua in Type IIB string theory on $T^6/(\mathbb{Z}_2\times \mathbb{Z}_2^\prime)$ orientifolds with an emphasis on the enhanced 
symmetry in the complex structure moduli space. 
Next, we focus on non-perturbative effects, which slightly 
break the enhanced symmetries in moduli space of toroidal orbifolds as discussed in section \ref{sec:delta}. 
Finally, we plot the deviation of the complex structure modulus from fixed points and the 
typical SUSY breaking scale in section \ref{sec:up}.

\subsection{Flux vacua with enhanced symmetries}
\label{sec:flux}

In Type IIB string theory on $T^6/(\mathbb{Z}_2\times \mathbb{Z}_2^\prime)$ orientifolds, the moduli K\"ahler potential and the flux-induced superpotential are given by\footnote{We follow the convention of Ref. \cite{Ishiguro:2020tmo}.}
\begin{align}
    K &= -\ln (-i(S -\Bar{S})) -2\ln {\cal V}(T) -3\ln\left(i (\tau -\bar{\tau}) \right),
    \label{eq:Keff}
\end{align}
where $S,T,\tau$ denote the dilaton, K\"ahler moduli (volume moduli) and the complex structure modulus, respectively. 
Here and in what follows, we adopt the reduced Planck mass unit unless we specify it, and we consider the isotropic torus $\tau=\tau_1=\tau_2=\tau_3$ to simplify our analysis. 
In Type IIB flux compactifications, one can consider the so-called Gukov-Vafa-Witten type superpotential \cite{Gukov:1999ya} induced by background three-form fluxes:
\begin{align}
    W_{\rm flux} &= a^0 \tau^3- 3a\tau^2- 3b \tau-b_0 -S\left(c^0\tau^3-3c\tau^2-3d\tau-d_0 \right),
\label{eq:Wsim}
\end{align}
where $\{a^0, a, b, b_0, c^0, c , d, d_0\}$ represent three-form flux quanta with the notation of Ref. \cite{Ishiguro:2020tmo}. 
These integers are now quantized in multiple of 8 on $T^6/(\mathbb{Z}_2\times \mathbb{Z}_2^\prime)$ geometry.  
In this paper, we analyze the SUSY minima:
\begin{align}
    \partial_S W = \partial_\tau W = W = 0\,,
\end{align}
at which the energy of scalar potential vanishes $V=e^K (K^{I\bar{J}}D_IW\overline{D_JW}-3|W|^2)=0$ with $D_IW= \partial_I W + W\partial_I K$. Here, we use the so-called no-scale structure for the K\"ahler moduli: $K^{i\bar{j}}K_i K_{\bar{j}}=3$ with $K_i = \partial_{T^i}K$ and $K^{i\bar{j}}$ being the inverse of K\"ahler metric. 
The SUSY conditions can be analytically solved by redefining the 
superpotential as
\begin{align}
    W_{\rm RR} = a^0 \tau^3- 3a\tau^2- 3b \tau-b_0 = (r \tau + s) P(\tau)\,, \\
    W_{\rm NS} = c^0\tau^3-3c\tau^2-3d\tau-d_0 = (u \tau + v) P(\tau)\,.
\end{align}
Here, we denote a quadratic (integer-coefficient) polynomial $P(\tau)$ 
with respect to $\tau$, and the minimum of $\tau$ is found by solving $P(\tau)=0$. 
Following Ref. \cite{Betzler:2019kon}, we write 
\begin{align}
    P(\tau) = l \tau^2 + m \tau + n\,,
\end{align}
under $m^2 - 4 l n < 0$, whose expression leads to the vacuum expectation value of $\tau$:
\begin{align}
    \langle \tau \rangle &= \frac{ - m + \sqrt{m^2 - 4 l n}}{2 l} \quad (l, n > 0)\,,
    \nonumber\\
    \langle \tau \rangle &= \frac{ - m - \sqrt{m^2 - 4 l n}}{2 l} \quad (l, n < 0)\,.
    \label{eq:tauvev}
\end{align}

The vacuum expectation value of the dilaton field is obtained by solving $\partial_{\tau} W =0$, i.e.,
\begin{align}
    P(\tau)\partial_\tau\{(r \tau + s) - S (u \tau + v) \} + \{(r \tau + s) - S (u \tau + v) \} \partial_\tau P(\tau) = 0\,.
\end{align}
Since the $\tau$ is now stabilized at Eq. (\ref{eq:tauvev}) determined by $P(\tau)=0$, we require 
\begin{align}
    \langle S \rangle=  \frac{ r \tau + s}{u \tau + v}\,.
    \label{eq:Svev}
\end{align}
Note that the condition $\partial_\tau P(\tau) = 0$ gives rise to a real $\tau$ and the dilaton cannot be stabilized anymore. 
At this stage, one cannot stabilize the K\"ahler moduli and requires additional sources such as non-perturbative effects. 

Before going into the detail of the volume moduli stabilization, we also review the structure of flux vacua on the toroidal orientifold. 
Remarkably, the background three-form fluxes induce a net D3-brane charge:
\begin{align}
    N_{\rm flux}&= \frac{1}{l_s^4}\int H_3\wedge F_3 = c^0b_0 -d_0a^0 +\sum_{i=1}^3 (c^ib_i -d_ia^i) = 
    c^0b_0 -d_0a^0 +3(cb -da)\,,
    \label{eq:nD3}
\end{align}
with the string length $l_s$, which should be canceled on a compact manifold. 
Taking into the contributions of D3/D7-branes and O3/O7-planes, 
the flux-induced D3-brane charge is constrained as
\begin{align}
0\leq N_{\rm flux}\leq N_{\rm flux}^{\rm max}={\cal O}(10^5)\,.
    \label{eq:Nflux}
\end{align}
Here, we admit the F-theory extension of the Type IIB orientifolds where a largest value of 
O3-plane contribution is given by $1820448$~\cite{Candelas:1997eh,Taylor:2015xtz}. 
Furthermore, $N_{\rm flux}$ should be in multiple of 192 due to the fact that $\{a^0, a, b, b_0, c^0, c , d, d_0\}\in 8 \mathbb{Z}$. 

\begin{figure}[H]
  \begin{center}
   \includegraphics[height=100mm]{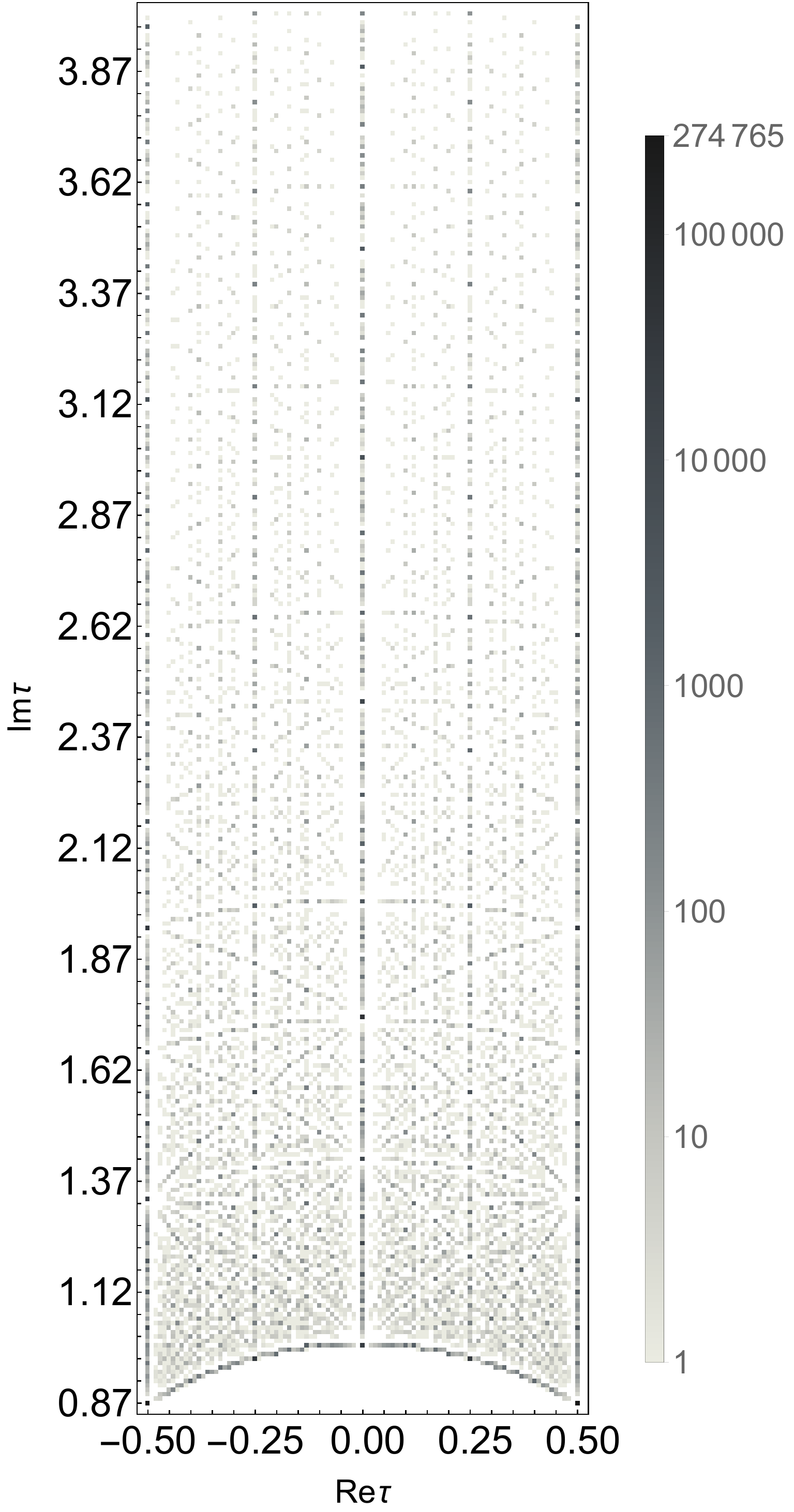}
  \end{center}
  \caption{The numbers of stable vacua on the fundamental domain of $\tau$ in the case of $N_{\rm flux}^{\rm max}=192 \times 1000$ \cite{Ishiguro:2020tmo}.}
\label{fig:1}
\end{figure}

For concreteness, we focus on the vacuum structure of $\tau$ whose fixed points in the moduli space are $\tau = i, w=\frac{-1+i\sqrt{3}}{2}, i\infty$, each of which corresponds to the $\mathbb{Z}_2$, $\mathbb{Z}_3$, $\mathbb{Z}_2$ fixed points, respectively. 
The $\tau = i, w$ fixed points are statistically favored in the flux landscape, as seen in 
Fig. \ref{fig:1}, where the higher the degeneracy of vacua, the darker the
color is. 
Note that one cannot realize $\tau = i\infty$ requiring the infinite value of flux quanta, and it is inconsistent 
with the charge cancellation condition of D3-brane charge (\ref{eq:Nflux}), namely the tadpole cancellation condition.
In this respect, we adopt $\tau = 2i$ as an approximation of $\tau=i\infty$. 
Such an approximation will often be used in the phenomenological analysis of modular flavor models. 

\subsection{Stabilization of volume moduli by non-perturbative effects}
\label{sec:delta}

In this section, we analyze the stabilization of volume moduli along the line of KKLT scenario \cite{Kachru:2003aw}. 
The stabilization of volume moduli is performed by the following non-perturbative effects:
\begin{align}
    W = W_{\rm flux}(\tau, S) + W_{\rm np}(S,T)\,,
\end{align}
where 
\begin{align}
    W_{\rm np} = \sum_m C_m e^{ia_m T+ ib_m S}
\end{align}
is supposed to be generated by D-brane instanton effects with $a_m, b_m = 2\pi$ or strong dynamics on D7-branes wrapping the rigid cycle with $a_m = 2\pi/N$ and $N$ being the rank of the gauge group. 
Here and in what follows, we consider a simple setup where 
the volume of internal manifold is determined by a single K\"ahler modulus $T$ whose K\"ahler potential is given by $K = -3 \ln (i (\bar{T}-T))$.

In the KKLT construction, the dilaton and the complex structure moduli are determined in the context of Type IIB flux compactifications as discussed in section \ref{sec:flux}. 
Note that the vacuum expectation value of flux superpotential vanishes in our analysis in the previous section; thereby the dilaton-dependent non-perturbative effects would induce the constant term in the effective superpotential:
\begin{align}
    W_{\rm eff} = W_{\rm np}(\langle S\rangle , T)\,,
\end{align}
which includes the following terms required in the KKLT construction:
\begin{align}
W_{\rm eff} \simeq \langle e^{ibS}\rangle + Ce^{iaT}\,.
\end{align}
Thus, the overall K\"ahler modulus is  stabilized at $T=T_0$ satisfying
\begin{align}
    D_T W_{\rm eff} = \partial_T W_{\rm eff} + K_T W_{\rm eff} = 0\,,
\end{align}
at which the minimum value of $T$ is estimated as
\begin{align}
    a\langle T \rangle \approx \ln (C/w_0)\,, 
    \label{eq:Tvev}
\end{align}
with $w_0 = \langle e^{ibS}\rangle$. 
Here, the origin of small superpotential $w_0$ relies on the dilaton-dependent non-perturbative effects. 
It is also possible to realize the small flux superpotential in Type IIB/F-theory flux compactifications (see, Refs.~\cite{Demirtas:2019sip,Honma:2021klo}, for the large complex structure regime). 
In what follows, the prefactor $C$ is assumed to be a constant, in particular, 1. 

So far, we have assumed that the dilaton and the complex structure moduli are stabilized in flux compactifications. However, the presence of non-perturbative effects slightly shifts their values. 
Indeed, the true vacuum is found by solving
\begin{align}
    D_I W=0\,,
\end{align}
which changes the moduli values obtained in the previous section. 
To find the slight difference from the fixed points of complex structure modulus, we utilize the perturbation method; the non-perturbative superpotential $W_{\rm np}$ causes the shift of the minima:
\begin{align}
    \tau &= \langle \tau \rangle + \delta \tau\,, \nonumber\\
    S &= \langle S \rangle + \delta S\,, \nonumber\\ 
    T &= \langle T \rangle + \delta T\,,    
\label{eq:deviation}
\end{align}
where the reference points $\{\langle \tau \rangle, \langle S \rangle, \langle T \rangle\}$ are given in Eqs. (\ref{eq:tauvev}), (\ref{eq:Svev}) and (\ref{eq:Tvev}), respectively.

Following Ref.\cite{Abe:2006xi}, we estimate the deviation up to a linear order. Let us consider the K\"ahler-invariant quantity $G = K + \ln |W|^2$ satisfying $G_A = \partial_A G= 0$ at the SUSY minima. 
Here and in what follows, the index $A$ denotes both the holomorphic and anti-holomorphic fields: $\{S,T,\tau, \bar{S},\bar{T},\bar{\tau}\}$. 
From the expansion (\ref{eq:deviation}), $G_A$ is expanded as
\begin{align}
    G_A = G_A\bigl|_{\langle \rangle} + \delta \phi^B G_{AB}\bigl|_{\langle \rangle} + ({\cal O}(\delta \phi)^2)\,,
\end{align}
where $\bigl|_{\langle \rangle}$ means $\bigl|_{\tau = \langle \tau \rangle, S = \langle S \rangle, T = \langle T \rangle}$. 
Under the assumption $a, b > 1$, we obtain
\begin{align}
    G_{IJ}, G_{\bar{I}\bar{J}} \gg G_{I\bar{J}}, G_{\bar{I}J}\,,
\end{align}
which implies that $G_{AB}$ and $G^{AB}$ are diagonalized only by the holomorphic and anti-holomorphic parts, respectively. 
As a result, we obtain
\begin{align}
    \delta \phi^I = G^{IJ}G_J\bigl|_{\langle \rangle} + ({\cal O}(\delta \phi)^2)\,,
\end{align}
whose explicit form is written by
\begin{align}
    \delta \tau &= W_{\rm eff} \left(-\frac{G_S}{W_{S\tau}}\right)\biggl|_{\langle \rangle} + {\cal O}(W_{\rm eff}^2)\,, 
    \nonumber\\
    \delta S &=  \frac{W_{\rm eff}}{W_{S\tau}}\left(\frac{W_{\tau\tau}}{W_{S\tau}}G_S -G_\tau \right)\biggl|_{\langle \rangle} + {\cal O}(W_{\rm eff}^2)\,,
    \nonumber\\    
    \delta T &= \left(-\frac{G_{ST}}{G_{TT}}\right)\biggl|_{\langle \rangle} \delta S\,.
\end{align}
Note that the internal volume should be larger than the string length, 
\begin{align}
    {\rm Im}(T) \gg 1\,,
\end{align} 
and the weak string coupling ${\rm Im}(S)>1$; thereby the magnitude of the flux superpotential is exponentially small:
\begin{align}
    \langle W_{\rm eff}\rangle \simeq w_0 + e^{ia T} < 10^{-3}\,.
\end{align}
Here and in the following numerical calculations, we take $a=b=2\pi$ for concreteness.

In this way, the deviation of the vacuum values $\{\delta \tau, \delta S, \delta T\}$ are determined by the non-perturbative effects, 
implying that the deviation is naturally suppressed with respect to the volume modulus. 
From the phenomenological point of view, such a small deviation of $\tau$ is useful for predicting the masses and mixing angles of quarks and leptons, as discussed in detail in section \ref{sec:A4}. 
Before going into a phenomenological analysis, we discuss the supersymmetry breaking effects in 
the next section.

\subsection{Moduli values at nearby fixed points}
\label{sec:up}

So far, we have analyzed the stabilization of the complex structure modulus, dilaton and K\"ahler moduli 
at SUSY minima. However, the obtained vacuum energy is negative, i.e., anti-de Sitter (AdS) vacuum. 
To realize a dS vacuum, it is required to uplift the AdS vacuum to the dS one. 
Among several proposals for the uplifting scenarios, we focus on the anti D3-brane as originally 
utilized in the KKLT scenario \cite{Kachru:2003aw}. 
The anti D3-brane induces the positive vacuum energy due to its SUSY breaking effect,
\begin{align}
    V_{\rm up}=\frac{D}{(i(\bar{T}-T))^3}\,,
\end{align}
where a constant $D$ is chosen to realize the present vacuum energy. 
Then, the effective scalar potential is described as
\begin{align}
    V = e^K\left( K^{I\bar{J}}D_I W \overline{D_JW} -3|W|^2 \right) + V_{\rm up}\,, 
\end{align}
indicating that the uplifting source further causes the shift of the moduli values 
obtained in the previous section. 

\begin{figure}[H]
\begin{minipage}{0.49\hsize}
  \begin{center}
  \includegraphics[height=4.17cm]{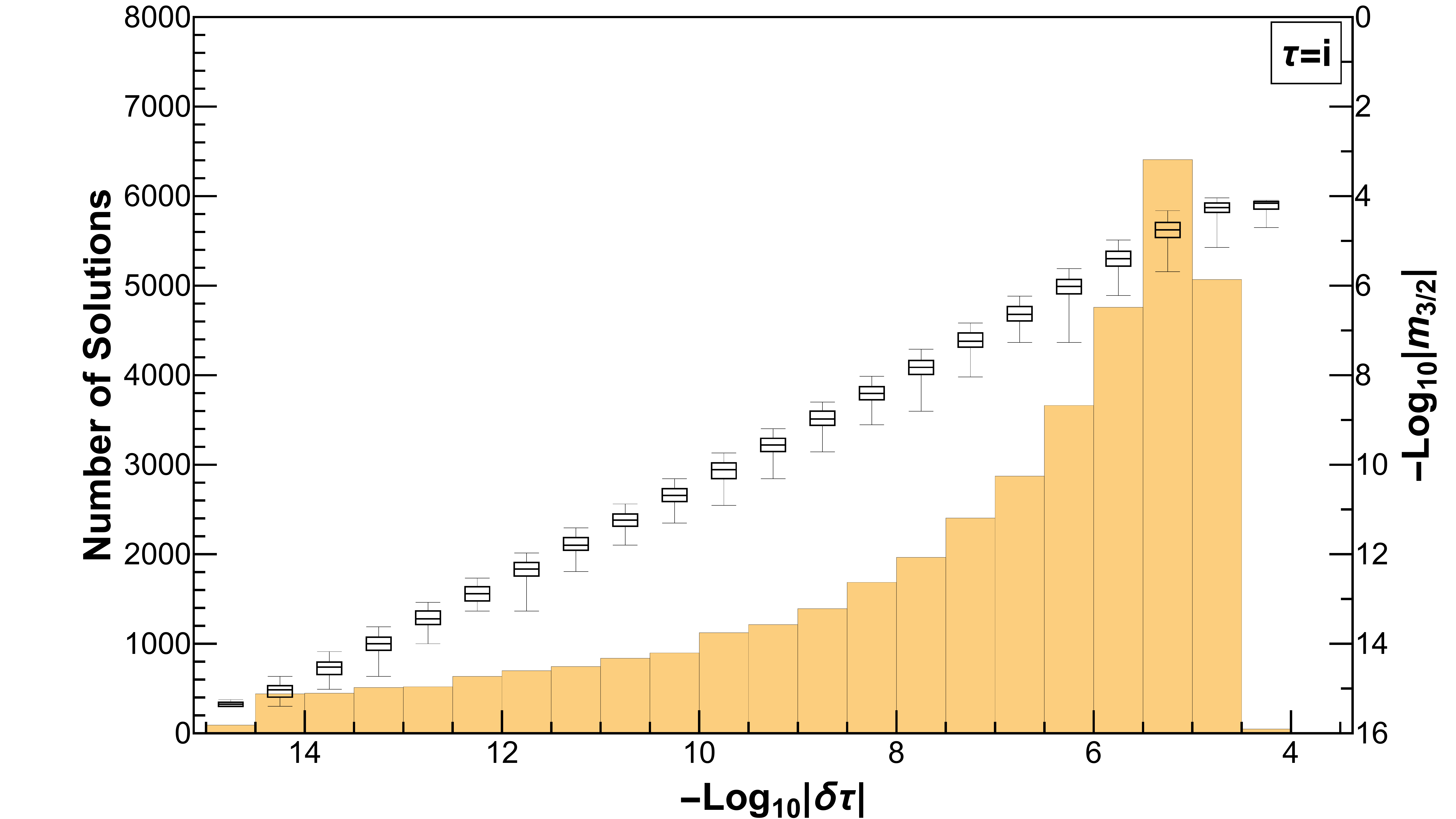}
  \end{center}
 \end{minipage}
 \begin{minipage}{0.49\hsize}
  \begin{center}
   \includegraphics[height=4.17cm]{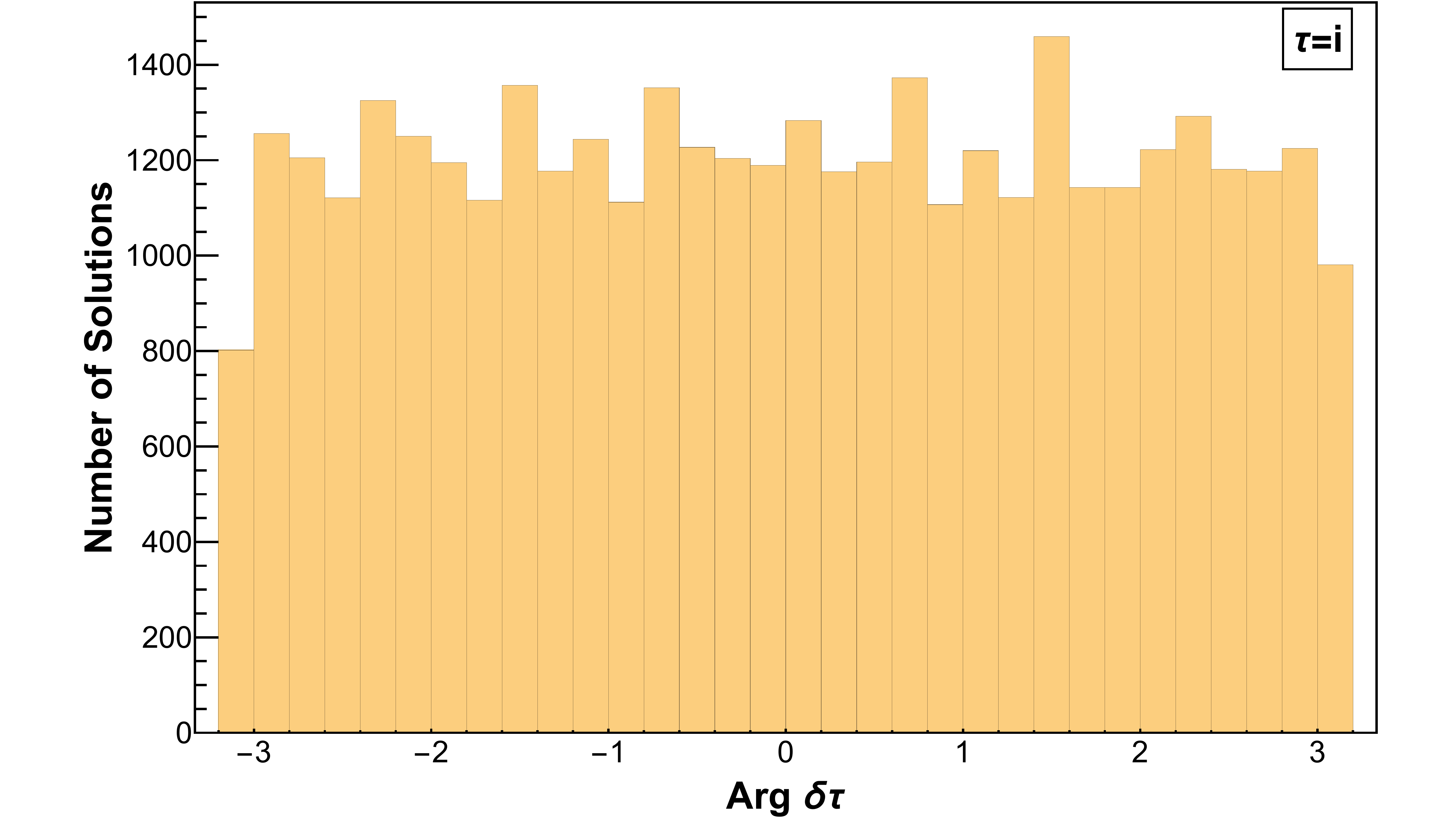}
  \end{center}
 \end{minipage}
  \caption{The number of vacua at nearby $\langle\tau\rangle=i$ as a function of $|\delta\tau|$ in the left panel and ${\rm Arg}(\delta\tau)$ in the right panel, respectively. In the left panel, the absolute value of gravitino mass is plotted as a function of $|\delta \tau|$.}
\label{fig:Z2_up}
\end{figure}

\begin{figure}[H]
\begin{minipage}{0.49\hsize}
  \begin{center}
  \includegraphics[height=4.17cm]{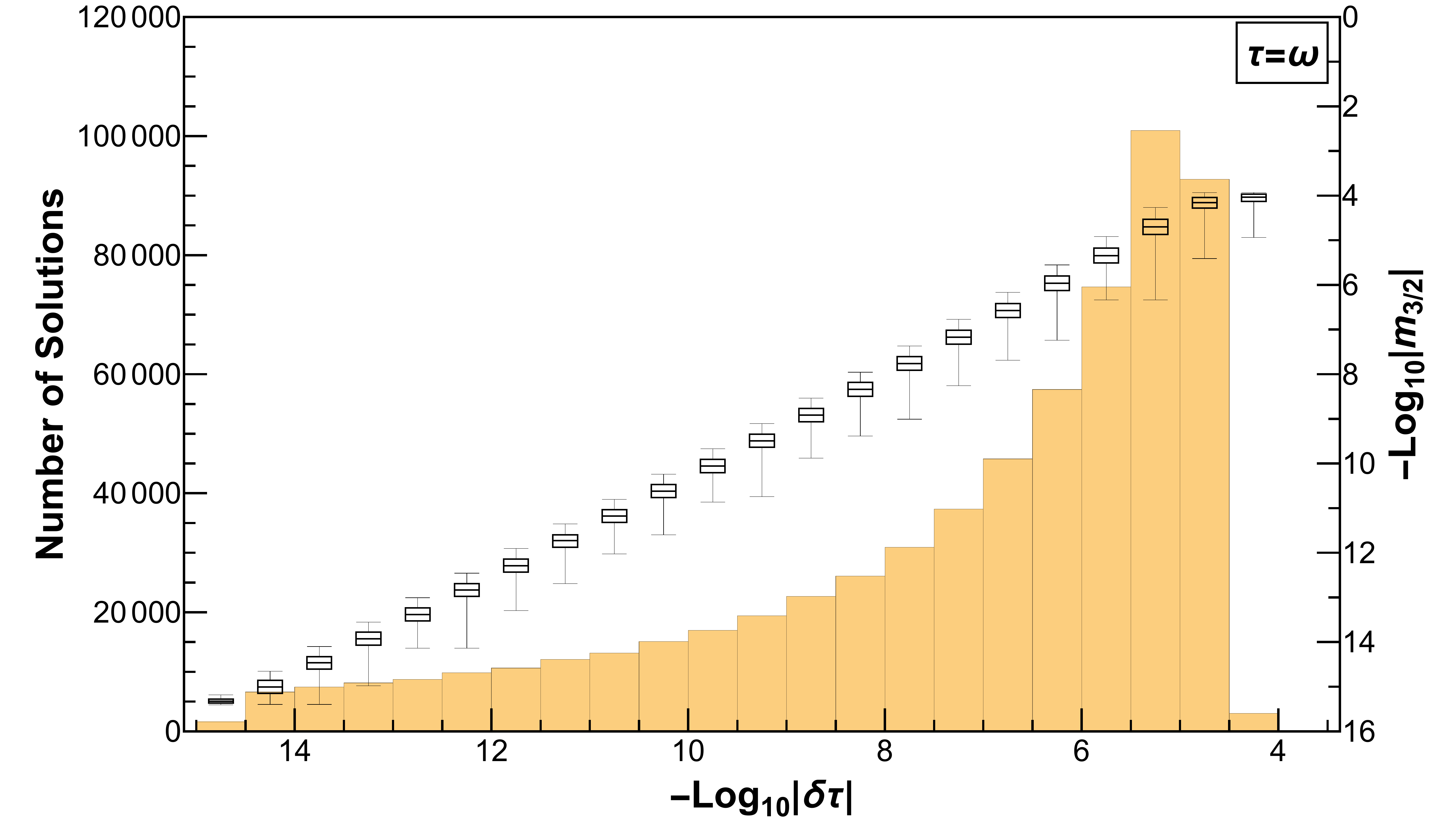}
  \end{center}
 \end{minipage}
 \begin{minipage}{0.49\hsize}
  \begin{center}
   \includegraphics[height=4.17cm]{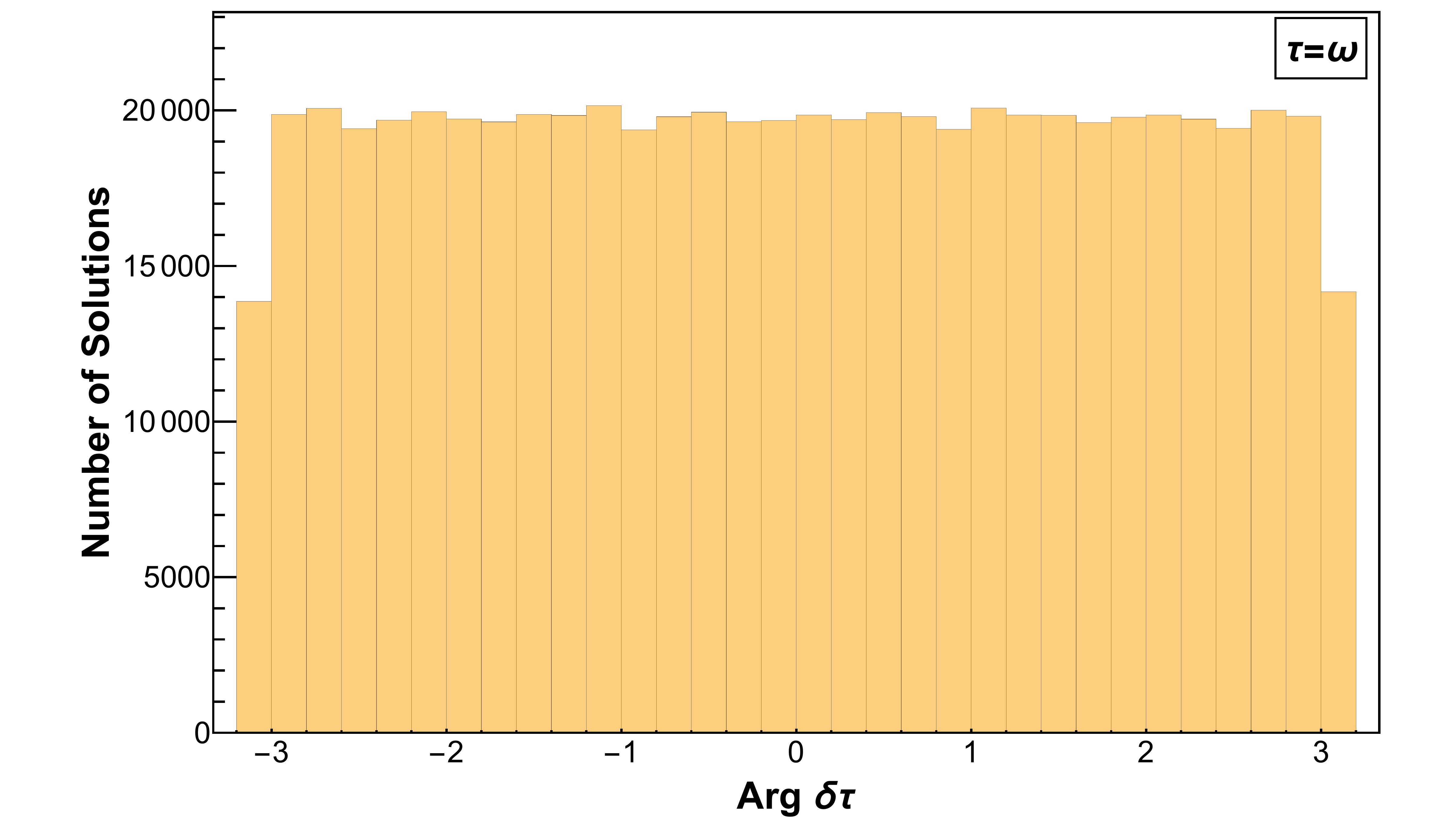}
  \end{center}
 \end{minipage}
  \caption{The number of vacua at nearby $\langle\tau\rangle=w$ as a function of $|\delta\tau|$ in the left panel and ${\rm Arg}(\delta\tau)$ in the right panel, respectively. In the left panel, the absolute value of gravitino mass is plotted as a function of $|\delta \tau|$.}
\label{fig:Z3_up}
\end{figure}

\begin{figure}[H]
\begin{minipage}{0.49\hsize}
  \begin{center}
  \includegraphics[height=4.17cm]{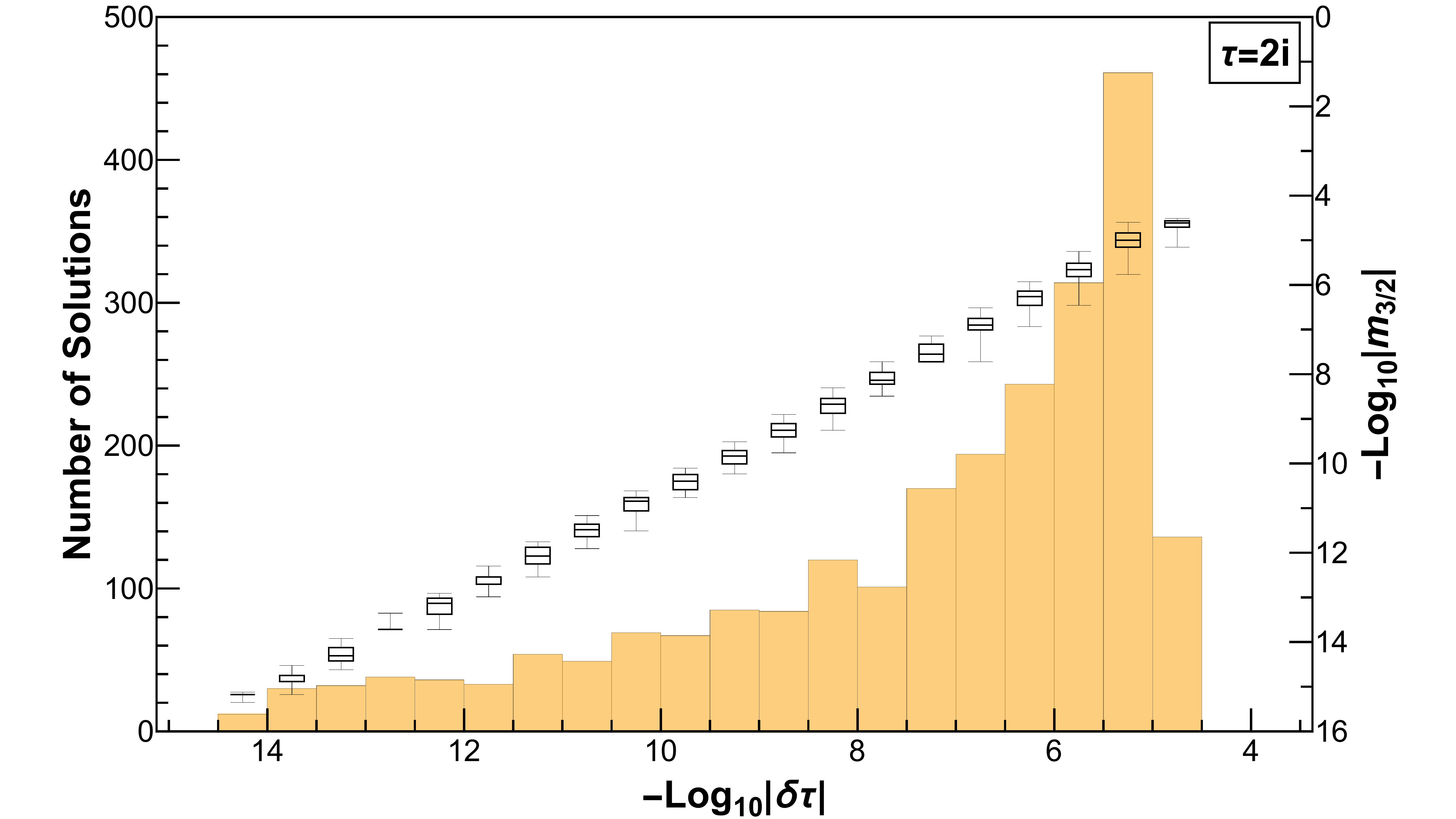}
  \end{center}
 \end{minipage}
 \begin{minipage}{0.49\hsize}
  \begin{center}
   \includegraphics[height=4.17cm]{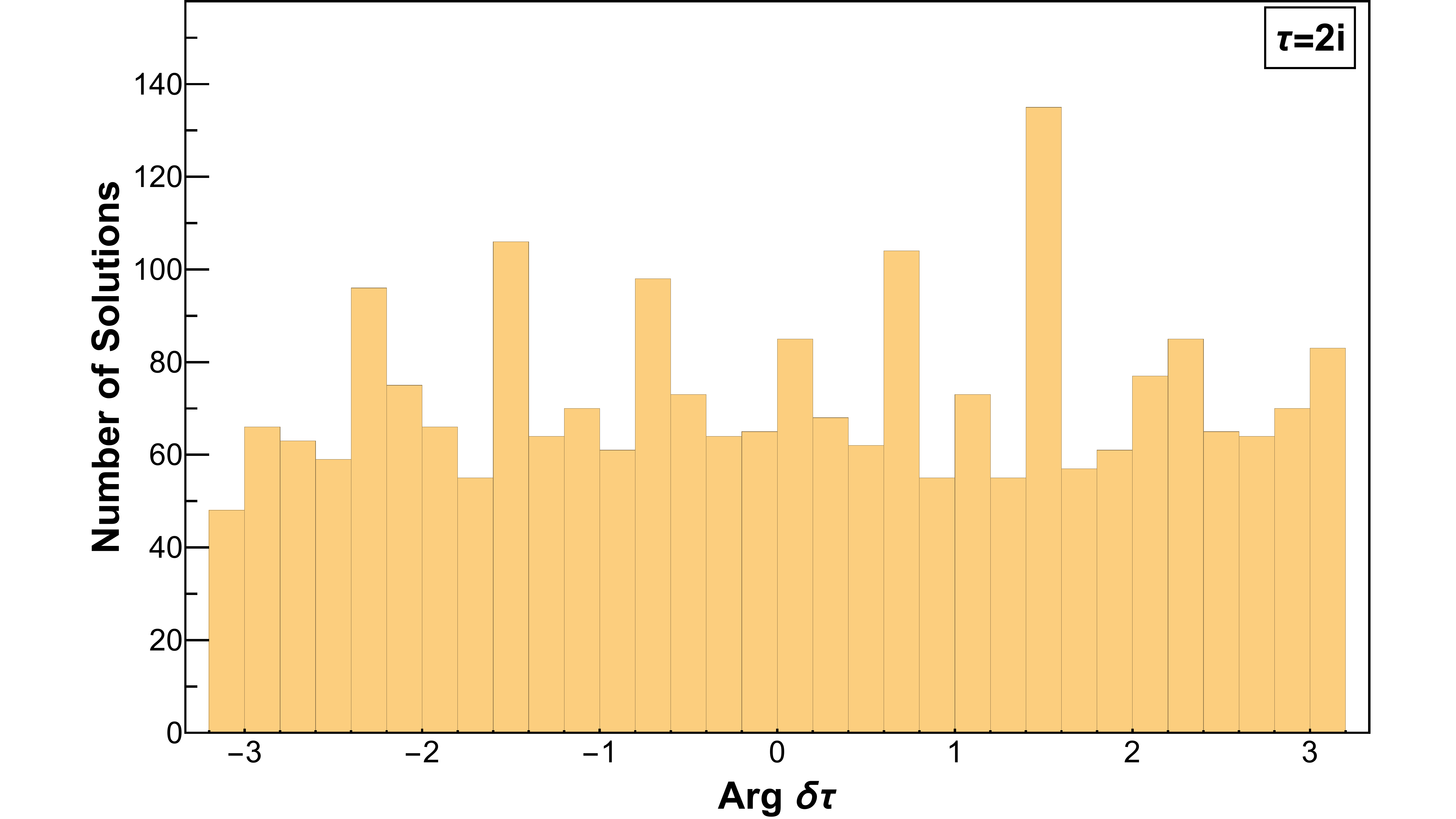}
  \end{center}
 \end{minipage}
  \caption{The number of vacua at nearby $\langle\tau\rangle=2i$ as a function of $|\delta\tau|$ in the left panel and ${\rm Arg}(\delta\tau)$ in the right panel, respectively. In the left panel, the absolute value of gravitino mass is plotted as a function of $|\delta \tau|$.}
\label{fig:2i_up}
\end{figure}

To see the deviation of complex structure moduli values from fixed points, 
we numerically calculate the deviation of $\tau$ from $\langle \tau \rangle = i, w, 2i$ by utilizing a 
finite number of flux vacua with $N_{\rm flux}^{\rm max}=1000$. 
By calculating the minimization condition of the full scalar potential $\partial_I V =0$ for $I=\tau,S,T$, 
we find deviations of the complex structure modulus from fixed points $\delta \tau \equiv \tau - \langle \tau \rangle$ as shown in Figs. \ref{fig:Z2_up}, \ref{fig:Z3_up} and \ref{fig:2i_up}. 
It turns out that the flux landscape prefers $|\delta \tau| \simeq 10^{-5}$ from fixed points $\langle \tau \rangle = i, w, 2i$, but there is no sizable difference in the phase direction. 
It means that the CP symmetry parametrized by $\tau \rightarrow -\bar{\tau}$ is broken in a generic moduli space. 
Furthermore, we plot the typical SUSY breaking scale, i.e., the gravitino mass $m_{3/2}=e^{K/2}W$, 
in the same figures. 
At the statistically favored moduli values $|\delta \tau| ={\cal O}(10^{-5})$, 
the gravitino mass is $m_{3/2} = {\cal O}(10^{-5})$ in the reduced Planck mass unit. 
Note that the small $\delta \tau$ is originating from non-perturbative effects and the uplifting source, both 
of which are the same order.

\section{$A_4$ modular flavor model}
\label{sec:A4}

To illustrate implications of distributions of moduli fields around fixed points, 
we study the phenomenology of lepton sector on a concrete $A_4$ modular flavor model.

\subsection{Setup}

\begin{table}[H]
\centering
\begin{tabular}{|c||c|c|c|c||c||}\hline\hline  
  & ~$L$~& ~$\{e^c,\mu^c,\tau^c\}$~ & ~$N^c$~& ~$H_u$~ & ~$H_d$~  
  \\\hline 
 $SU(2)_L$ & $\bf{2}$  & $\bf{1}$ & $\bf{1}$ & $\bf{2}$  & $\bf{2}$   \\\hline 
$U(1)_Y$ & $-\frac{1}{2}$ & $1$ & $0$ & $\frac{1}{2}$ & $-\frac{1}{2}$  \\\hline
 $A_4$ & ${\bf 3}$ & $\{\bf 1,1',1'\}$ & ${\bf 3}$ & ${\bf 1}$ & ${\bf 1}$      \\\hline
 $-k_I$ & $-2$ & $\{-2,-4,-4\}$ & $-2$ & $0$ & $0$\\\hline
\end{tabular}
\caption{Charge assignments under $SU(2)_L\times U(1)_Y\times A_{4}$ in the lepton and Higgs sectors, where $k_I$ denotes the modular weight of matter superfields $\Phi_I$.}
\label{tab:fields}
\end{table}

For concreteness, we specify charge assignments under $SU(2)_L\times U(1)_Y\times A_{4}$ for the lepton and Higgs sectors as summarized in Tab. \ref{tab:fields}. 
Here, the $A_4$ group belongs to the $SL(2,\mathbb{Z})$ modular group parametrized by the modulus $\tau$. 
The Yukawa couplings are constructed in a modular invariant way. (For more details, see, Appendix \ref{app}.) 
Then, we can write down the modular invariant superpotential:
\begin{align}
    W &= y_e (Y_{\bf 3}^{(4)} L)_1 H_d e^c + \sum_{\bf r = 3, 3^\prime} y_{\mu}^{({\bf r})}  (Y_{\bf r}^{(6)} L)_1 H_d \mu^c + \sum_{\bf r = 3, 3^\prime} y_{\tau}^{({\bf r})}  (Y_{\bf r}^{(6)} L)_1 H_d \tau^c
    \nonumber\\
    &+ \sum_{\bf r = 1, 1^\prime} y_{d}^{({\bf r})}  (Y_{\bf r}^{(4)} L H_u N^c)_1 
    +  y_{d}^{({\bf 3_S})}  (Y_{\bf 3}^{(4)} H_u (LN^c)_{\bf 3_S})_1 
        +  y_{d}^{({\bf 3_A})}  (Y_{\bf 3}^{(4)}  H_u (LN^c)_{\bf 3_A})_1 
    \nonumber\\
    &+ \sum_{\bf r = 1, 1^\prime, 3} M^{({\bf r})}  (Y_{\bf r}^{(4)}  N^c N^c)_1,
\end{align}
where $Y_{\bf r}^{(k)}$ denotes the holomorphic modular form with weight $k$ for ${\bf r}$ representations under the $A_4$ group, and  $\{y_e,y_{\mu}^{({\bf r})}, y_{\tau}^{({\bf r})}, y_{d}^{({\bf r})}\}$ are parameters.
Here, we introduce the Majorana mass terms to realize small neutrino masses. 

In the following, we enumerate the mass matrix of the lepton sector. 

\begin{enumerate}
    \item Charged-lepton mass matrix
    
After the electroweak symmetry breaking, charged-lepton mass matrix is written as
\begin{align}
    (m_l)_{LR} = \frac{v_d}{\sqrt{2}}
    \begin{pmatrix}
       Y_{1}^{(4)} & Y_3^{(6)} +\epsilon_\mu Y_{3^\prime}^{(6)} & Y_3^{(6)} +\epsilon_\tau Y_{3^\prime}^{(6)}\\
       Y_{3}^{(4)} & Y_2^{(6)} +\epsilon_\mu Y_{2^\prime}^{(6)} & Y_2^{(6)} +\epsilon_\tau Y_{2^\prime}^{(6)}\\
       Y_{2}^{(4)} & Y_1^{(6)} +\epsilon_\mu Y_{1^\prime}^{(6)} & Y_1^{(6)} +\epsilon_\tau Y_{1^\prime}^{(6)}\\
    \end{pmatrix}
    \times
    \begin{pmatrix}
       y_e & 0 & 0\\
       0 & y_\mu^{(\bf 3)} & 0\\
       0 & 0 & y_\tau^{(\bf 3)}\\
    \end{pmatrix}
,
\end{align}
where we introduce
\begin{align}
\langle H_d \rangle = v_d,\quad
\epsilon_\mu = \frac{y_\mu^{(\bf 3^\prime)}}{y_\mu^{({\bf 3})}},\quad
\epsilon_\tau = \frac{y_\tau^{(\bf 3^\prime)}}{y_\tau^{({\bf 3})}},\quad
Y_{\bf 3}^{(k)} = 
\begin{pmatrix}
   Y_1^{(k)}\\
   Y_2^{(k)}\\
   Y_3^{(k)}\\
\end{pmatrix}
,\quad
Y_{\bf 3^\prime}^{(k)} = 
\begin{pmatrix}
   Y_{1^\prime}^{(k)}\\
   Y_{2^\prime}^{(k)}\\
   Y_{3^\prime}^{(k)}\\
\end{pmatrix}
.
\end{align}
The explicit modular forms are listed in Appendix \ref{app}. 
Then the charged-lepton mass square eigenstate can be found by
${\rm diag}( |m_e|^2 , |m_\mu|^2 , |m_\tau|^2)\equiv V^\dag_{l_L} m_l^\dag m_l V_{l_L}$.
We numerically determine the three parameters $y_e,y_\mu^{(\bf 3)},y_\tau^{(\bf 3)}$ to fit the three charged-lepton masses by applying the relations:
\begin{align}
&{\rm Tr}[m_l^\dag m_l] = |m_e|^2 + |m_\mu|^2 + |m_\tau|^2,\\
&{\rm Det}[m_l^\dag m_l] = |m_e|^2  |m_\mu|^2  |m_\tau|^2,\\
&({\rm Tr}[m_l^\dag m_l])^2 -{\rm Tr}[(m_l^\dag m_l)^2] =2( |m_e|^2  |m_\nu|^2 + |m_\mu|^2  |m_\tau|^2+ |m_e|^2  |m_\tau|^2 ).
\end{align}
Therefore, input parameters are $\epsilon_\mu,\ \epsilon_\tau$ 
in the charged-lepton sector.

\item Dirac Yukawa mass matrix

\begin{align}
    (m_D)_{LN} &= \frac{v_u}{\sqrt{2}}
    \Biggl[
     \frac{y_d^{({\bf 3_S})}}{3}
    \begin{pmatrix}
2Y_1^{(4)} & -Y_3^{(4)} & -Y_2^{(4)}\\
-Y_3^{(4)} & 2Y_2^{(4)} & -Y_1^{(4)}\\
-Y_2^{(4)} & -Y_1^{(4)} & 2Y_3^{(4)}\\
    \end{pmatrix}
+
     \frac{y_d^{({\bf 3_A})}}{2}
    \begin{pmatrix}
0 & Y_3^{(4)} & -Y_2^{(4)}\\
-Y_3^{(4)} & 0 & Y_1^{(4)}\\
Y_2^{(4)} & -Y_1^{(4)} & 0\\
    \end{pmatrix}
\nonumber\\
&\hspace{40pt}+
y_d^{({\bf 1})} Y_{\bf 1}^{(4)}
    \begin{pmatrix}
1 & 0 & 0\\
0 & 0 & 1\\
0 & 1 & 0\\
    \end{pmatrix}
+
y_d^{({\bf 1^\prime})} Y_{\bf 1^\prime}^{(4)}
    \begin{pmatrix}
0 & 0 & 1\\
0 & 1 & 0\\
1 & 0 & 0\\
    \end{pmatrix}
\Biggl]
\nonumber\\
&=m_{d_0}
    \Biggl[
     \begin{pmatrix}
2Y_1^{(4)} & (-1+g_D)Y_3^{(4)} & -(1+g_D)Y_2^{(4)}\\
-(1+g_D)Y_3^{(4)} & 2Y_2^{(4)} & (-1+g_D)Y_1^{(4)}\\
(-1+g_D)Y_2^{(4)} & -(1+g_D)Y_1^{(4)} & 2Y_3^{(4)}\\
    \end{pmatrix}
\nonumber\\
&\hspace{40pt}
+h_1
    \begin{pmatrix}
1 & 0 & 0\\
0 & 0 & 1\\
0 & 1 & 0\\
    \end{pmatrix}
+
h_2    \begin{pmatrix}
0 & 0 & 1\\
0 & 1 & 0\\
1 & 0 & 0\\
    \end{pmatrix}
\Biggl]
\nonumber\\
&\equiv m_{d_0}  \tilde m_D
,
\end{align}
where we define
\begin{align}
\langle H_u \rangle = v_u, 
\quad
m_{d_0} \equiv \frac{y_d^{(\bf 3_S)}}{3\sqrt2}v_u,
\quad
g_D = \frac{3y_d^{(\bf 3_A)}}{2y_d^{(\bf 3_S)}},
\quad
h_1 = \frac{3 y_d^{({\bf 1})} Y_{\bf 1}^{(4)}}{y_d^{(\bf 3_S)}},
\quad
h_2 = \frac{3 y_d^{({\bf 1'})} Y_{\bf 1'}^{(4)}}{y_d^{(\bf 3_S)}}.
\end{align}

\item Majorana mass matrix

\begin{align}
    M_N &= \frac{M_1}{3}
    \begin{pmatrix}
2Y_1^{(4)} & -Y_3^{(4)} & -Y_2^{(4)}\\
-Y_3^{(4)} & 2Y_2^{(4)} & -Y_1^{(4)}\\
-Y_2^{(4)} & -Y_1^{(4)} & 2Y_3^{(4)}\\
    \end{pmatrix}
+
M_2 Y_{\bf 1}^{(4)}
    \begin{pmatrix}
1 & 0 & 0\\
0 & 0 & 1\\
0 & 1 & 0\\
    \end{pmatrix}
+
M_3 Y_{\bf 1^\prime}^{(4)}
    \begin{pmatrix}
0 & 0 & 1\\
0 & 1 & 0\\
1 & 0 & 0\\
    \end{pmatrix}
    \nonumber\\
&=M_{0}
    \Biggl[
    \begin{pmatrix}
2Y_1^{(4)} & -Y_3^{(4)} & -Y_2^{(4)}\\
-Y_3^{(4)} & 2Y_2^{(4)} & -Y_1^{(4)}\\
-Y_2^{(4)} & -Y_1^{(4)} & 2Y_3^{(4)}\\
    \end{pmatrix}
+
f_1
    \begin{pmatrix}
1 & 0 & 0\\
0 & 0 & 1\\
0 & 1 & 0\\
    \end{pmatrix}
+
f_2
    \begin{pmatrix}
0 & 0 & 1\\
0 & 1 & 0\\
1 & 0 & 0\\
    \end{pmatrix}
\Biggl]
\nonumber\\
&\equiv M_{0}  \tilde  M_N 
,
\end{align}
where we define
\begin{align}
M_0 \equiv \frac{M_1}{3},
\quad
f_1 = \frac{3 Y_{\bf 1}^{(4)} M_2}{M_1},
\quad
f_2 = \frac{3 Y_{\bf 1'}^{(4)} M_3}{M_1}.
\end{align}
%
%
Then, the active neutrino mass matrix is given by
\begin{align}
m_\nu&\approx 
-m_D^T M_N^{-1}   m_D
=-\kappa \tilde m_D^T \tilde M_N^{-1} \tilde   m_D=-\kappa \tilde m_\nu,
\label{eq:Mnu}
\end{align}
where the mass dimensional parameter $\kappa$ is defined by $m_{d_0}^2/M_0$.
$\tilde m_\nu$ is diagonalized by applying a unitary matrix as $V^\dag_\nu (\tilde m_\nu^\dag \tilde m_\nu)V_\nu=(\tilde m_{1}^2,\tilde m_{2}^2,\tilde m_{3}^2)$. 
In this case, $\kappa$ is determined by 
\begin{align}
({\rm NH}):\  \kappa^2= \frac{|\Delta m_{\rm atm}^2|}{\tilde m_3^2-\tilde m_1^2},
\quad
({\rm IH}):\  \kappa^2= \frac{|\Delta m_{\rm atm}^2|}{\tilde m_{2}^2-\tilde m_{3}^2},
 \end{align}
where $\Delta m_{\rm atm}^2$ is atmospheric neutrino mass square difference, and NH and IH stand for normal and inverted hierarchies, respectively.
The solar mass square difference is also found in terms of $\kappa$ as follows:
\begin{align}
\Delta m_{\rm sol}^2= {\kappa^2}({\tilde m_{2}^2-\tilde m_{1}^2}).
 \end{align}
In our numerical analysis, we regard $\Delta m_{\rm atm}^2$ as an input parameter from experiments so that $\Delta m_{\rm sol}^2$ be output parameter giving numerical $(\tilde m_{1}^2,\tilde m_{2}^2,\tilde m_{3}^2)$.
Then, one finds $U_{\rm PMNS}=V^\dag_{l_L} V_\nu$, and 
it is parametrized by three mixing angles $\theta_{ij} (i,j=1,2,3; i < j)$, one CP violating Dirac phase $\delta_{\rm CP}$,
and two Majorana phases $\{\alpha_{21}, \alpha_{32}\}$ as follows:
\small
\begin{equation}
U_{\rm PMNS} = 
\begin{pmatrix} c_{12} c_{13} & s_{12} c_{13} & s_{13} e^{-i \delta_{\rm CP}} \\ 
-s_{12} c_{23} - c_{12} s_{23} s_{13} e^{i \delta_{\text{CP}}} & c_{12} c_{23} - s_{12} s_{23} s_{13} e^{i \delta_{\text{CP}}} & s_{23} c_{13} \\
s_{12} s_{23} - c_{12} c_{23} s_{13} e^{i \delta_{\text{CP}}} & -c_{12} s_{23} - s_{12} c_{23} s_{13} e^{i \delta_{\text{CP}}} & c_{23} c_{13} 
\end{pmatrix}
\begin{pmatrix} 1 & 0 & 0 \\ 0 & e^{i \frac{\alpha_{21}}{2}} & 0 \\ 0 & 0 & e^{i \frac{\alpha_{31}}{2}} \end{pmatrix},
\end{equation}
\normalsize
where $c_{ij}$ and $s_{ij}$ stand for $\cos \theta_{ij}$ and $\sin \theta_{ij}$, respectively. 
These mixings are rewritten in terms of the component of $U_{\rm PMNS}$ as follows:
\begin{align}
\sin^2\theta_{13}=|(U_{\rm PMNS})_{13}|^2,\quad 
\sin^2\theta_{23}=\frac{|(U_{\rm PMNS})_{23}|^2}{1-|(U_{\rm PMNS})_{13}|^2},\quad 
\sin^2\theta_{12}=\frac{|(U_{\rm PMNS})_{12}|^2}{1-|(U_{\rm PMNS})_{13}|^2}.
\end{align}
In addition, we can compute the Jarlskog invariant, $\delta_{\text{CP}}$ from PMNS matrix elements $(U_{\rm PMNS})_{\alpha i}\equiv U_{\alpha i}$:
\begin{equation}
J_{\rm CP} = \text{Im} [U_{e1} U_{\mu 2} U_{e 2}^* U_{\mu 1}^*] = s_{23} c_{23} s_{12} c_{12} s_{13} c^2_{13} \sin \delta_{\rm CP},
\end{equation}
and the Majorana phases are also estimated in terms of other invariants $I_1$ and $I_2$ constructed by PMNS matrix elements:
\begin{equation}
I_1 = \text{Im}[U^*_{e1} U_{e2}] = c_{12} s_{12} c_{13}^2 \sin \left( \frac{\alpha_{21}}{2} \right), \
I_2 = \text{Im}[U^*_{e1} U_{e3}] = c_{12} s_{13} c_{13} \sin \left( \frac{\alpha_{31}}{2}  - \delta_{\rm CP} \right).
\end{equation}
Furthermore, the effective mass for the neutrinoless double beta decay is given by
\begin{align}
\langle m_{ee}\rangle=\kappa|\tilde D_{\nu_1} c^2_{12} c^2_{13}+\tilde D_{\nu_2} s^2_{12} c^2_{13}e^{i\alpha_{21}}+\tilde D_{\nu_3} s^2_{13}e^{i(\alpha_{31}-2\delta_{\rm CP})}|\,,
\end{align}
where its observed value could be measured by KamLAND-Zen experiment in future~\cite{KamLAND-Zen:2016pfg}. 
In our numerical analysis below, we will do $\Delta\chi$ square analysis referring to Ref.~\cite{Esteban:2020cvm}.

\end{enumerate}

\subsection{Numerical analysis}
In this section, we show the allowed region with $\chi$ square analysis to satisfy the current neutrino oscillation data, where we randomly select within the following ranges of input parameters,
\begin{align}
&|\delta\tau|  \in [10^{-20},0.1], \ \{ \epsilon_\mu, \epsilon_\tau, g_D, f_1,f_2, h_1,h_2\}   \in [10^{-4},2],
\end{align}
where we assume all the parameters (except $\tau$) are real for simplicity.
 We also take Yukawa couplings of the SM charged leptons 
at the GUT scale $2\times 10^{16}$ GeV and $\Delta m_{\rm atm}^2$ as input parameters,  where $\tan\beta=5$ is taken
as a bench mark~\cite{Bjorkeroth:2015ora}:
\begin{align}
&y_e=(1.97\pm 0.024) \times 10^{-6}, \ 
y_\mu=(4.16\pm 0.050) \times 10^{-4}, \ 
y_\tau=(7.07\pm 0.073) \times 10^{-3},\\
&|\Delta m_{\rm atm}^2|
=(2.431-2.598)\times10^{-21}\ {\rm eV^2}\,\,\ {\rm for}\ {\rm NH},\\
&|\Delta m_{\rm atm}^2|
=(2.412-2.583)\times10^{-21}\ {\rm eV^2}\,\,\ {\rm for}\ {\rm IH},
\end{align}
where the charged-lepton masses are within 1$\sigma$ region, while $\Delta m_{\rm atm}^2$ is within 3$\sigma$ region.
Here, the lepton masses are  defined by $m_\ell=y_\ell v_H$ with $v_H=174$ GeV.
Then, we pick the output data up only when the $\chi$ square is within 5$\sigma $ considering five measured neutrino oscillation data; $(\Delta m^2_{\rm sol},\ \sin^2\theta_{13},\ \sin^2\theta_{23},\ \sin^2\theta_{12}$)~\cite{Esteban:2018azc}.
Here, we do not include $\delta_{\rm CP}$ in the $\chi$ square analysis due to the large ambiguity of experimental results in $3\sigma$ interval.  
In general, IH case is more difficult to accumulate more data to satisfy the neutrino oscillation data, because the minimum $\chi$ square is $2.7$ in Nufit 5.0~\cite{Esteban:2018azc}.

\subsubsection{Nearby $\tau = i$}

\begin{figure}[H]
\begin{minipage}{0.49\hsize}
  \begin{center}
  \includegraphics[scale=0.25]{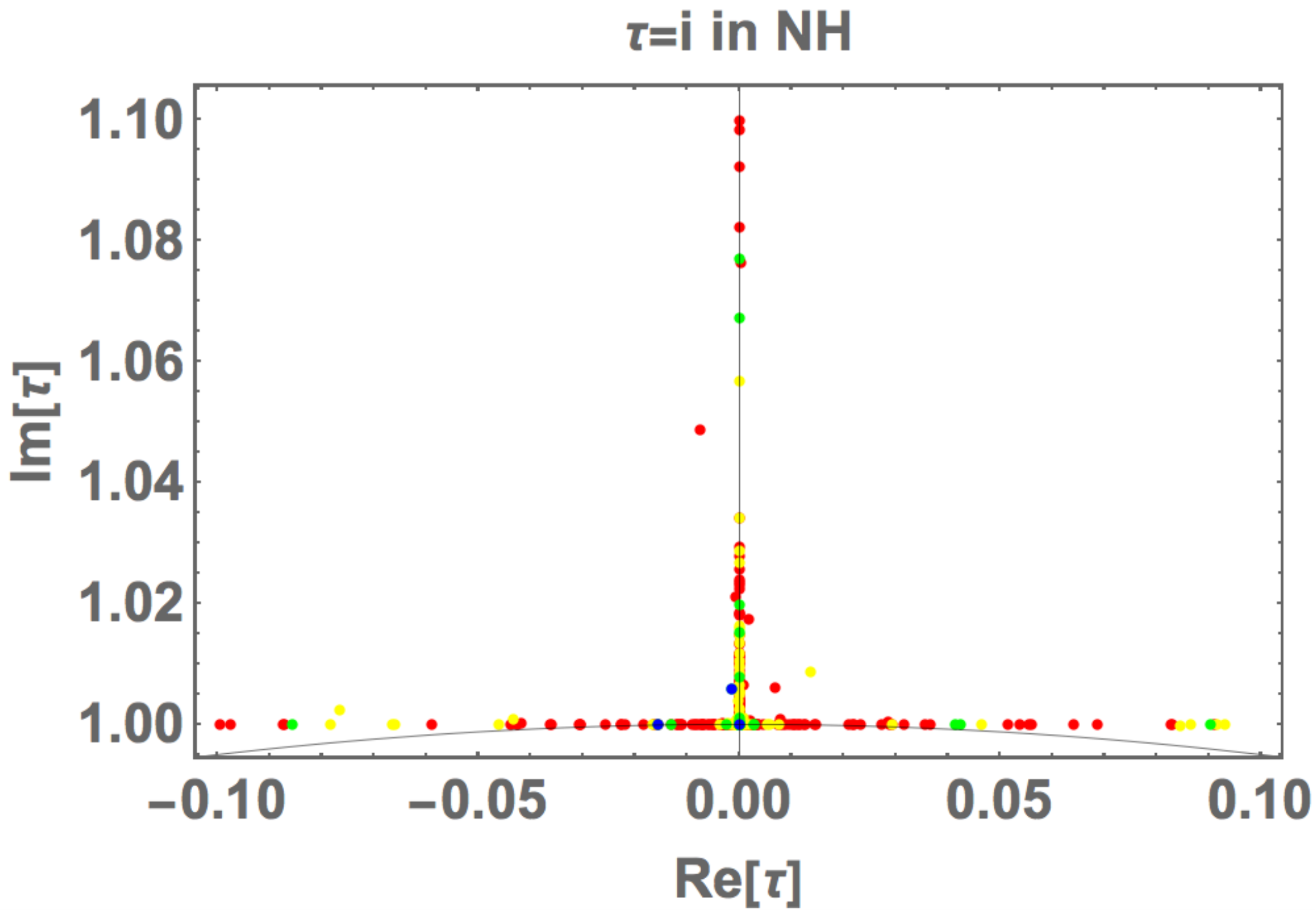}
  \end{center}
 \end{minipage}
\begin{minipage}{0.49\hsize}
  \begin{center}
  \includegraphics[scale=0.3]{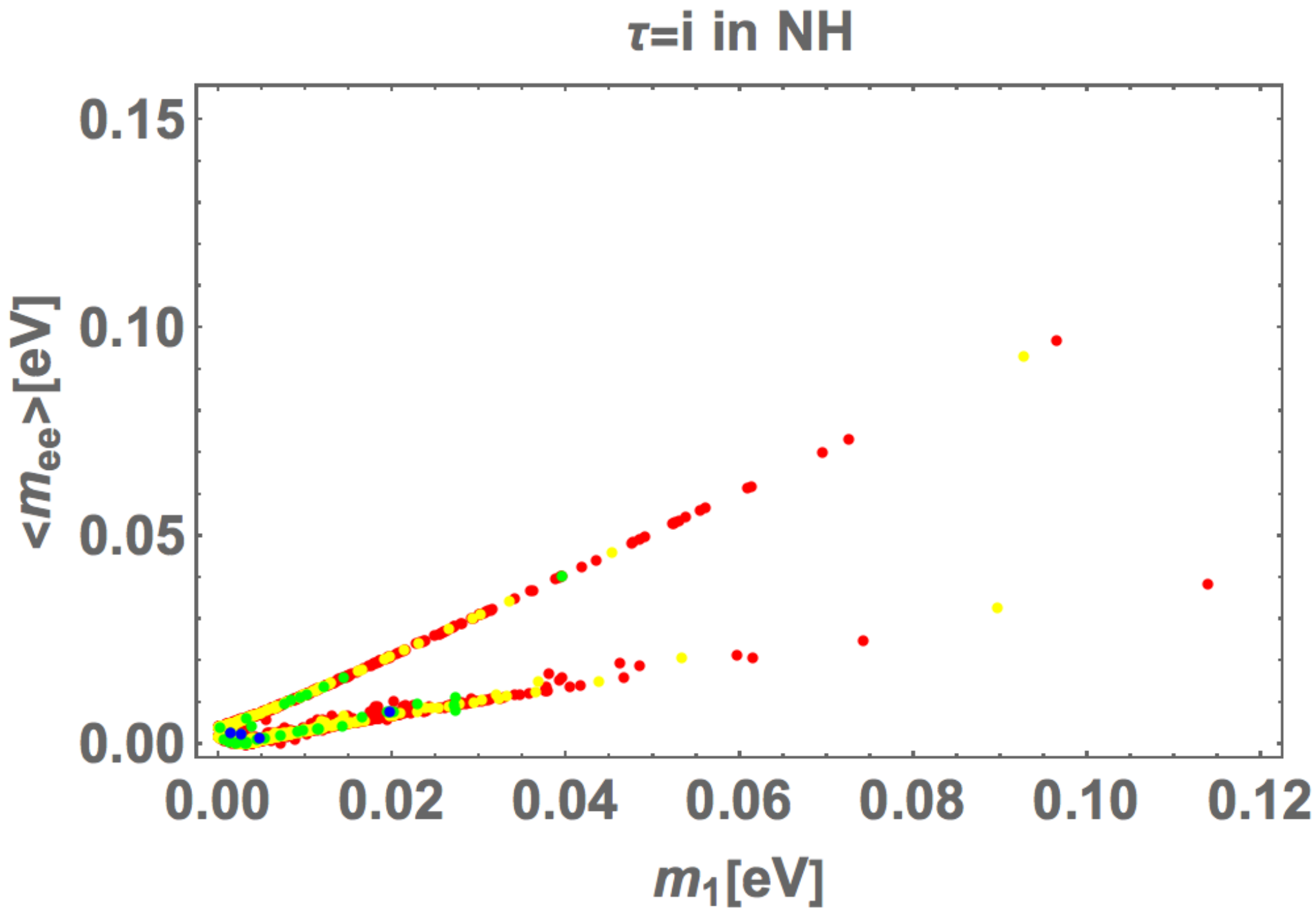}
  \end{center}
 \end{minipage}
\begin{minipage}{0.49\hsize}
  \begin{center}
  \includegraphics[scale=0.3]{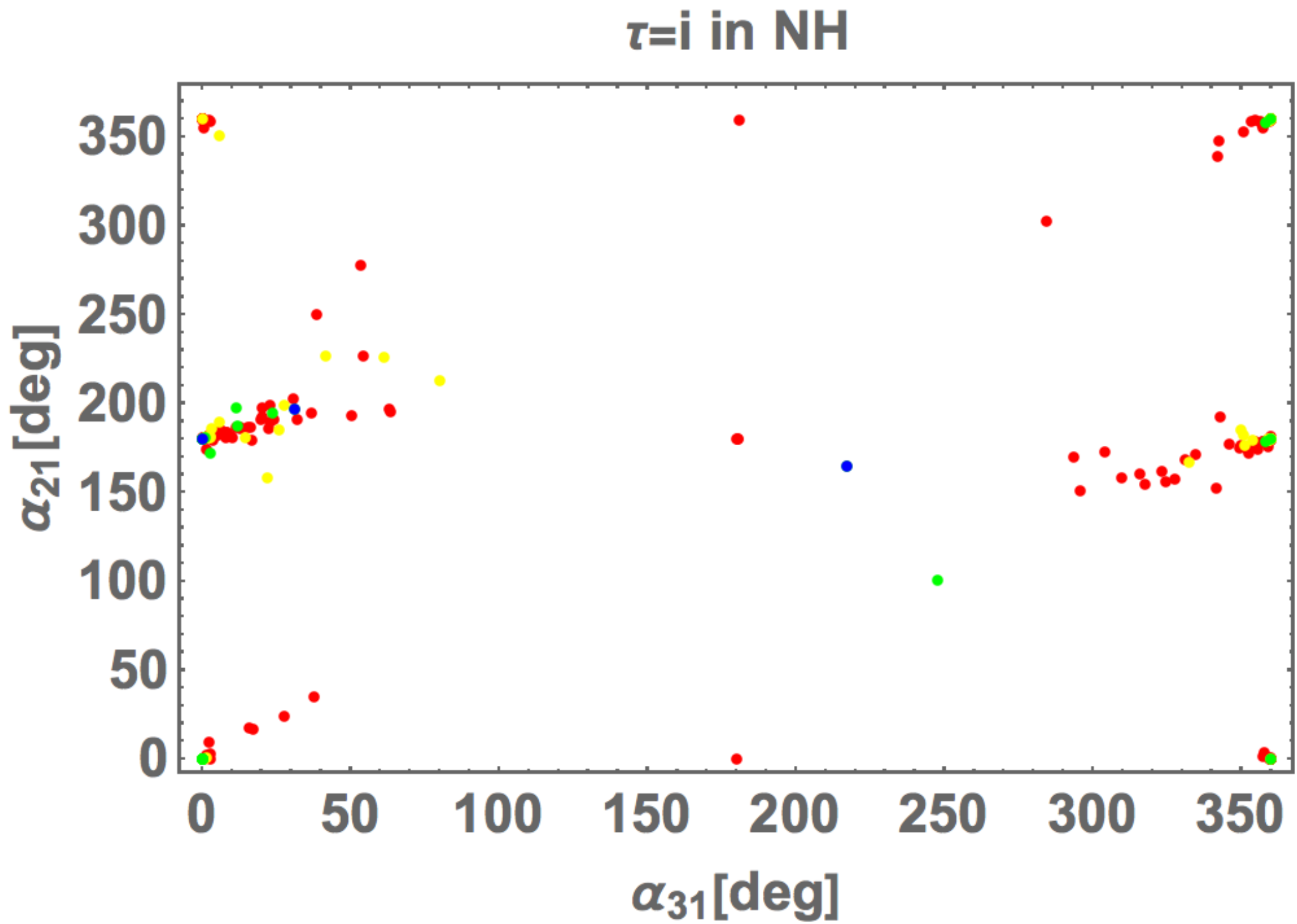}
  \end{center}
 \end{minipage}
 \begin{minipage}{0.49\hsize}
  \begin{center}
   \includegraphics[scale=0.3]{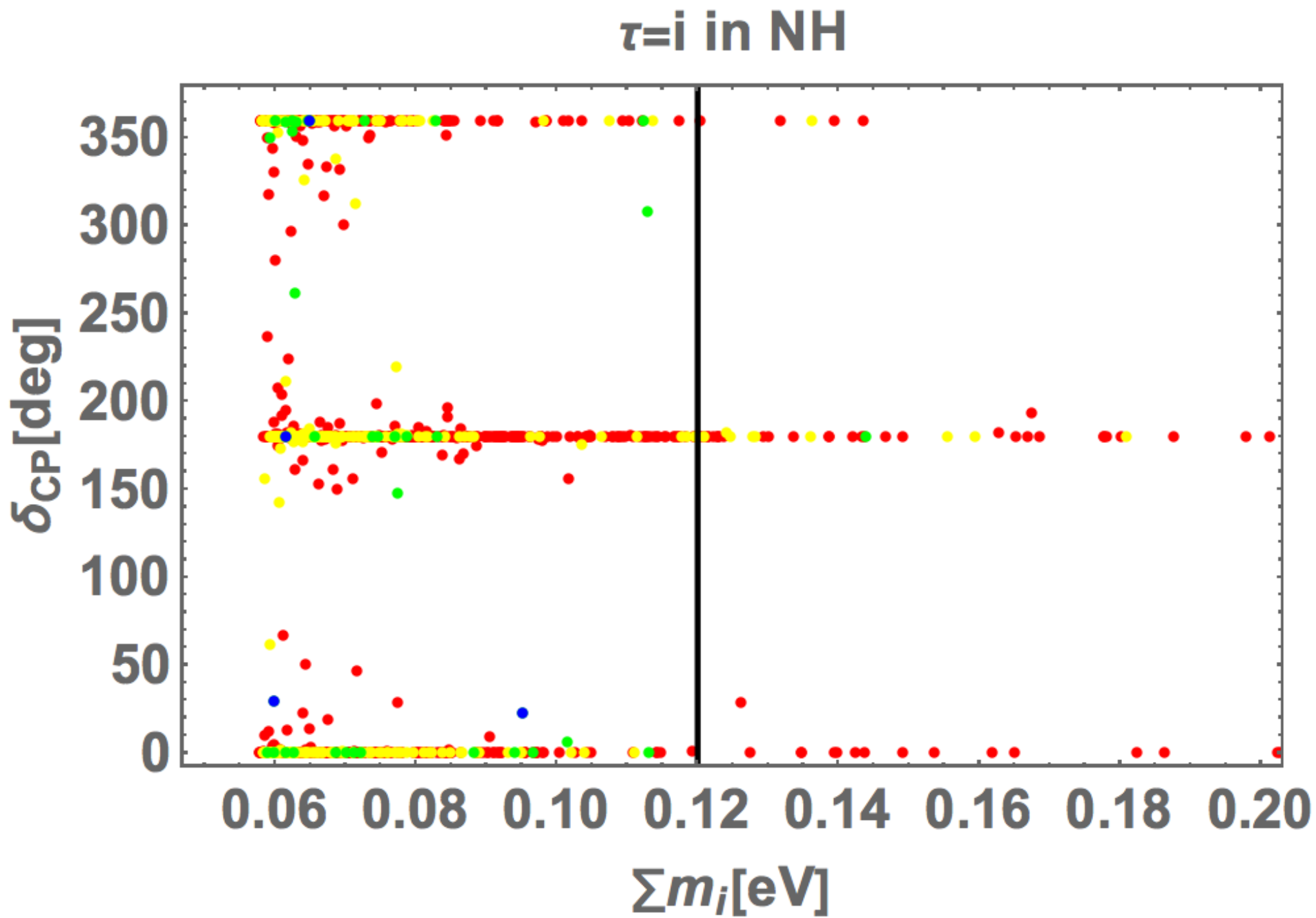}
  \end{center}
 \end{minipage}
 \caption{Each of color represents ${\rm blue} \le1\sigma$, $1\sigma< {\rm green}\le 2\sigma$, $2\sigma< {\rm yellow}\le 3\sigma$, $3\sigma<{\rm red}\le5\sigma$.}
\label{fig.chi-tau=i_nh}
\end{figure}
In Fig.~\ref{fig.chi-tau=i_nh}, we show our several allowed regions on $\tau$ at nearby $\tau=i$  in case of NH, where
each of color represents ${\rm blue} \le1\sigma$, $1\sigma< {\rm green}\le 2\sigma$, $2\sigma< {\rm yellow}\le 3\sigma$, $3\sigma<{\rm red}\le5\sigma$.
The up-left one represents the allowed region of the imaginary part of $\tau$ in terms of the real part of $\tau$. 
The up-right one demonstrates the allowed region of neutrinoless double beta decay $\langle m_{ee}\rangle$ in terms of the lightest active neutrino mass $m_1$.
There are two linear correlations between them. 
Furthermore, the smaller $\chi$ square is localized at nearby their smaller masses. 
The down-left one shows the allowed region of Majorana phases $\alpha_{21}$ and $\alpha_{31}$.
Since we take all input parameters (except $\tau$) to be 
real values, both the allowed regions are localized at nearby by $0^\circ,\ 180^\circ$.
The down-right one depicts the allowed region of Dirac phase $\delta_{\text{CP}}$ in terms of  the sum of neutrino masses $\sum m_i$.
The vertical line is the upper bound on cosmological constraint $\sum m_i < 0.12$ eV \cite{Planck:2018vyg}. There is an intriguing tendency that allowed region of smaller $\chi$ square is localized at smaller $\sum m_i$ that is within the cosmological bound.
Another feature is that the best fit value of Dirac CP phase $\sim195^\circ$ would be reproduced when we allow up to $5\sigma$ interval.

\begin{figure}[H]
\begin{minipage}{0.49\hsize}
  \begin{center}
  \includegraphics[scale=0.3]{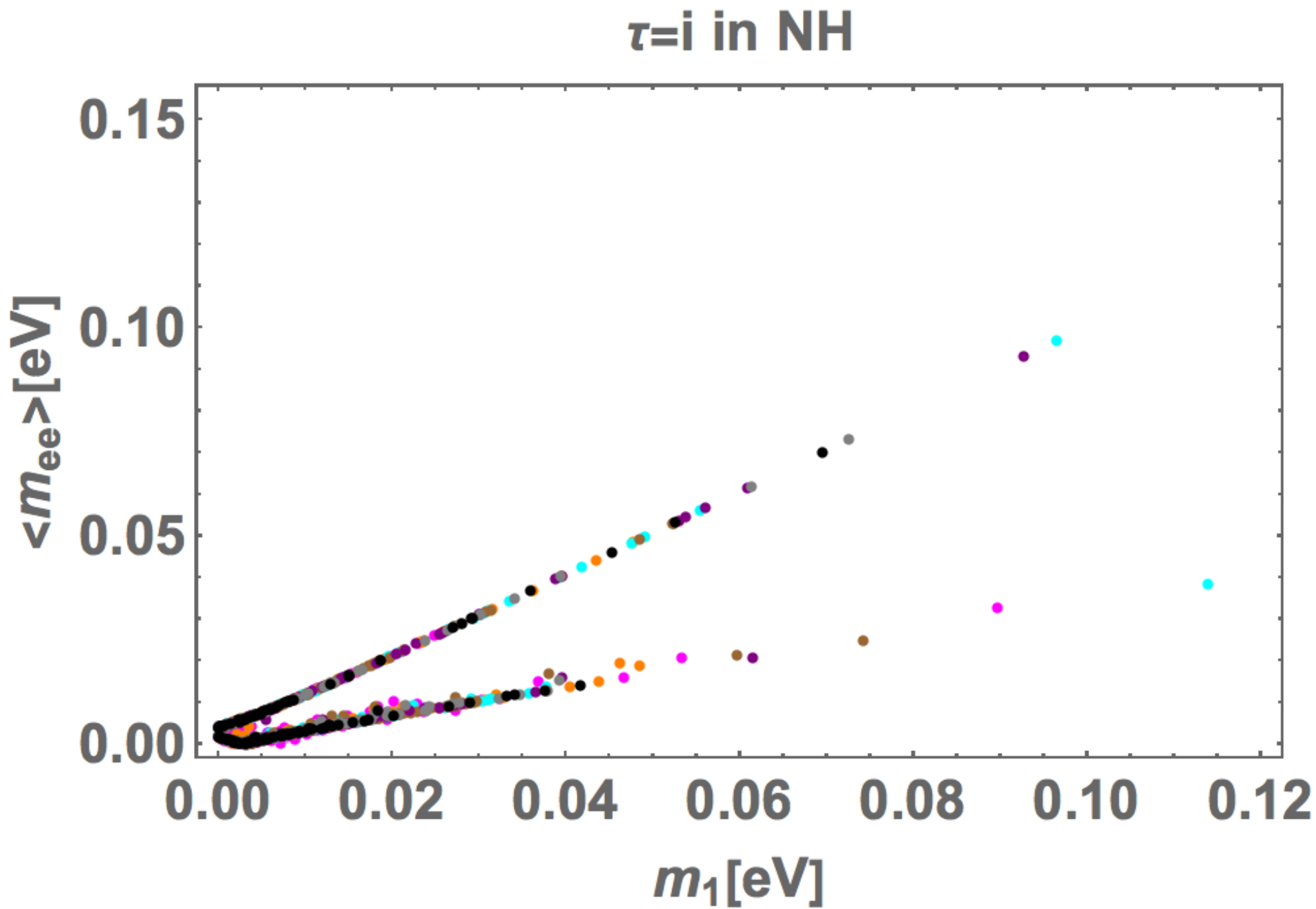}
  \end{center}
 \end{minipage}
\begin{minipage}{0.49\hsize}
  \begin{center}
  \includegraphics[scale=0.3]{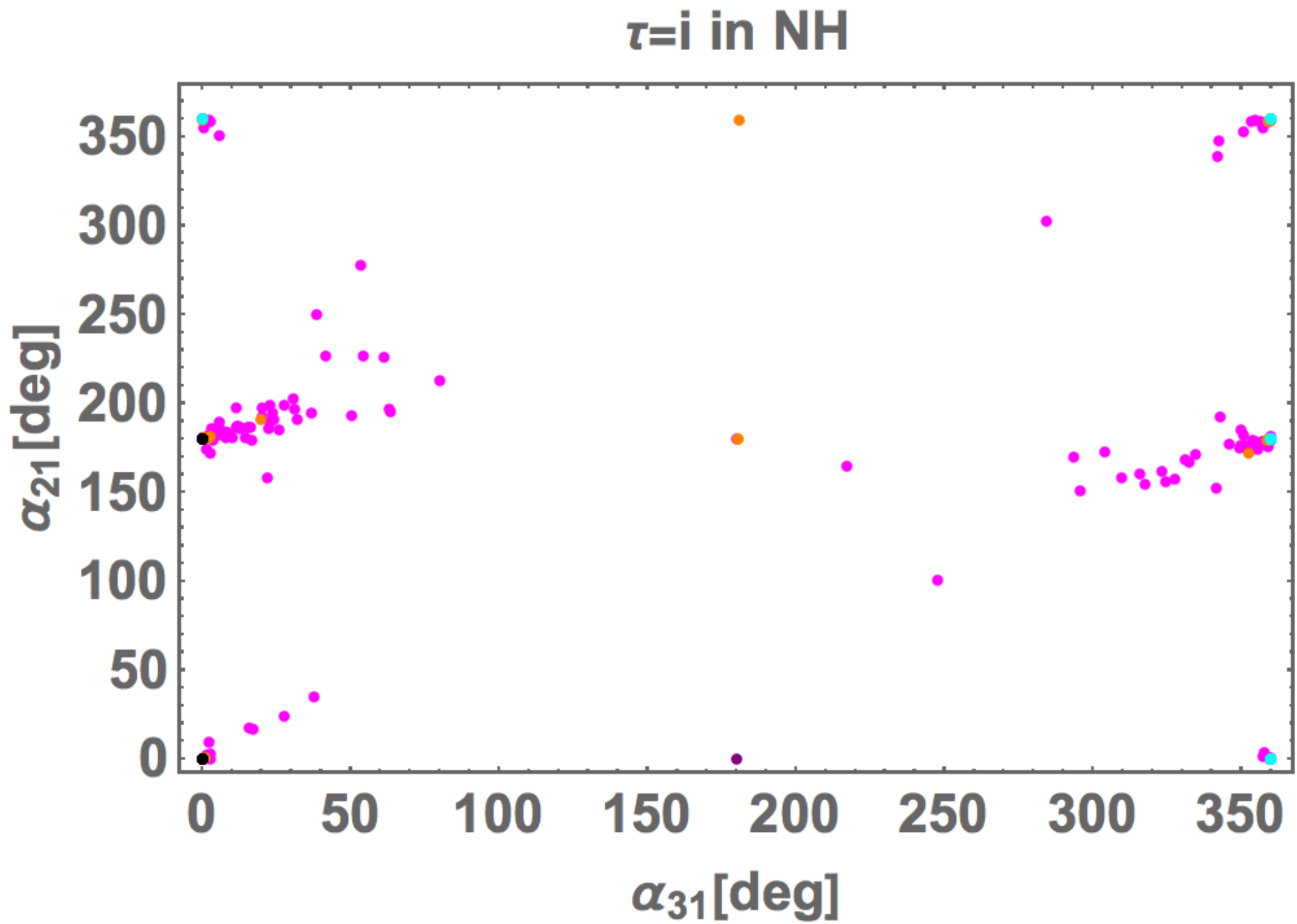}
  \end{center}
 \end{minipage}
\begin{minipage}{0.49\hsize}
  \begin{center}
  \includegraphics[scale=0.3]{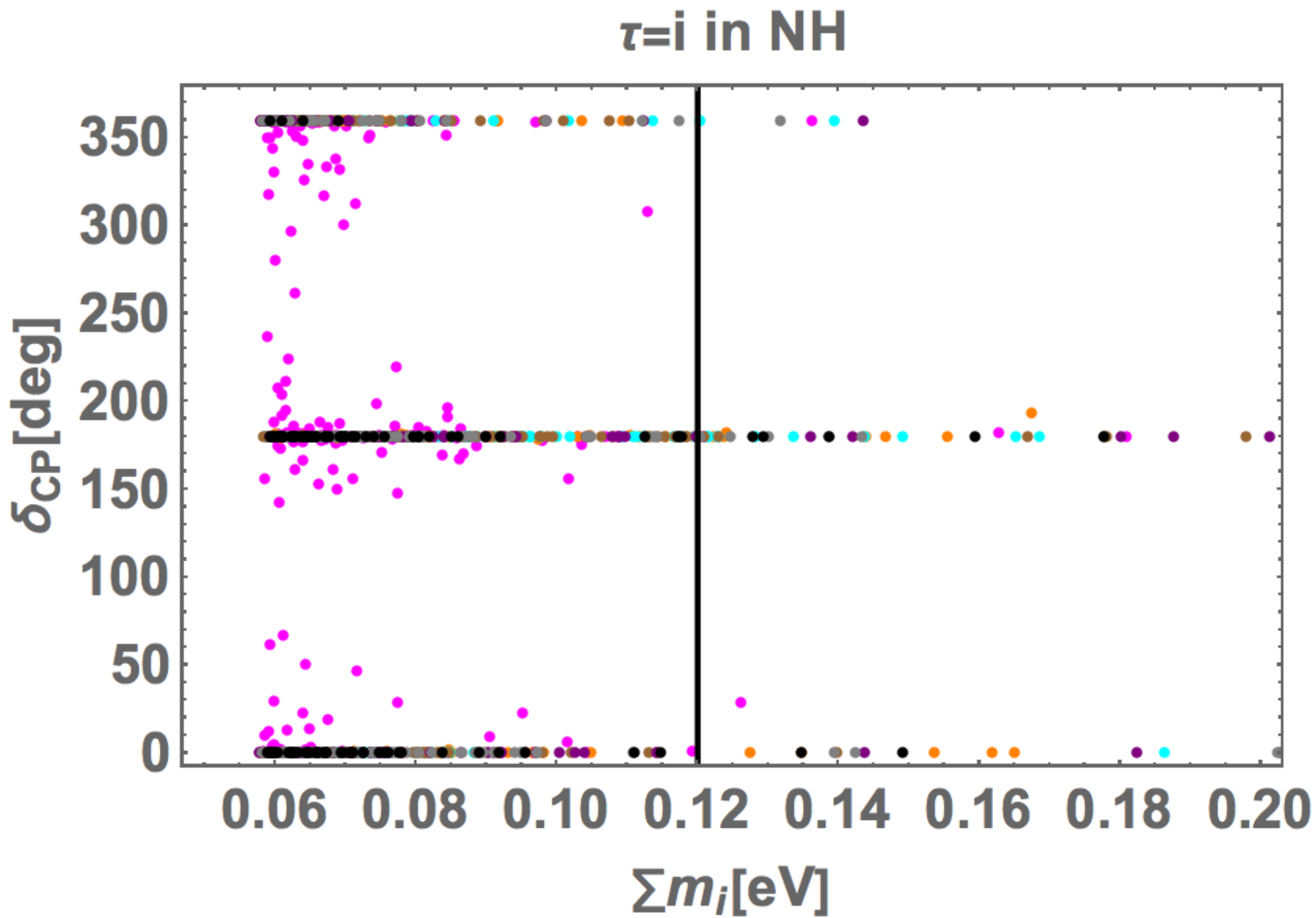}
  \end{center}
 \end{minipage}
 \caption{$|\delta \tau| < 10^{-15}$ for black, $10^{-15} \leq |\delta \tau| < 10^{-12}$ for gray, $10^{-12} \leq |\delta \tau| < 10^{-10}$ for purple, $10^{-10} \leq |\delta \tau| < 10^{-7}$ for brown, $10^{-7} \leq |\delta \tau| < 10^{-5}$ for blue green, $10^{-5} \leq |\delta \tau| < 10^{-3}$ for orange, and $10^{-3} \leq |\delta \tau| < 10^{-1}$ for magenta.}
 \label{fig.dev-tau=i_nh}
\end{figure}
In Fig.~\ref{fig.dev-tau=i_nh}, we show the several figures in terms of deviation from $\tau=i$ in the same case of Fig.~\ref{fig.chi-tau=i_nh} at 5 $\sigma$ interval, where
each of color represents $|\delta \tau| < 10^{-15}$ for black, $10^{-15} \leq |\delta \tau| < 10^{-12}$ for gray, $10^{-12} \leq |\delta \tau| < 10^{-10}$ for purple, $10^{-10} \leq |\delta \tau| < 10^{-7}$ for brown, $10^{-7} \leq |\delta \tau| < 10^{-5}$ for blue green, $10^{-5} \leq |\delta \tau| < 10^{-3}$ for orange, and $10^{-3} \leq |\delta \tau| < 10^{-1}$ for magenta.
The up-left one is the same as the case of up-right one in Fig.~\ref{fig.chi-tau=i_nh}.
It implies that smaller deviations $|\delta\tau|$ tend to be localized at nearby their smaller masses.
The up-right one is the same as the case of down-left one in Fig.~\ref{fig.chi-tau=i_nh}.
This figure would show rather trivial.
Therefore, the smaller deviation is localized at $0^\circ$ and $180^\circ$, while the larger deviation deviates from these two points. It directly follows from our phase source is $\tau$ only. 
The down-left one is the same as the case of down-right one in Fig.~\ref{fig.chi-tau=i_nh}.
The smaller deviation would be favored in the point of view of the bound on cosmological constraint.

\begin{figure}[H]
\begin{minipage}{0.49\hsize}
  \begin{center}
  \includegraphics[scale=0.3]{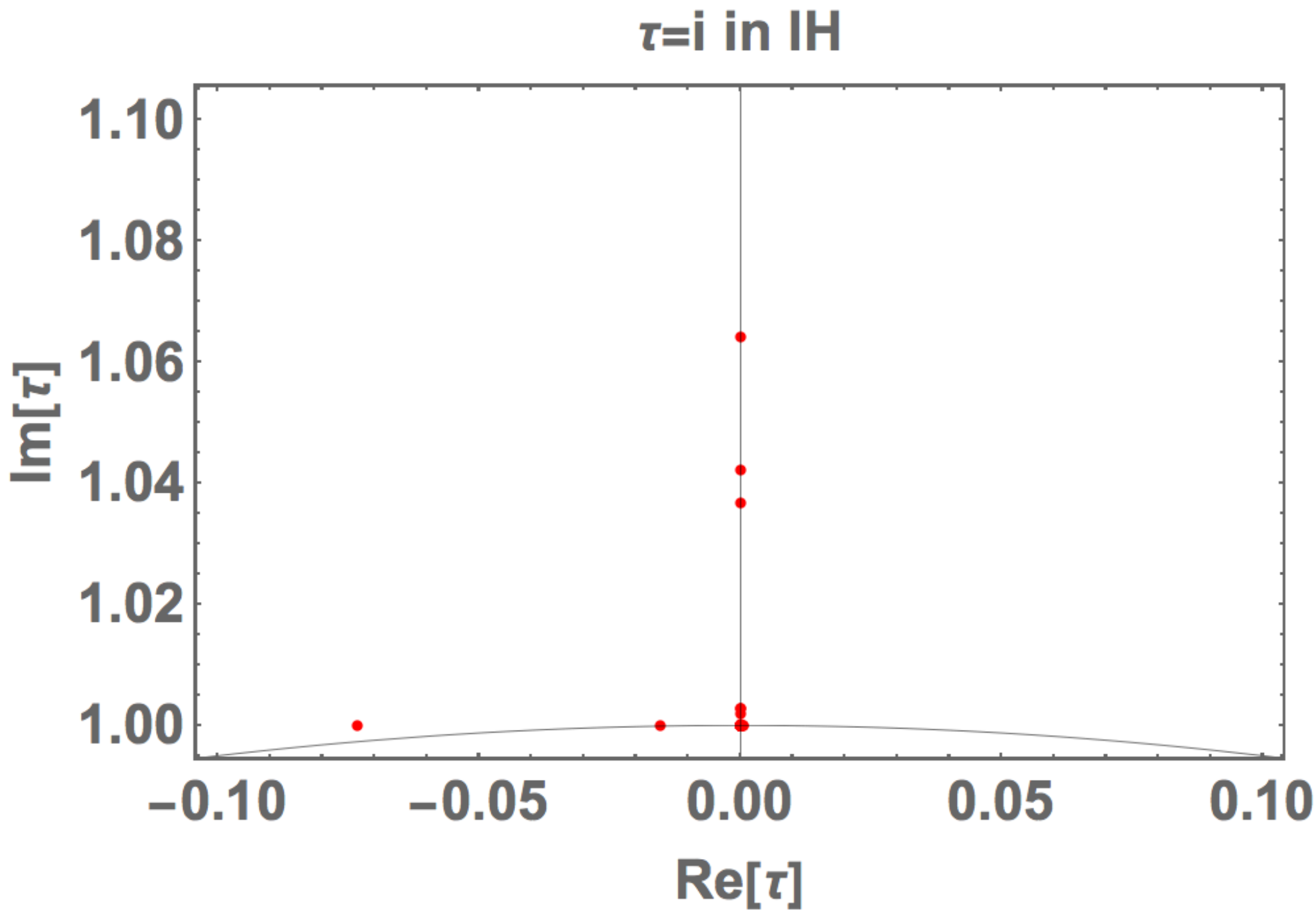}
  \end{center}
 \end{minipage}
\begin{minipage}{0.49\hsize}
  \begin{center}
  \includegraphics[scale=0.3]{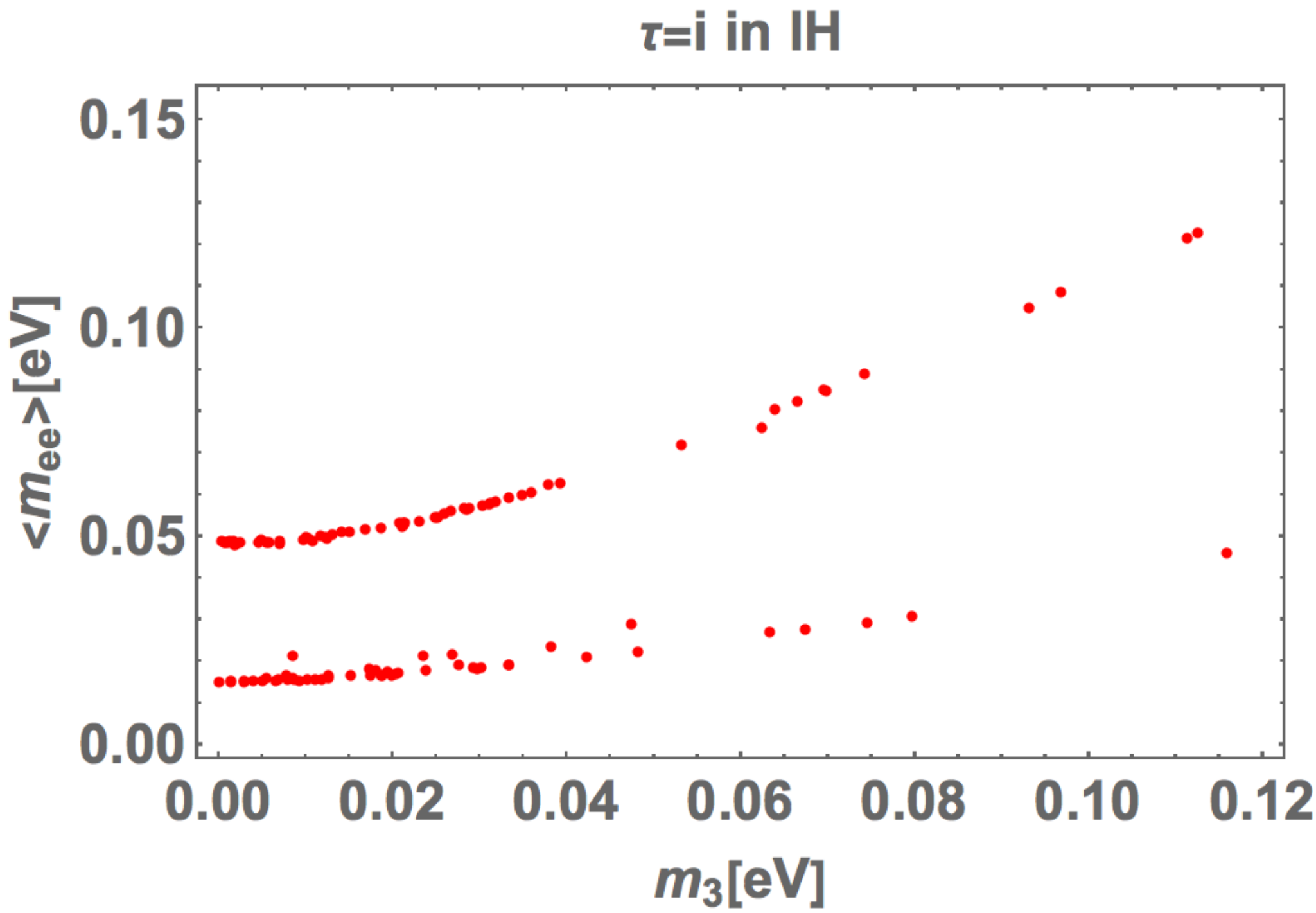}
  \end{center}
 \end{minipage}
\begin{minipage}{0.49\hsize}
  \begin{center}
  \includegraphics[scale=0.3]{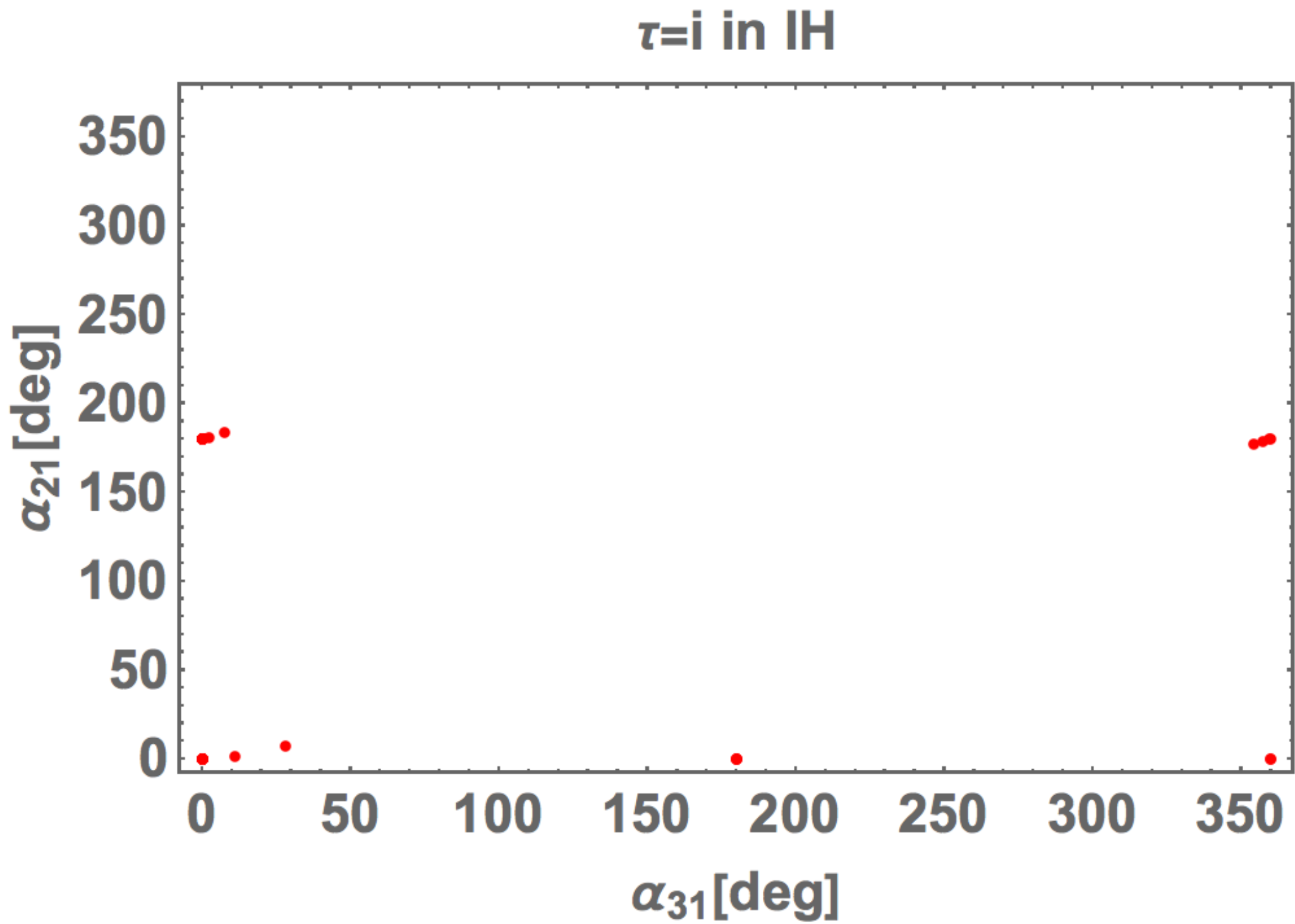}
  \end{center}
 \end{minipage}
 \begin{minipage}{0.49\hsize}
  \begin{center}
   \includegraphics[scale=0.3]{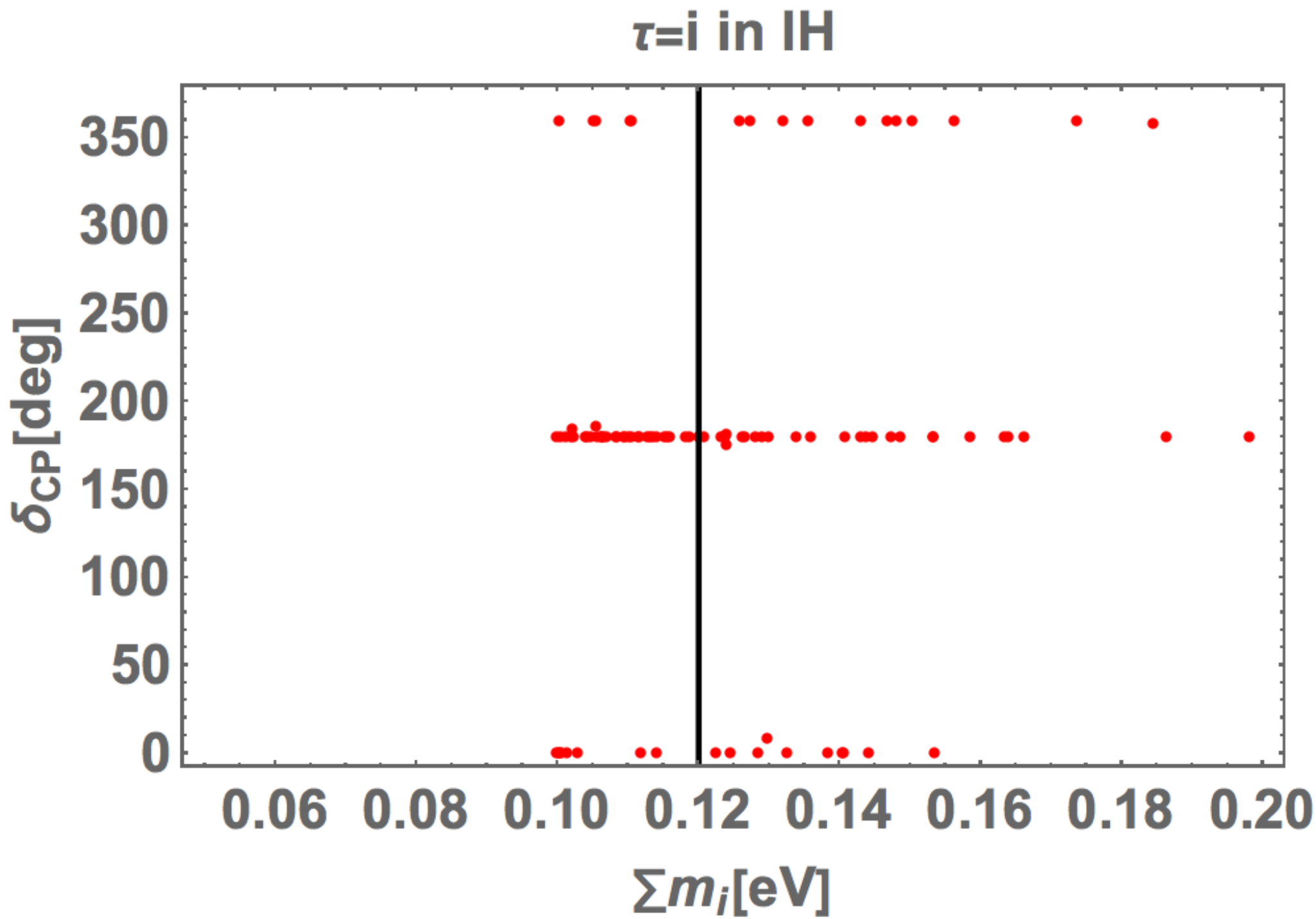}
  \end{center}
 \end{minipage}
 \caption{Each of color represents ${\rm blue} \le1\sigma$, $1\sigma< {\rm green}\le 2\sigma$, $2\sigma< {\rm yellow}\le 3\sigma$, $3\sigma<{\rm red}\le5\sigma$.}
\label{fig.chi-tau=i_ih}
\end{figure}
In Fig.~\ref{fig.chi-tau=i_ih}, we show our several allowed regions on $\tau$ at nearby $\tau=i$ in case of IH, where color legends are the same as the one of Fig.~\ref{fig.chi-tau=i_nh}. Therefore, we have found only the allowed region of $3\sigma-5\sigma$.
The up-left one represents the allowed region of imaginary part of $\tau$ in terms of real part of $\tau$. 
The up-right one demonstrates the allowed region of neutrinoless double beta decay $\langle m_{ee}\rangle$ in terms of the lightest active neutrino mass $m_3$.
There are two correlations between them; one is a linear line and another is a slightly curved one.
The solutions tend to be localized at nearby smaller mass of $m_3$ with $\langle m_{ee}\rangle=0.015,\ 0.05$ eV.
The down-left one shows the allowed region of Majorana phases $\alpha_{21}$ and $\alpha_{31}$.
Both the allowed regions are localized at nearby by $0^\circ,\ 180^\circ$ similar to the case of NH.
The down-right one depicts the allowed region of the sum of neutrino masses $\sum m_i$ in terms of Dirac phase $\delta_{\text{CP}}$.
The vertical line is the upper bound on cosmological constraint. $\delta_{\rm CP}$ is allowed at the points $0^\circ$ and $180^\circ$. While a large part of $\sum m_i$
would be ruled out by the cosmological bound.
Therefore, we would predict a narrow range of $0.1~{\rm eV}\le\sum m_i\le0.12~{\rm eV}$ in this case.

\begin{figure}[H]
\begin{minipage}{0.49\hsize}
  \begin{center}
  \includegraphics[scale=0.3]{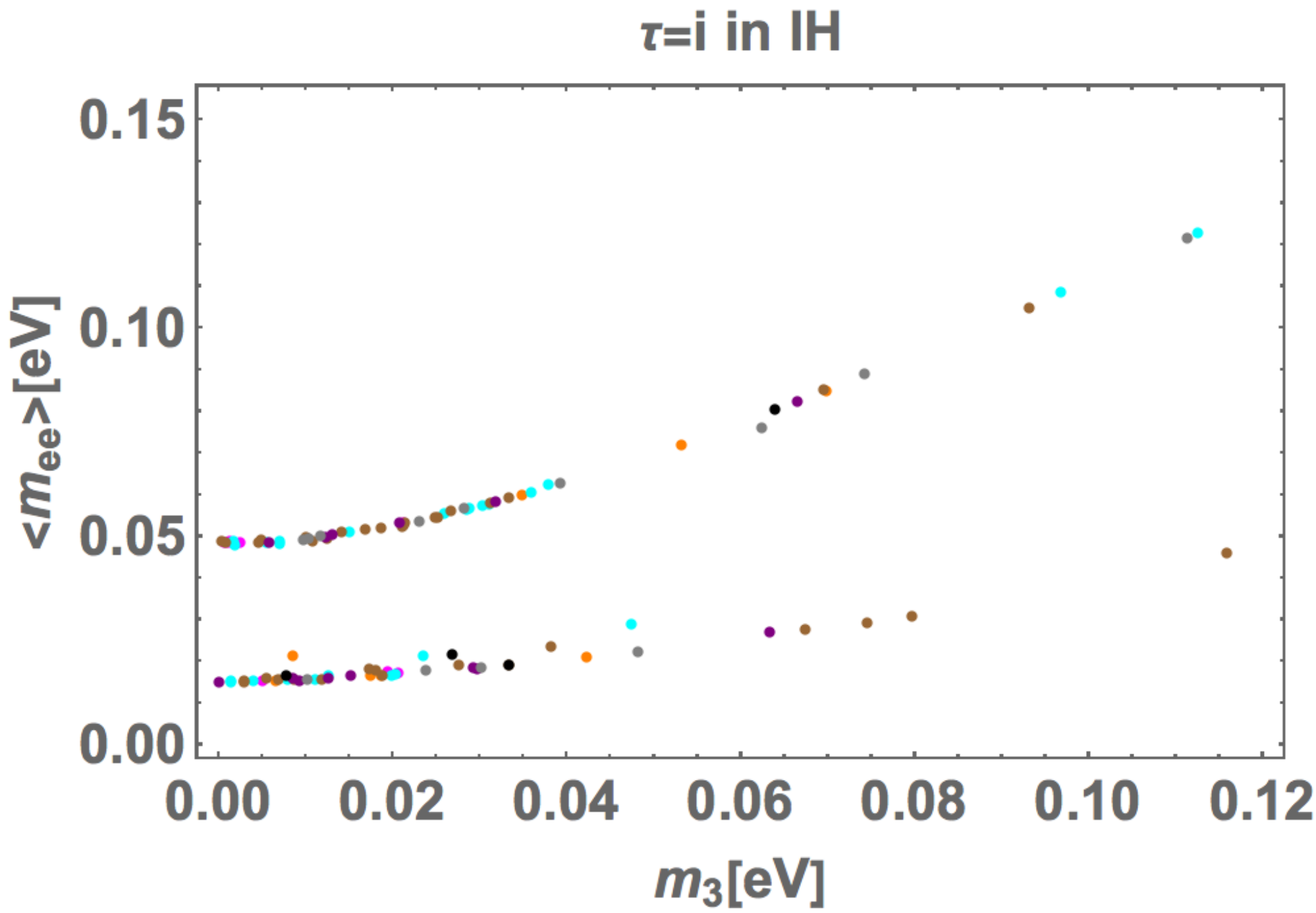}
  \end{center}
 \end{minipage}
\begin{minipage}{0.49\hsize}
  \begin{center}
  \includegraphics[scale=0.3]{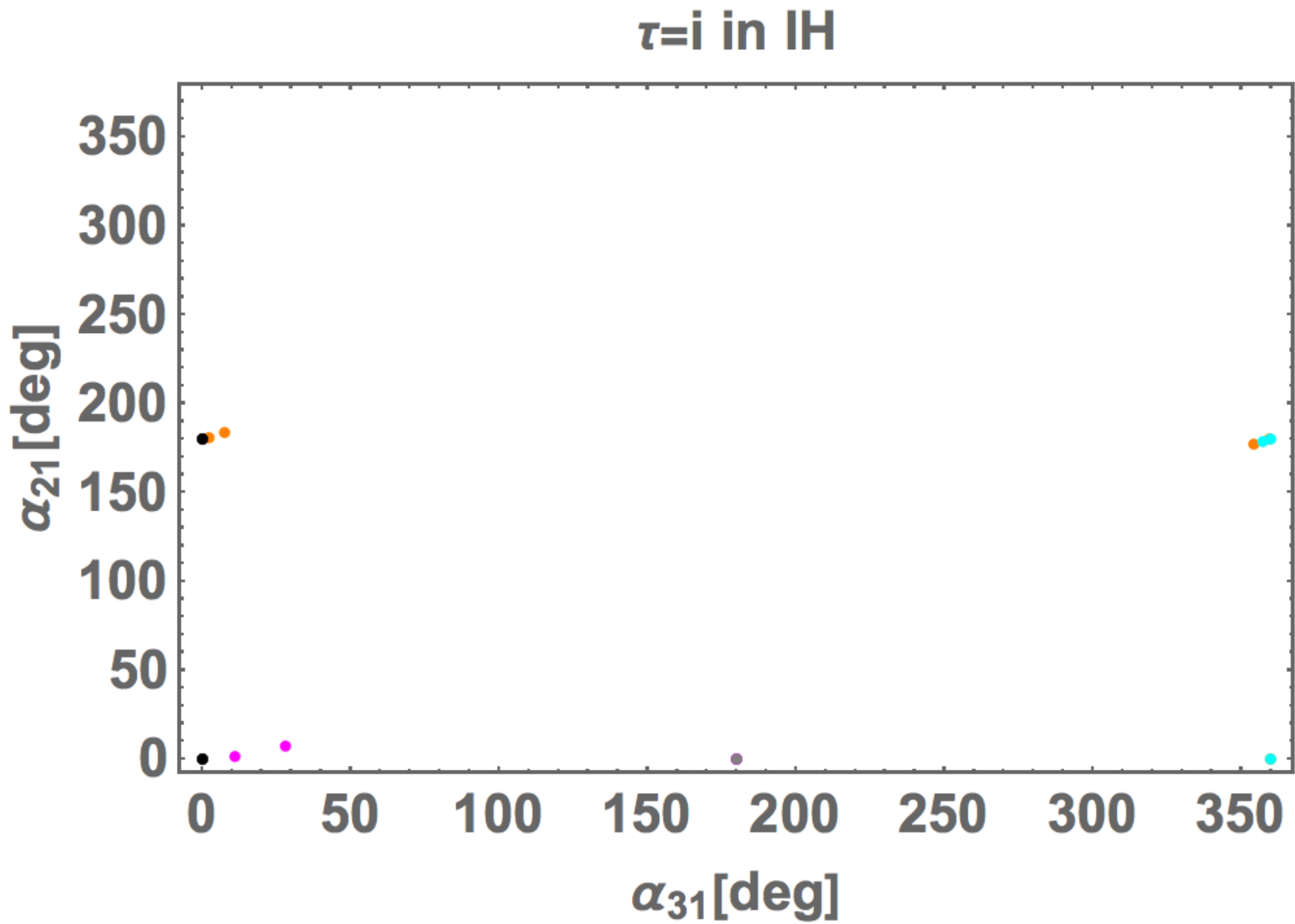}
  \end{center}
 \end{minipage}
\begin{minipage}{0.49\hsize}
  \begin{center}
  \includegraphics[scale=0.3]{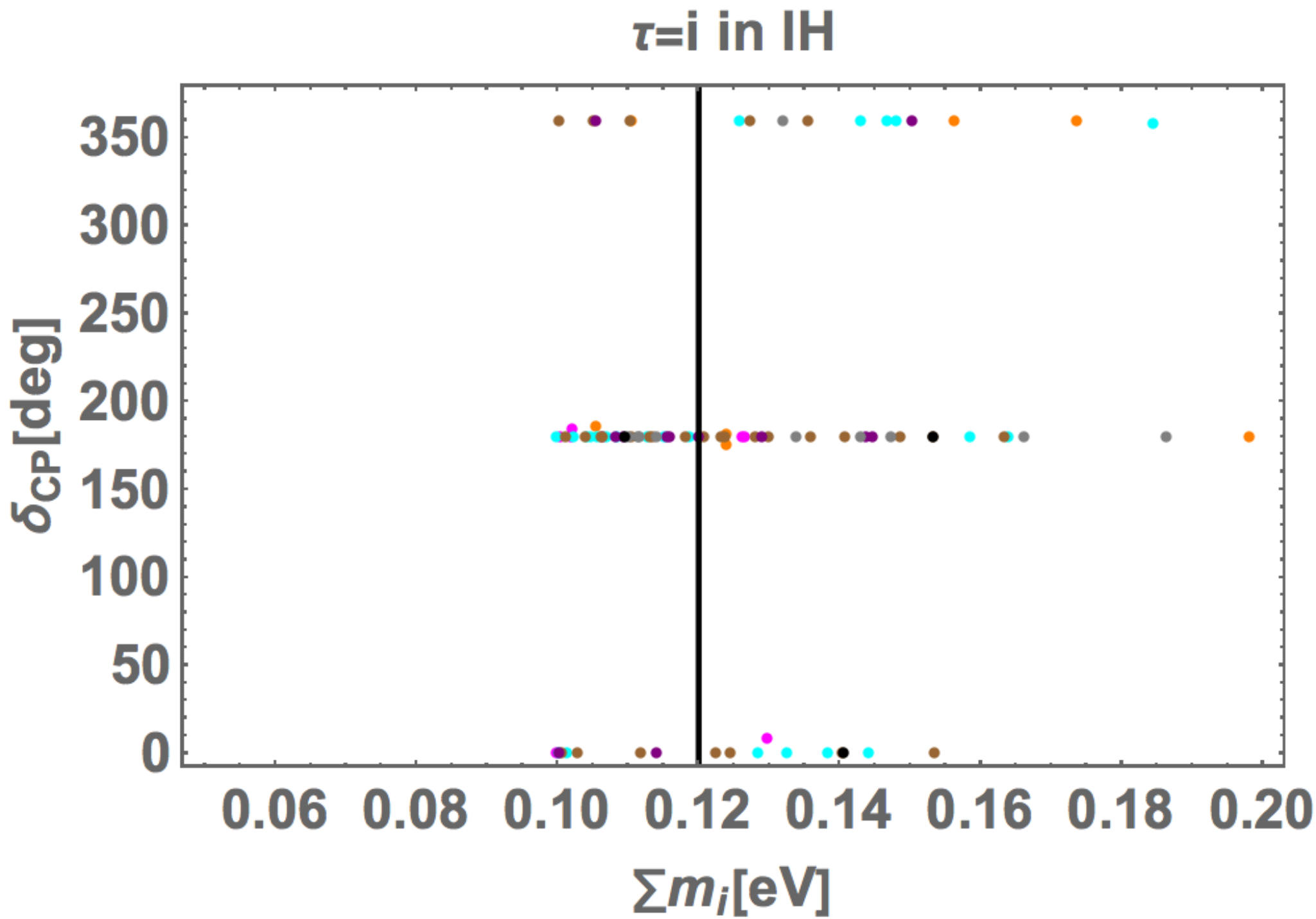}
  \end{center}
 \end{minipage}
 \caption{$|\delta \tau| < 10^{-15}$ for black, $10^{-15} \leq |\delta \tau| < 10^{-12}$ for gray, $10^{-12} \leq |\delta \tau| < 10^{-10}$ for purple, $10^{-10} \leq |\delta \tau| < 10^{-7}$ for brown, $10^{-7} \leq |\delta \tau| < 10^{-5}$ for blue green, $10^{-5} \leq |\delta \tau| < 10^{-3}$ for orange, and $10^{-3} \leq |\delta \tau| < 10^{-1}$ for magenta.}
 \label{fig.dev-tau=i_ih}
\end{figure}
%
In Fig.~\ref{fig.dev-tau=i_ih}, we show the several figures in terms of deviation from $\tau=i$ where the color legends are the same as the one in Fig.~\ref{fig.dev-tau=i_nh}.
The up-left one corresponds to the case of up-right one in Fig.~\ref{fig.chi-tau=i_ih}.
It implies that smaller deviations $|\delta\tau|$ tend to be localized at nearby their smaller masses $m_3$.
The up-right one corresponds to the case of down-left one in Fig.~\ref{fig.chi-tau=i_ih}.
This figure would show rather trivial.
Therefore, the smaller deviation is localized at $0^\circ$ and $180^\circ$, while the larger deviation deviates from these two points. It directly follows from the fact that our phase source is $\tau$ only. 
The down-left one corresponds to the case of down-right one in Fig.~\ref{fig.chi-tau=i_ih}.
The smaller deviation would be favored in the point of view of the bound on cosmological constraint. 

\begin{figure}[H]
\begin{minipage}{0.49\hsize}
  \begin{center}
  \includegraphics[scale=0.15]{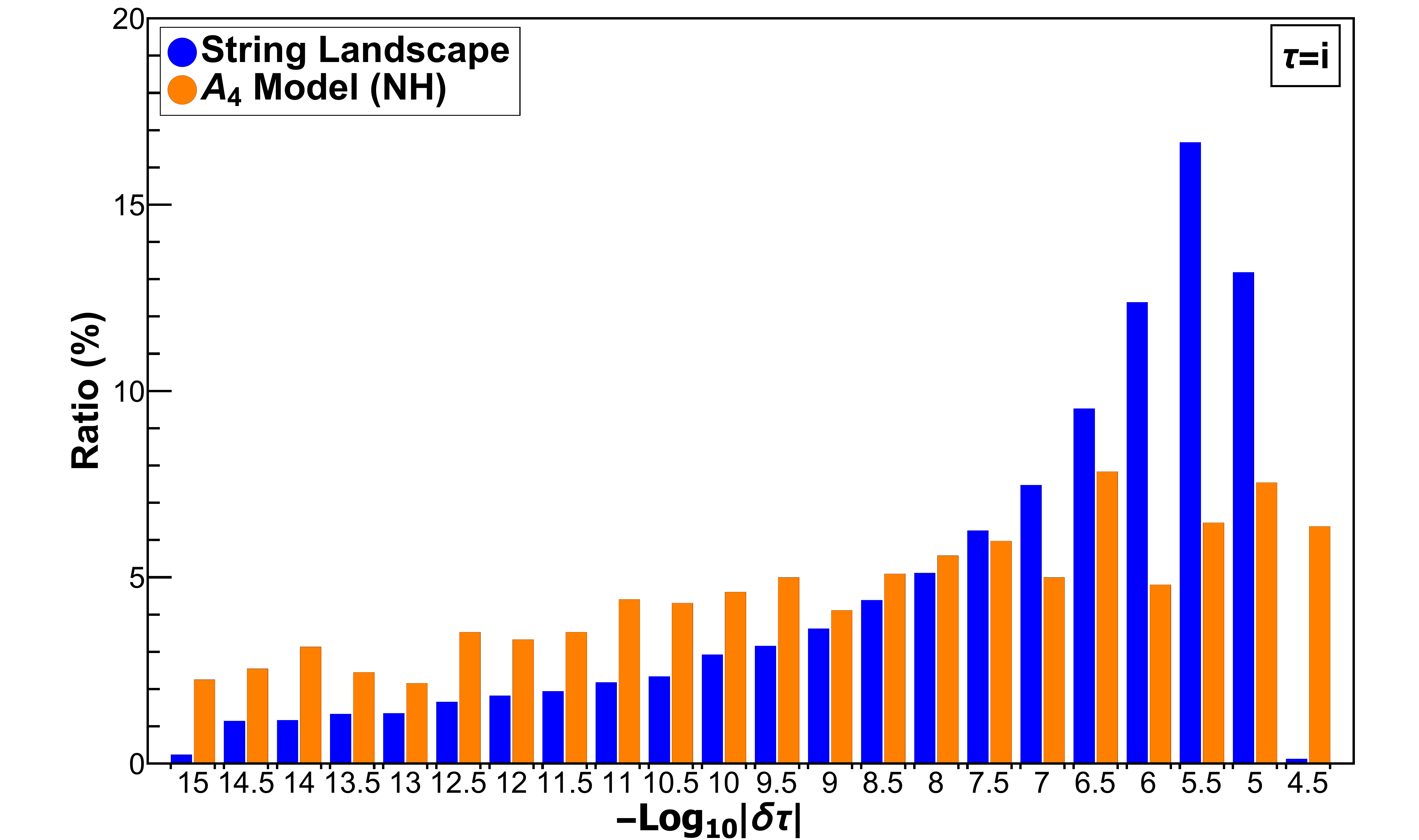}
  \end{center}
 \end{minipage}
\begin{minipage}{0.49\hsize}
  \begin{center}
  \includegraphics[scale=0.15]{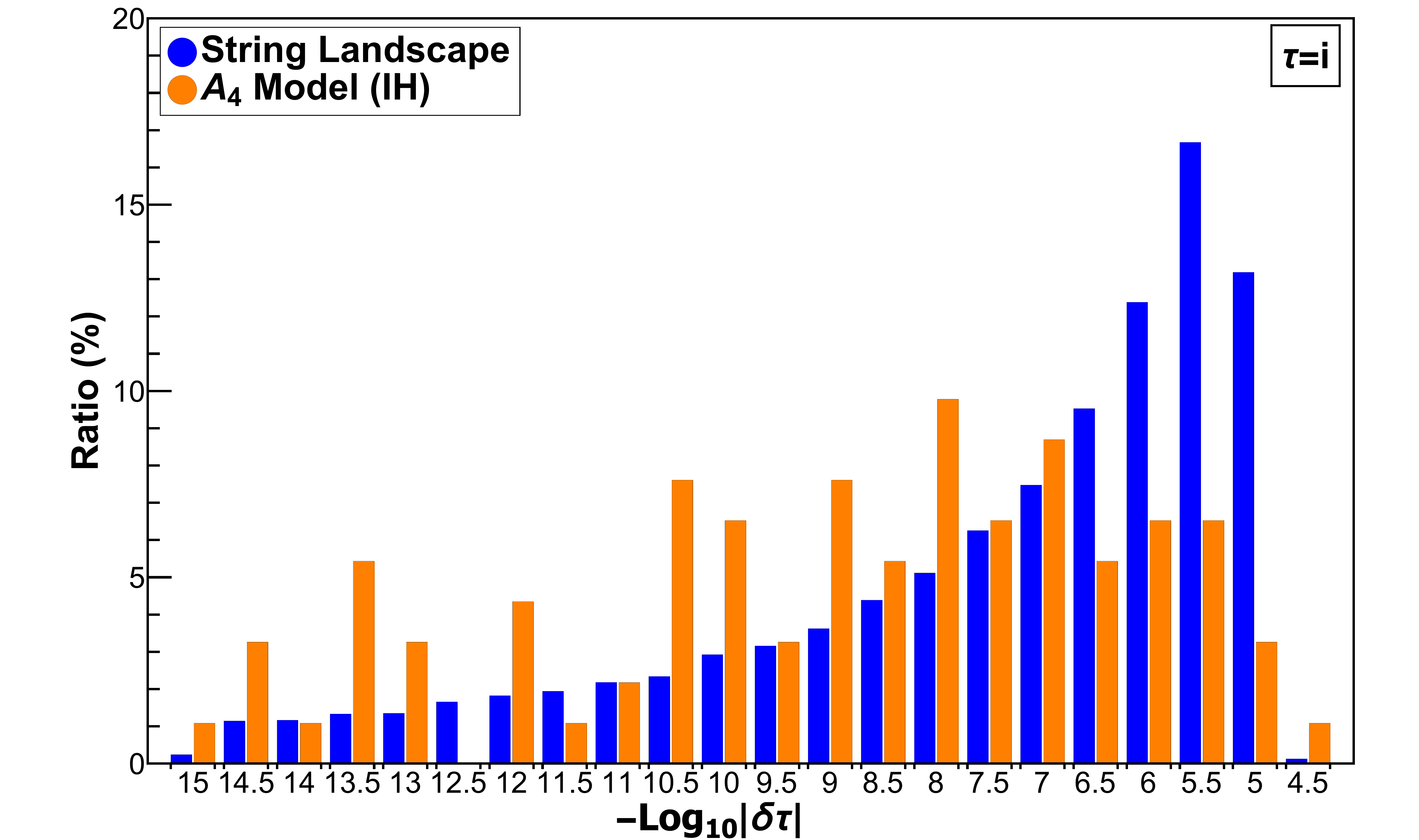}
  \end{center}
 \end{minipage}
 \caption{Ratios of the number of solutions in a corresponding range of $-{\rm Log}_{10} |\delta \tau|$ to the number of whole solutions for both the string landscape in Fig. \ref{fig:Z2_up} and the $A_4$ model within $5\sigma$. We present the NH and the IH in the left and right panels, respectively.}
  \label{fig.ratio-tau=i}
\end{figure}

Finally, we discuss ratios of the number of solutions in a corresponding range of $-{\rm Log}_{10} |\delta \tau|$ to the number of whole solutions for both the string landscape in Fig. \ref{fig:Z2_up} and the $A_4$ model within $5\sigma$. 
Fig. \ref{fig.ratio-tau=i} indicates both the distributions of $A_4$ model with NH and the moduli fields in the string landscape peak around $|\delta \tau|= {\cal O}(10^{-5})$, but such a signal will not be found in the IH case.

\subsubsection{Nearby $\tau = \omega$}

\begin{figure}[H]
\begin{minipage}{0.49\hsize}
  \begin{center}
  \includegraphics[scale=0.3]{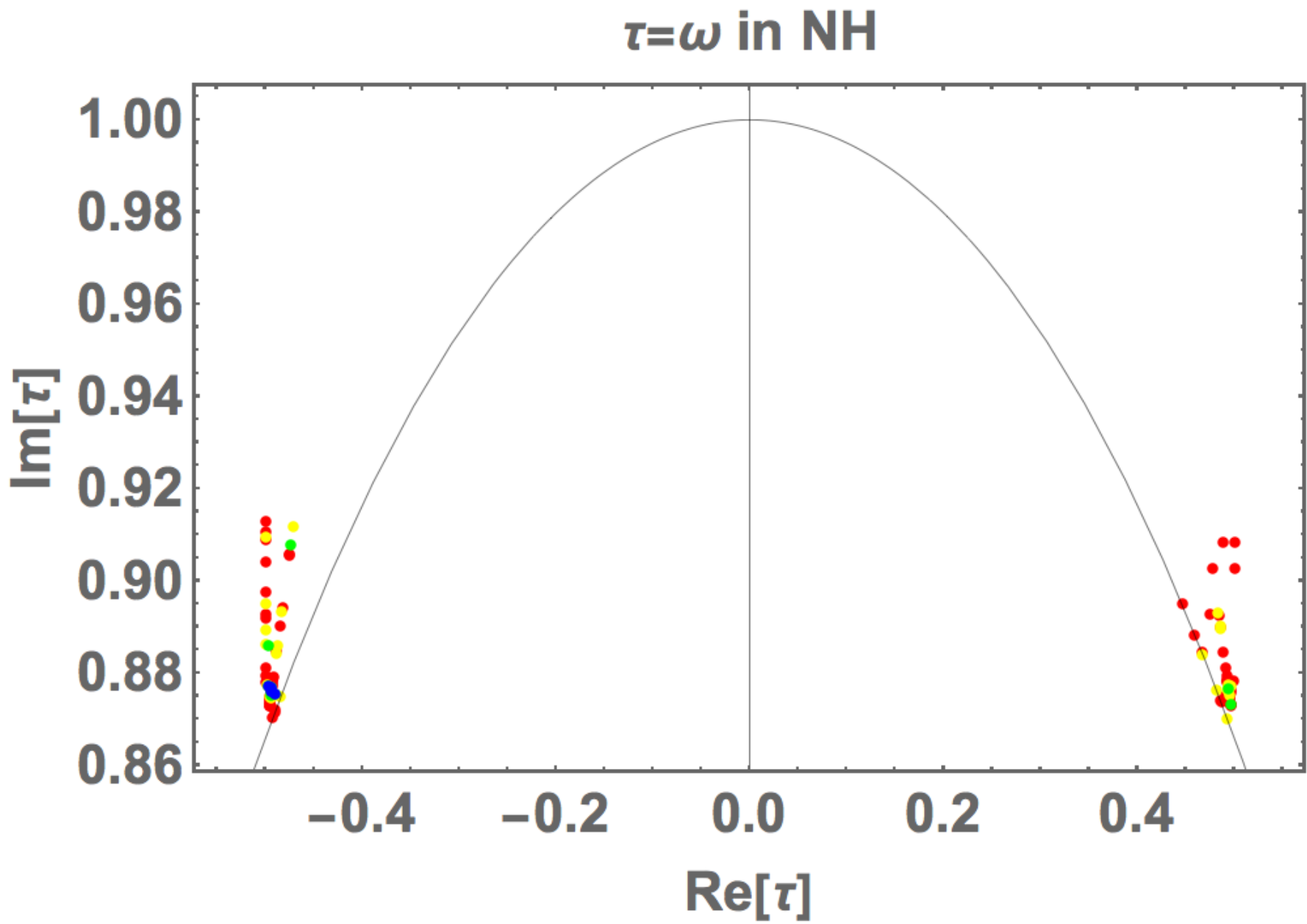}
  \end{center}
 \end{minipage}
\begin{minipage}{0.49\hsize}
  \begin{center}
  \includegraphics[scale=0.3]{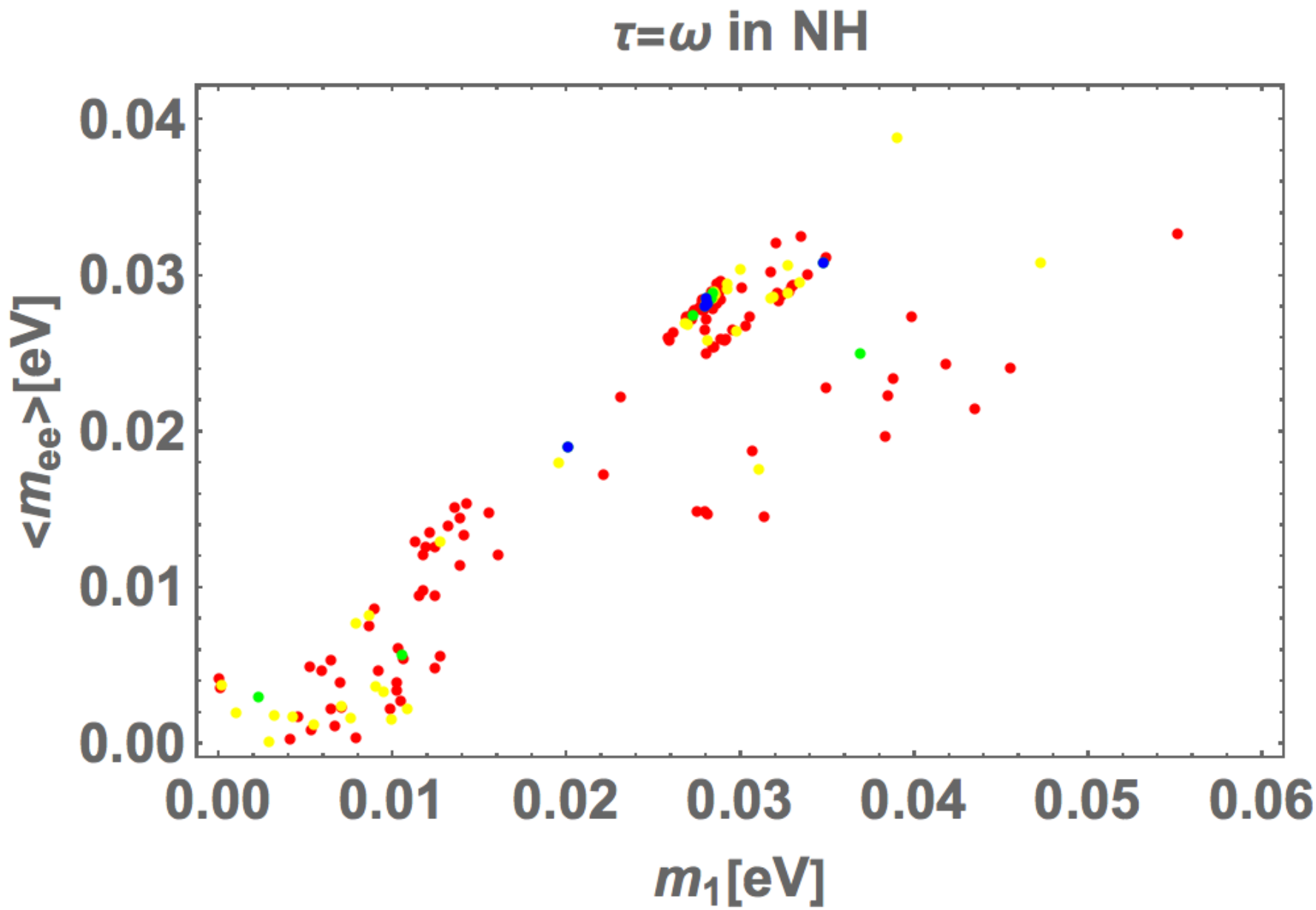}
  \end{center}
 \end{minipage}
\begin{minipage}{0.49\hsize}
  \begin{center}
  \includegraphics[scale=0.3]{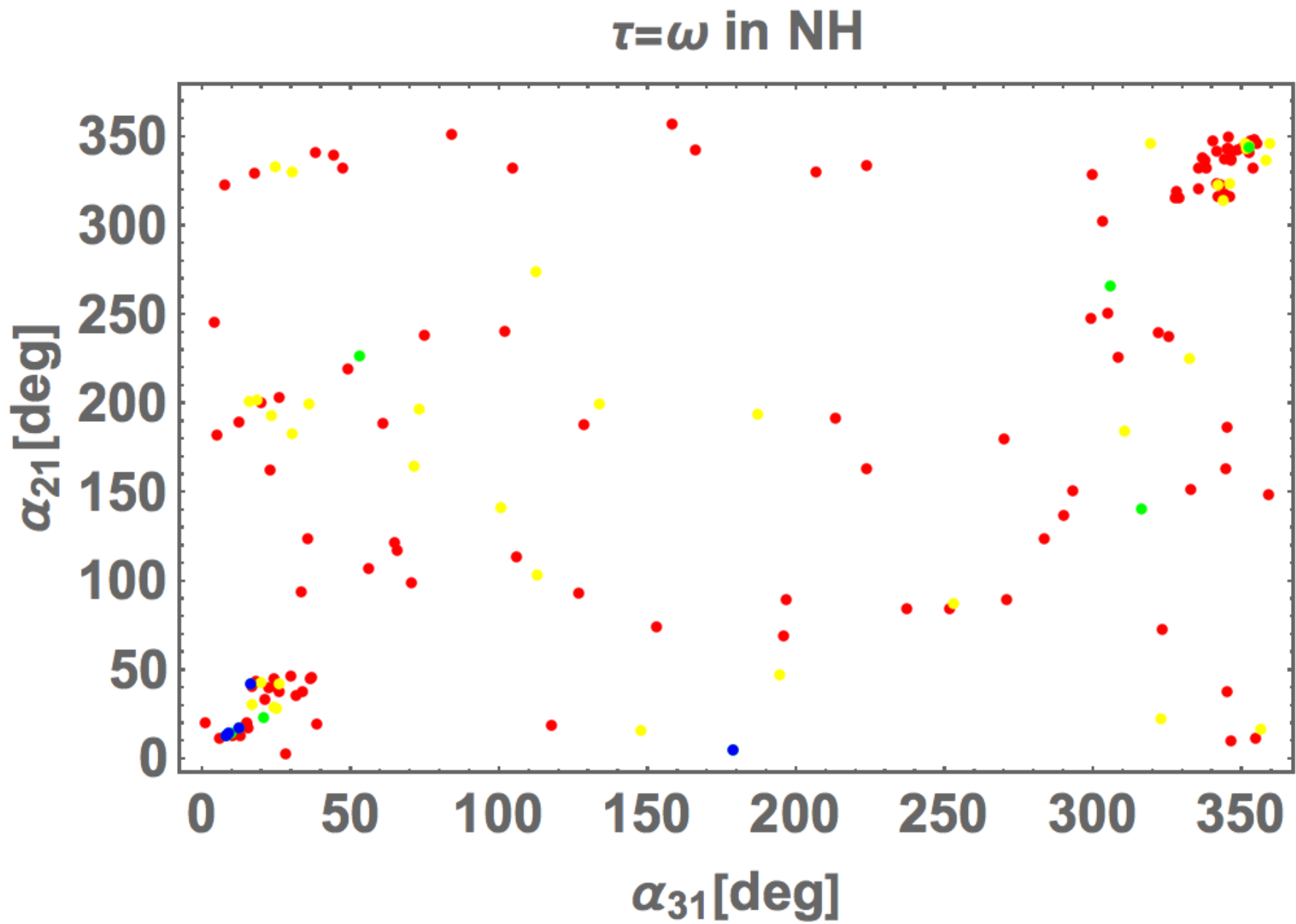}
  \end{center}
 \end{minipage}
 \begin{minipage}{0.49\hsize}
  \begin{center}
   \includegraphics[scale=0.3]{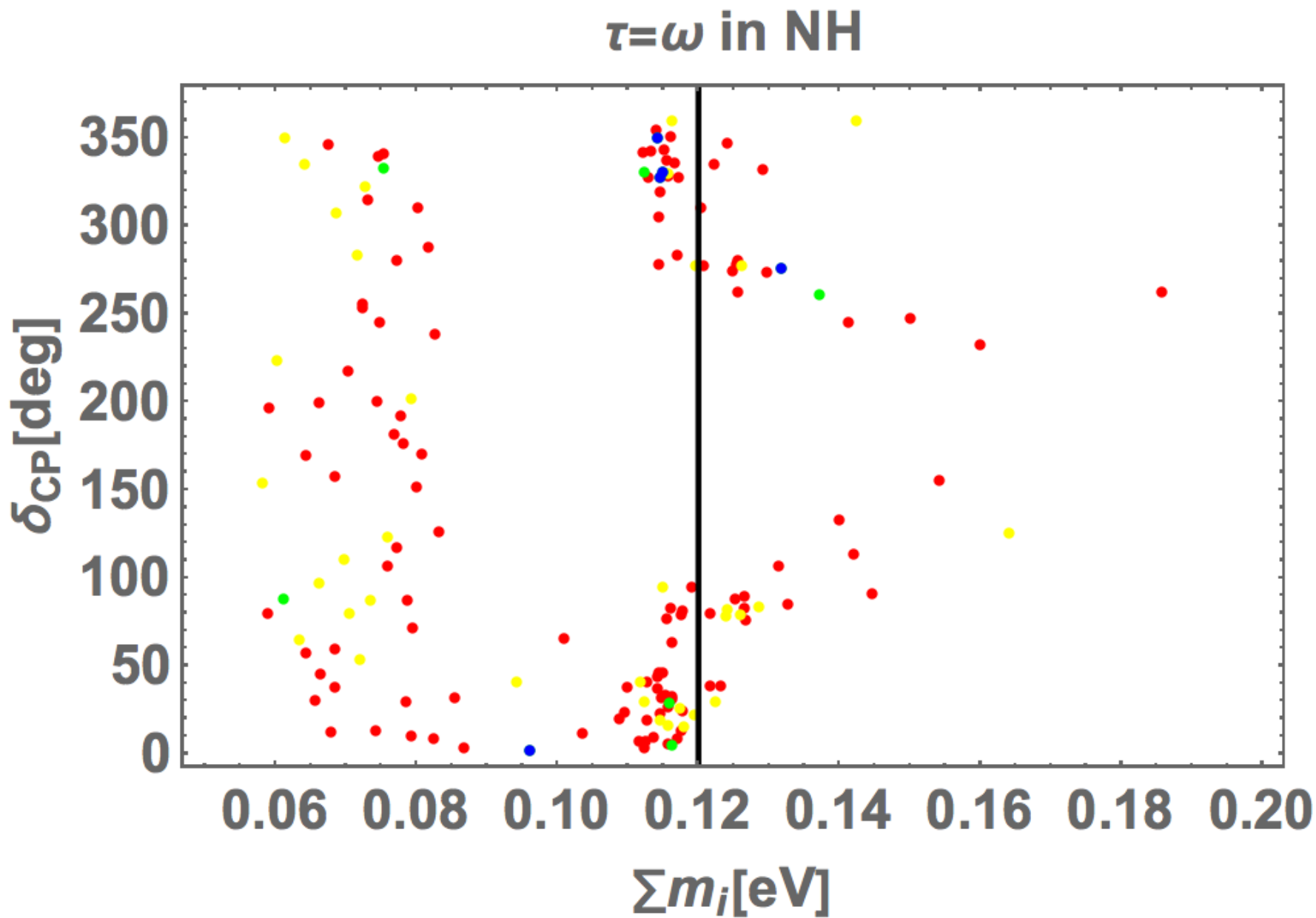}
  \end{center}
 \end{minipage}
 \caption{Each of color represents ${\rm blue} \le1\sigma$, $1\sigma< {\rm green}\le 2\sigma$, $2\sigma< {\rm yellow}\le 3\sigma$, $3\sigma<{\rm red}\le5\sigma$.}
 \label{fig.chi-tau=omega_nh}
\end{figure}
In Fig.~\ref{fig.chi-tau=omega_nh}, we show our several allowed regions on $\tau$ at nearby $\tau=\omega$ in case of NH,  where the color legends are the same as the one in Fig.~\ref{fig.dev-tau=i_nh}.
The up-left one represents the allowed region of the imaginary part of $\tau$ in terms of the real part of $\tau$. The smaller $\chi$ square denoted by blue color is closest to the fixed point of $\tau=\omega$,
which would be a good tendency.
The up-right one demonstrates the allowed region of neutrinoless double beta decay $\langle m_{ee}\rangle$ in terms of the lightest active neutrino mass $m_1$.
There is a linear correlation with width between them. 
Furthermore, all the regions of $\chi$ square tend to run the whole range. 
The down-left one shows the allowed region of Majorana phases $\alpha_{21}$ and $\alpha_{31}$.
Even though the whole region is allowed, there exist two islands at around $-50^\circ\le\alpha_{21},\alpha_{31}\le50^\circ$.
The down-right one depicts the allowed region of Dirac phase $\delta_{\text{CP}}$ in terms of the sum of neutrino masses $\sum m_i$.
The vertical line is the upper bound on cosmological constraint. Below this bound, the whole region is allowed for $\delta_{\rm CP}$. At nearby this bound,
$\delta_{\rm CP}$ is allowed by $0^\circ-100^\circ$
and $270^\circ-360^\circ$. Furthermore, the smaller $\chi$ square tends to be localized at nearby the cosmological bound, and its testability would be enhanced.

\begin{figure}[H]
\begin{minipage}{0.49\hsize}
  \begin{center}
  \includegraphics[scale=0.3]{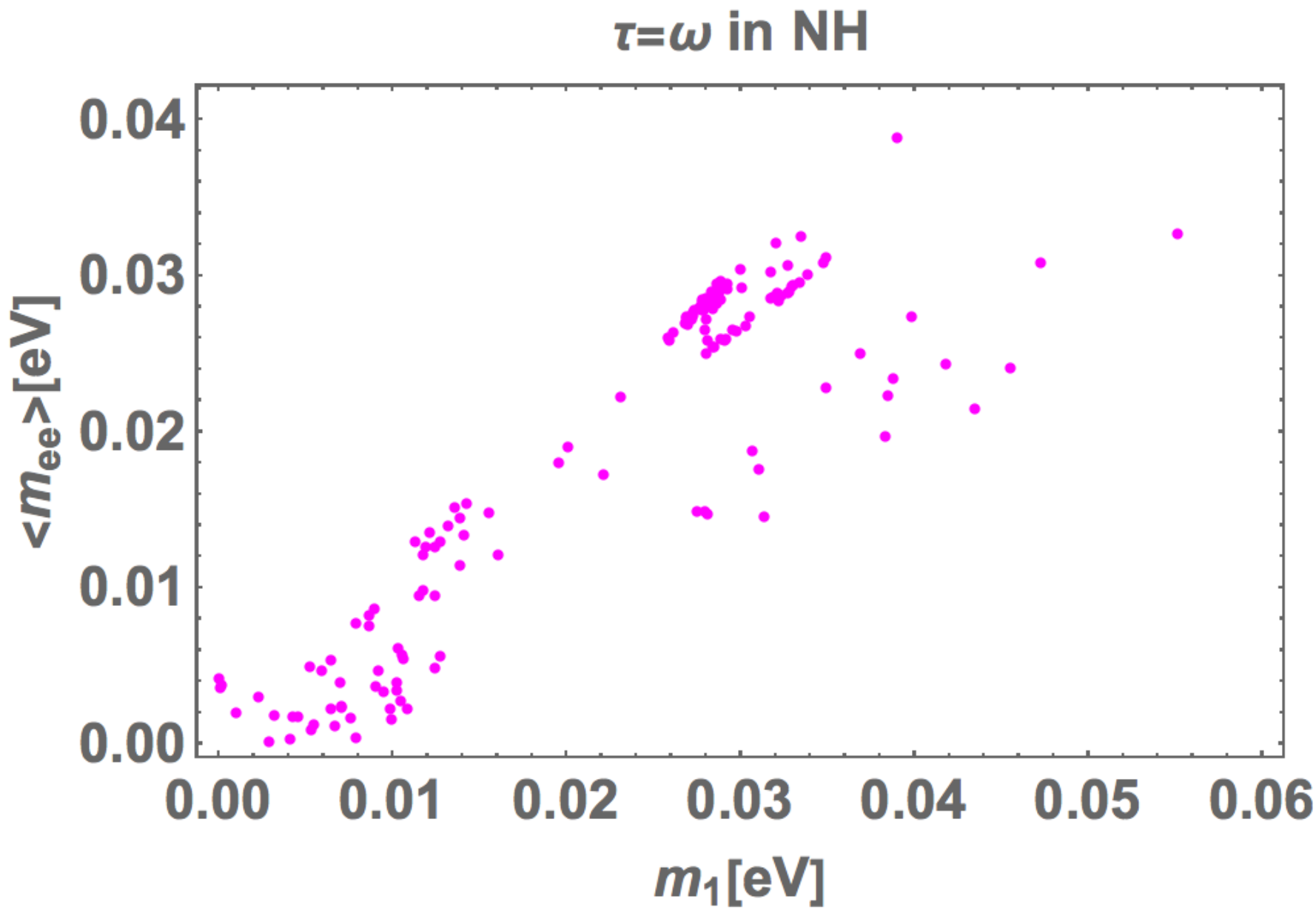}
  \end{center}
 \end{minipage}
\begin{minipage}{0.49\hsize}
  \begin{center}
  \includegraphics[scale=0.3]{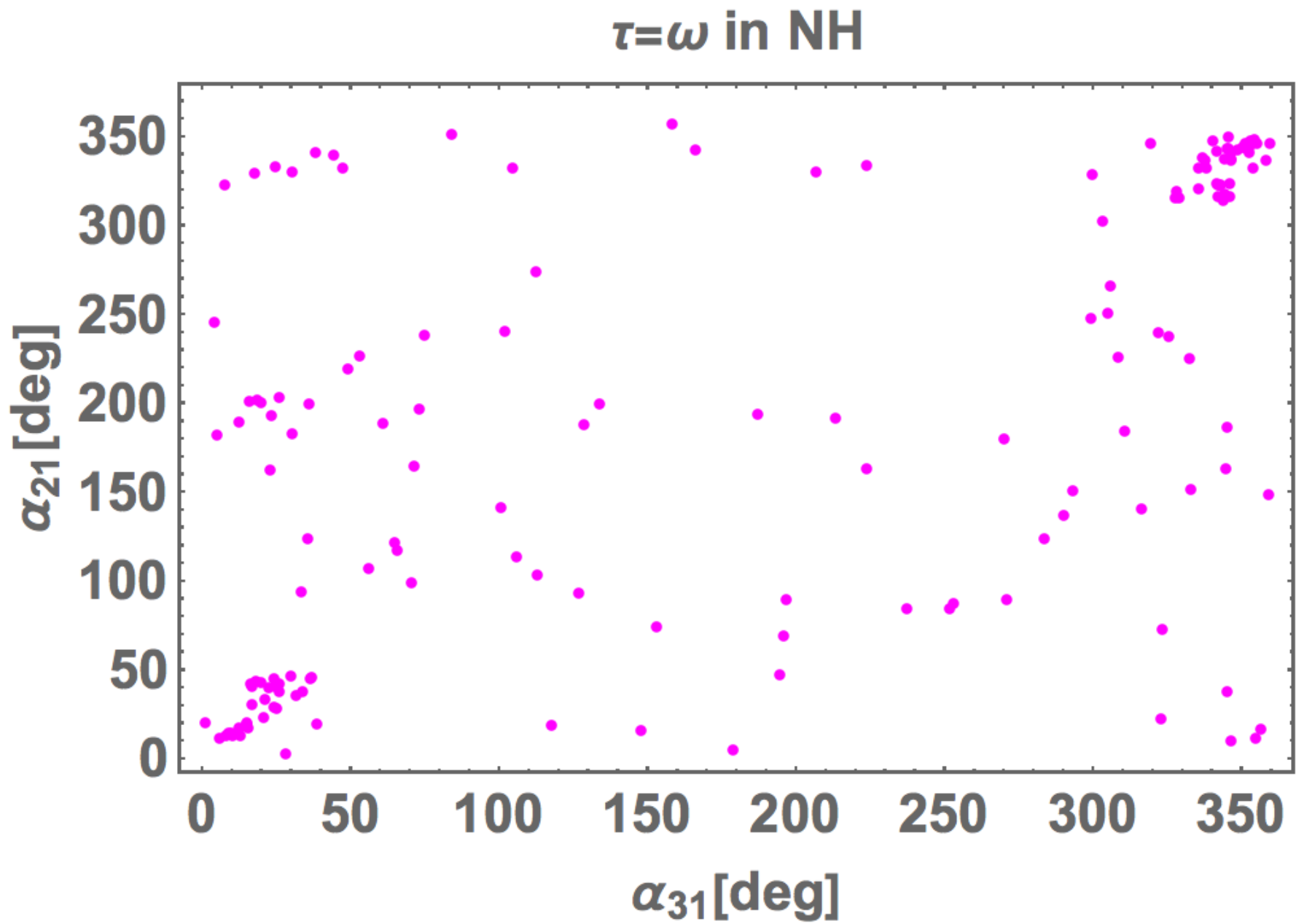}
  \end{center}
 \end{minipage}
\begin{minipage}{0.49\hsize}
  \begin{center}
  \includegraphics[scale=0.3]{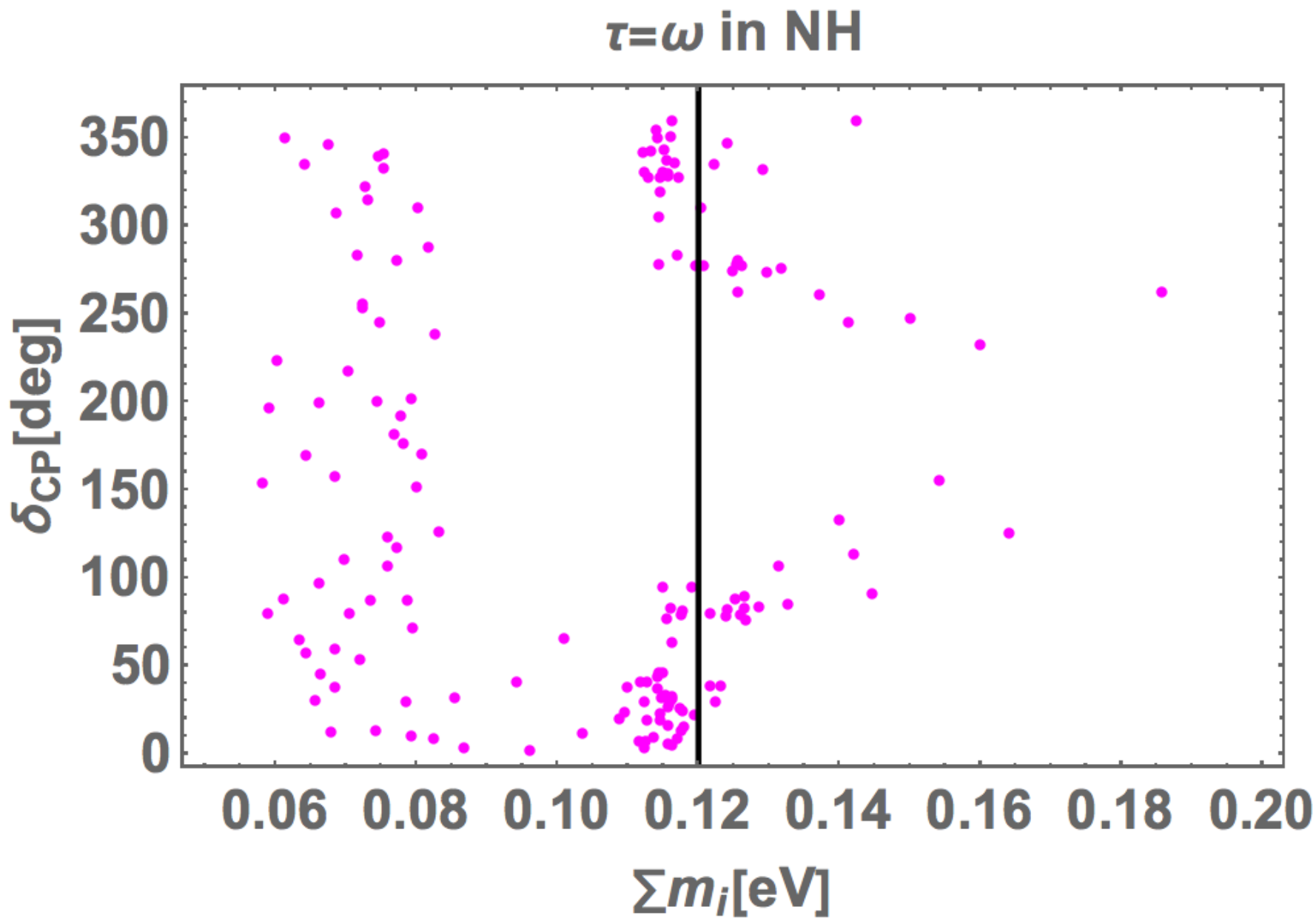}
  \end{center}
 \end{minipage}
 \caption{$|\delta \tau| < 10^{-15}$ for black, $10^{-15} \leq |\delta \tau| < 10^{-12}$ for gray, $10^{-12} \leq |\delta \tau| < 10^{-10}$ for purple, $10^{-10} \leq |\delta \tau| < 10^{-7}$ for brown, $10^{-7} \leq |\delta \tau| < 10^{-5}$ for blue green, $10^{-5} \leq |\delta \tau| < 10^{-3}$ for orange, and $10^{-3} \leq |\delta \tau| < 10^{-1}$ for magenta.}
 \label{fig.dev-tau=omega_nh}
\end{figure}
%
In Fig.~\ref{fig.dev-tau=omega_nh}, we show the several figures in terms of deviation from $\tau=\omega$ where the color legends are the same as the one in Fig.~\ref{fig.dev-tau=i_nh}.
The up-left one corresponds to the case of up-right one in Fig.~\ref{fig.chi-tau=omega_nh}.
The up-right one corresponds to the case of down-left one in Fig.~\ref{fig.chi-tau=omega_nh}.
The down-left one corresponds to the case of down-right one in Fig.~\ref{fig.chi-tau=omega_nh}.
These figures show us that larger deviation; $10^{-3} \leq |\delta \tau| < 10^{-1}$, is requested when the neutrino oscillations are satisfied.
It is not favored by the theoretical point of view as we already discussed in Sec.~\ref{sec:moduli}.

\begin{figure}[H]
\begin{minipage}{0.49\hsize}
  \begin{center}
  \includegraphics[scale=0.3]{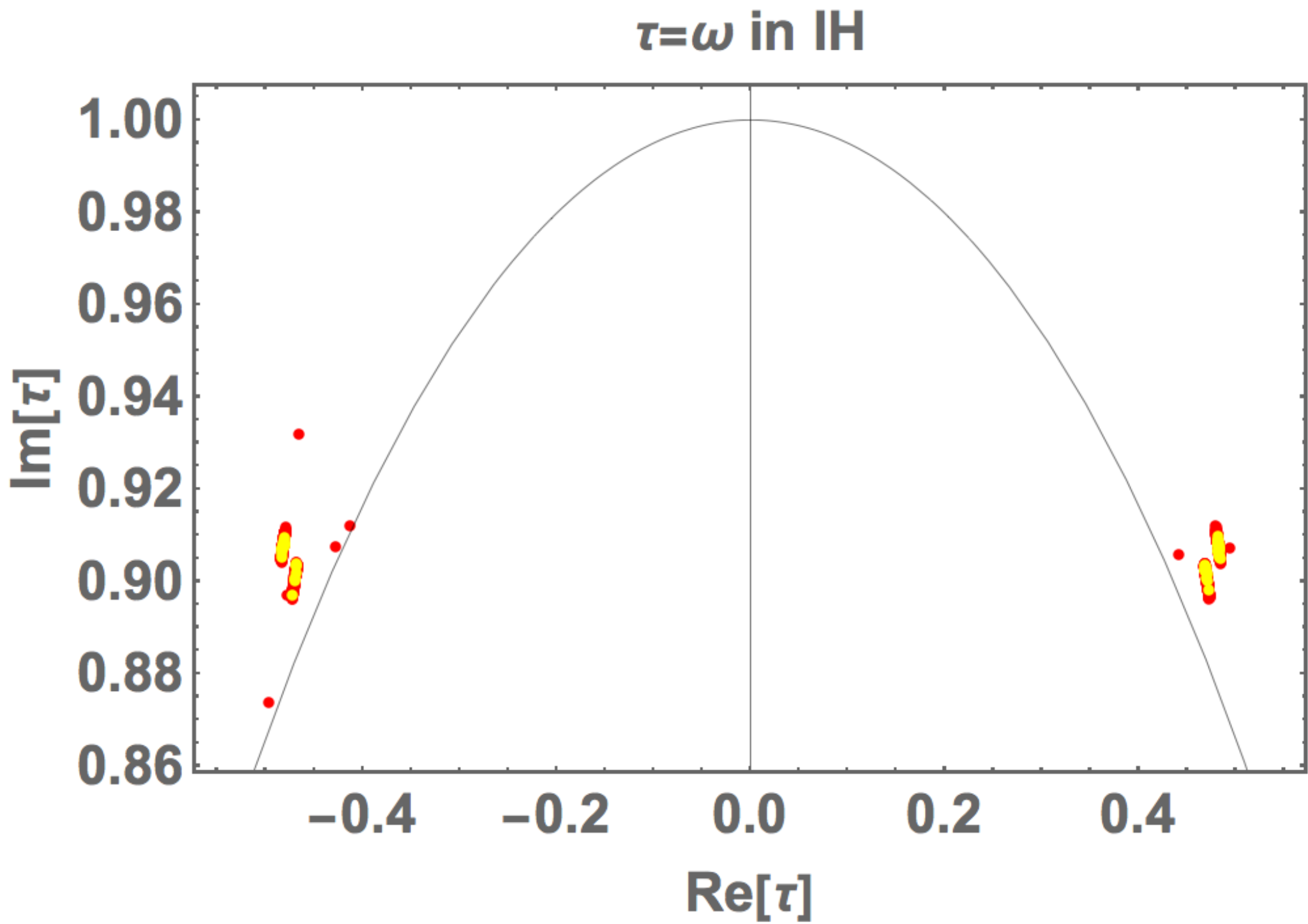}
  \end{center}
 \end{minipage}
\begin{minipage}{0.49\hsize}
  \begin{center}
  \includegraphics[scale=0.3]{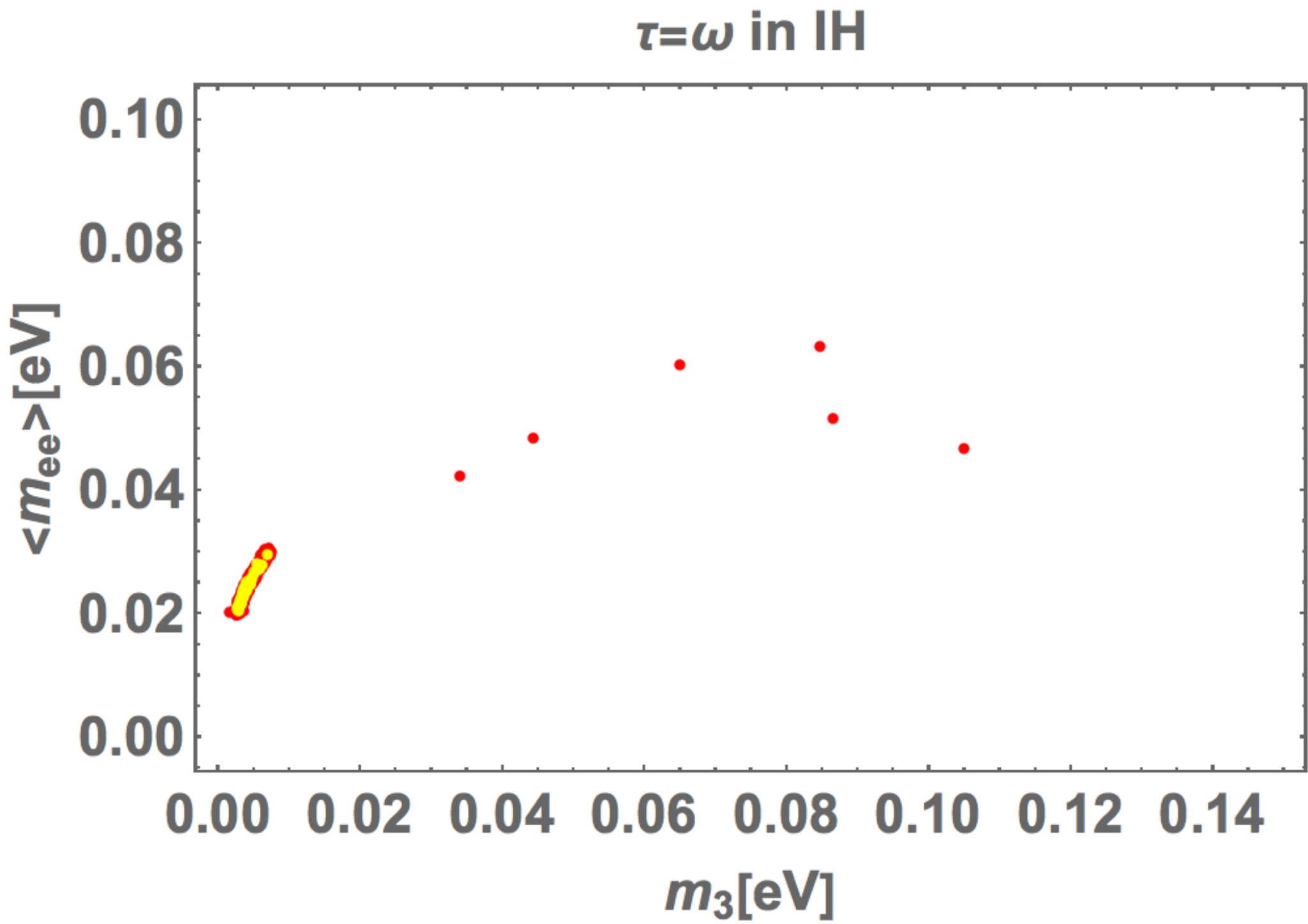}
  \end{center}
 \end{minipage}
\begin{minipage}{0.49\hsize}
  \begin{center}
  \includegraphics[scale=0.3]{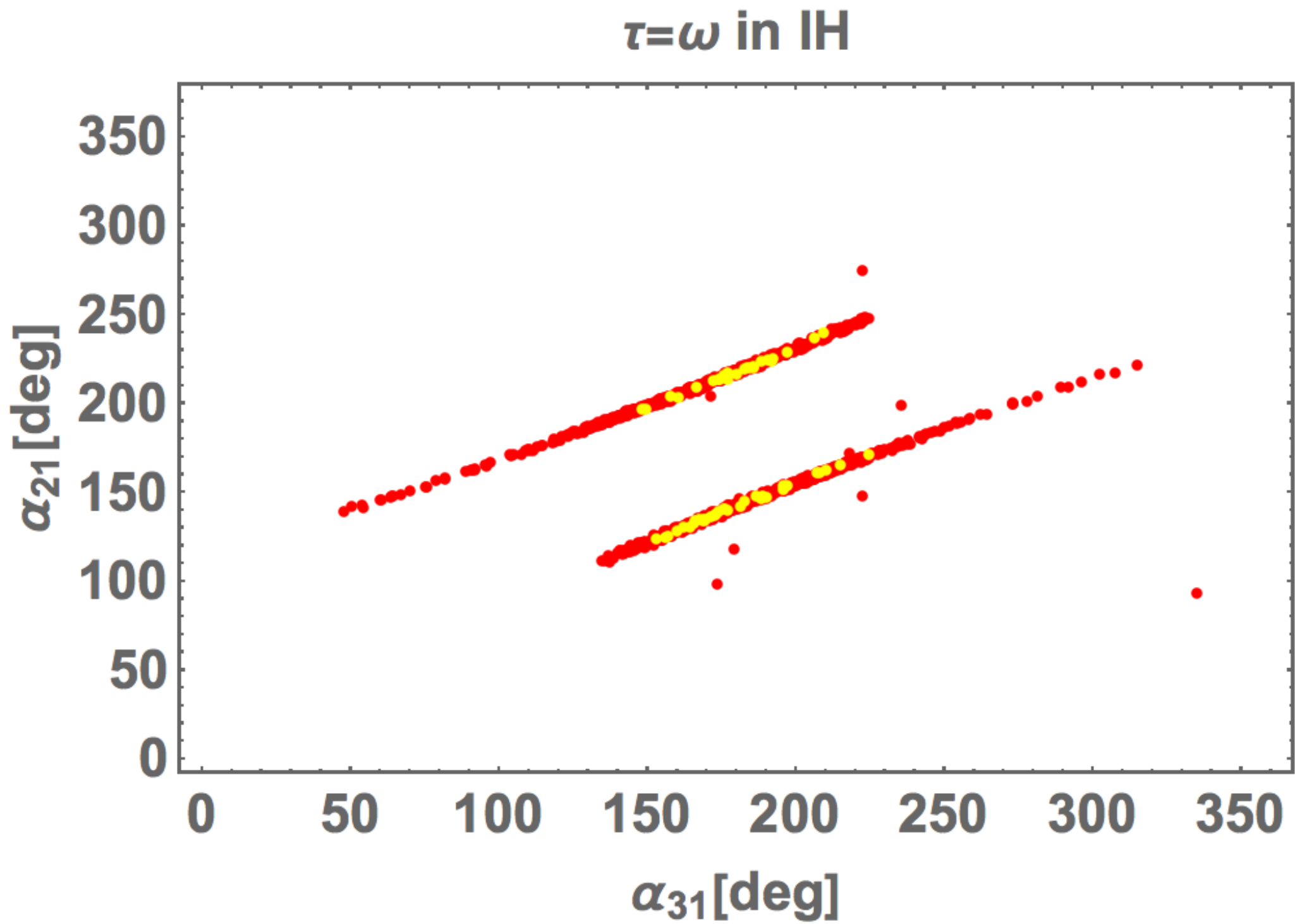}
  \end{center}
 \end{minipage}
 \begin{minipage}{0.49\hsize}
  \begin{center}
   \includegraphics[scale=0.3]{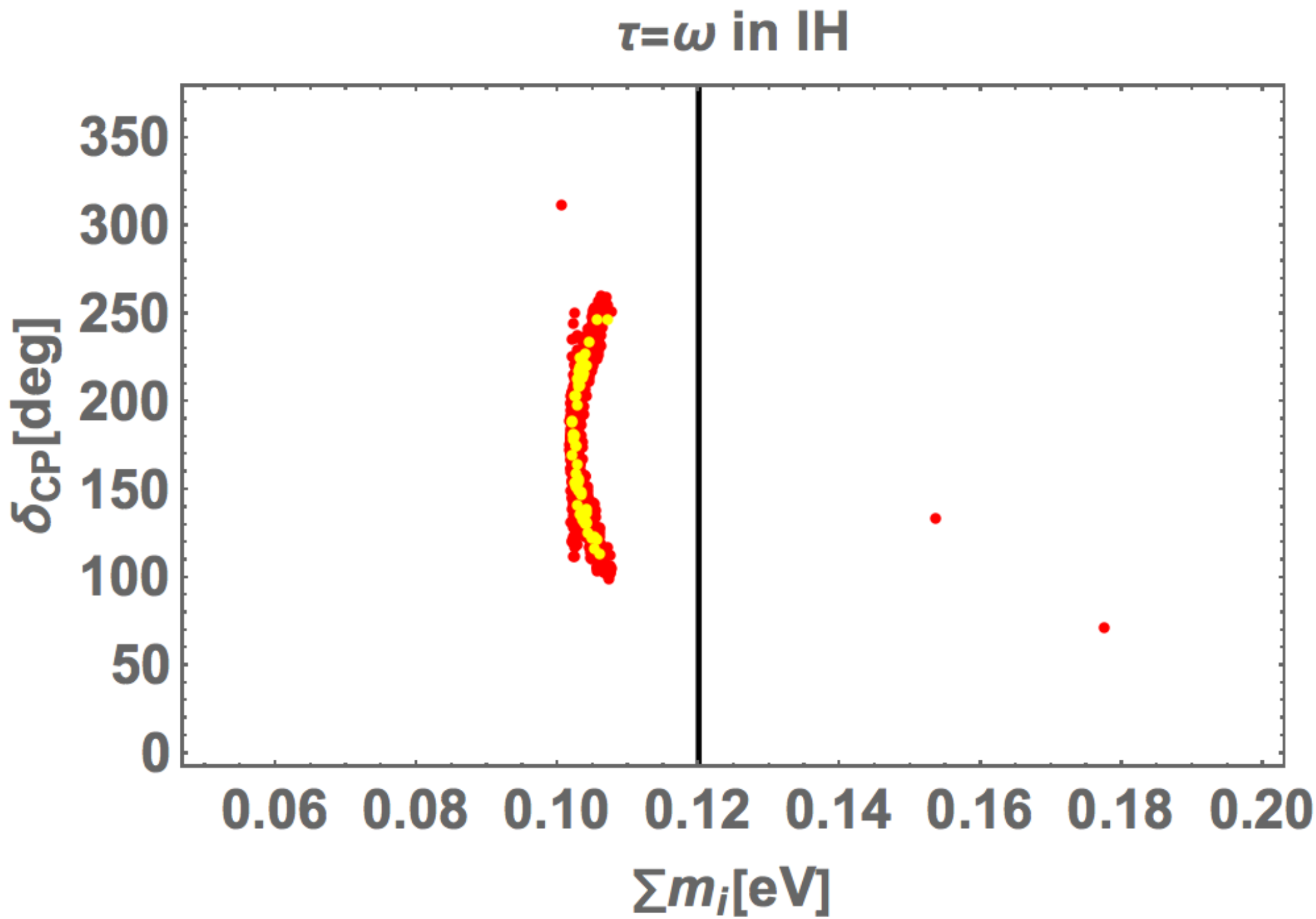}
  \end{center}
 \end{minipage}
 \caption{Each of color represents ${\rm blue} \le1\sigma$, $1\sigma< {\rm green}\le 2\sigma$, $2\sigma< {\rm yellow}\le 3\sigma$, $3\sigma<{\rm red}\le5\sigma$.}
 \label{fig.chi-tau=omega_ih}
\end{figure}
In Fig.~\ref{fig.chi-tau=omega_ih}, we show our several allowed regions on $\tau$ at nearby $\tau=\omega$  in case of IH,  where the color legends are the same as the one in Fig.~\ref{fig.dev-tau=i_nh}.
The up-left one represents the allowed region of the imaginary part of $\tau$ in terms of the real part of $\tau$.
We have found only the allowed region of $2\sigma-5\sigma$.
The up-right one demonstrates the allowed region of neutrinoless double beta decay $\langle m_{ee}\rangle$ in terms of the lightest active neutrino mass $m_3$.
There seems to be a linear correlation between them, and 0.02 eV$\lesssim\langle m_{ee}\rangle\lesssim$ 0.06 eV up to $5\sigma$, but the allowed regions are localized at nearby small masses up to $2\sigma$.
The down-left one shows the allowed region of Majorana phases $\alpha_{21}$ and $\alpha_{31}$. 
We find the allowed regions $100^\circ\lesssim\alpha_{21}\lesssim280^\circ$ and $50^\circ\lesssim\alpha_{31}\lesssim340^\circ$.
The down-right one depicts the allowed region of  Dirac phase $\delta_{\text{CP}}$ in terms of the sum of neutrino masses $\sum m_i$.
The allowed region at yellow plots; $\sum m_i\simeq0.11$ eV, is totally within the cosmological constraint.
This implies that $m_3$ is almost zero combined with the up-right figure.

\begin{figure}[H]
\begin{minipage}{0.49\hsize}
  \begin{center}
  \includegraphics[scale=0.3]{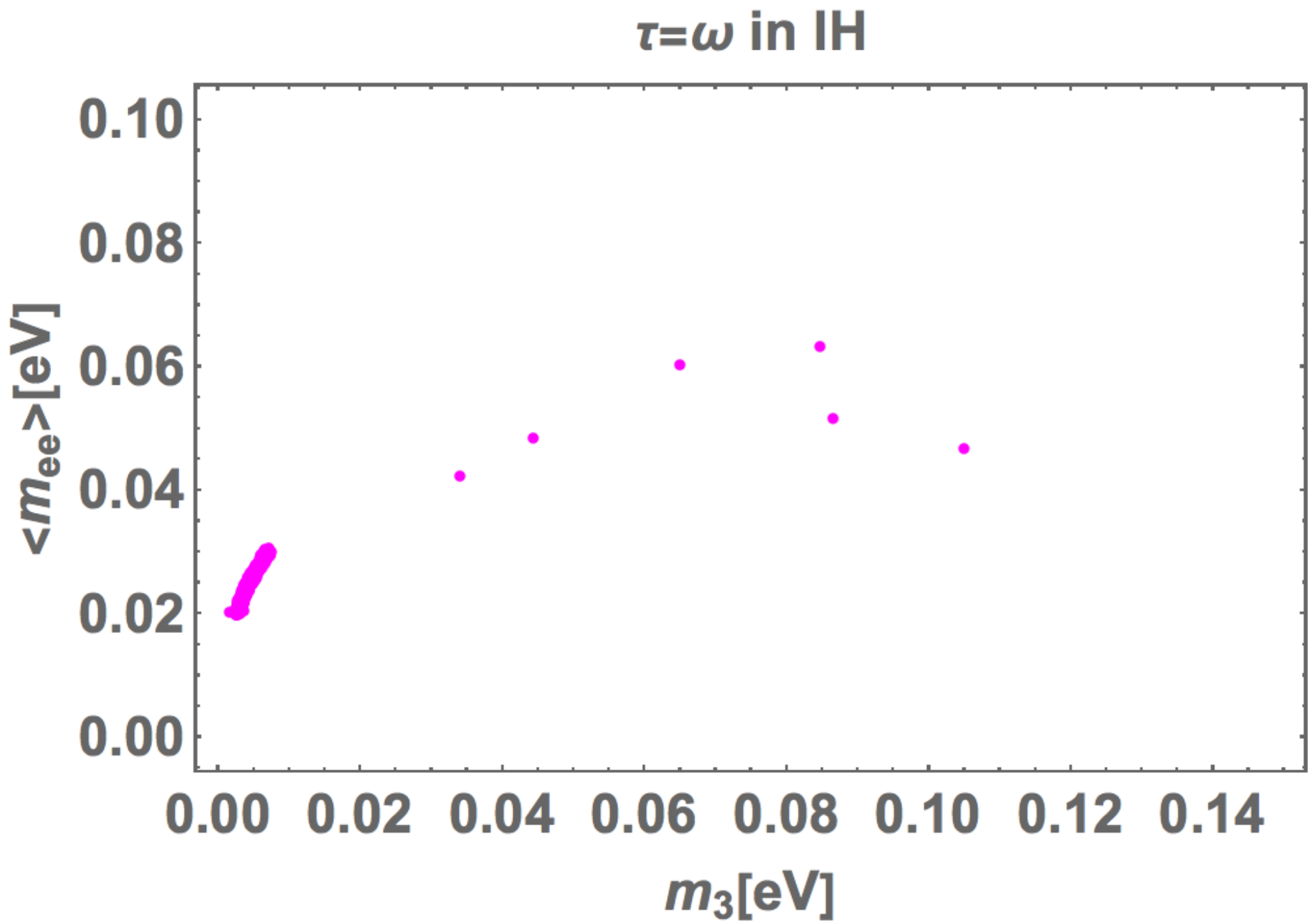}
  \end{center}
 \end{minipage}
\begin{minipage}{0.49\hsize}
  \begin{center}
  \includegraphics[scale=0.3]{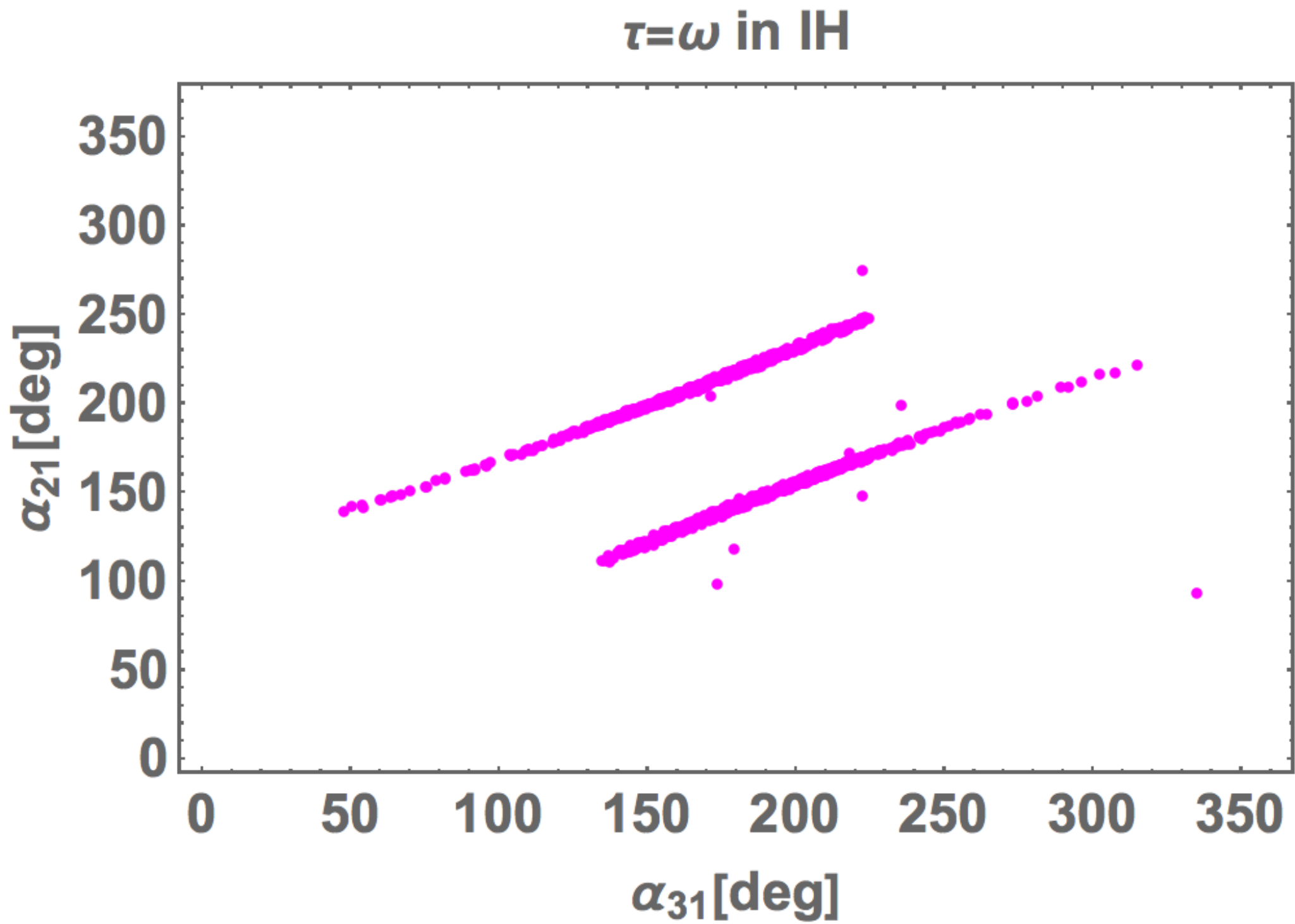}
  \end{center}
 \end{minipage}
\begin{minipage}{0.49\hsize}
  \begin{center}
  \includegraphics[scale=0.3]{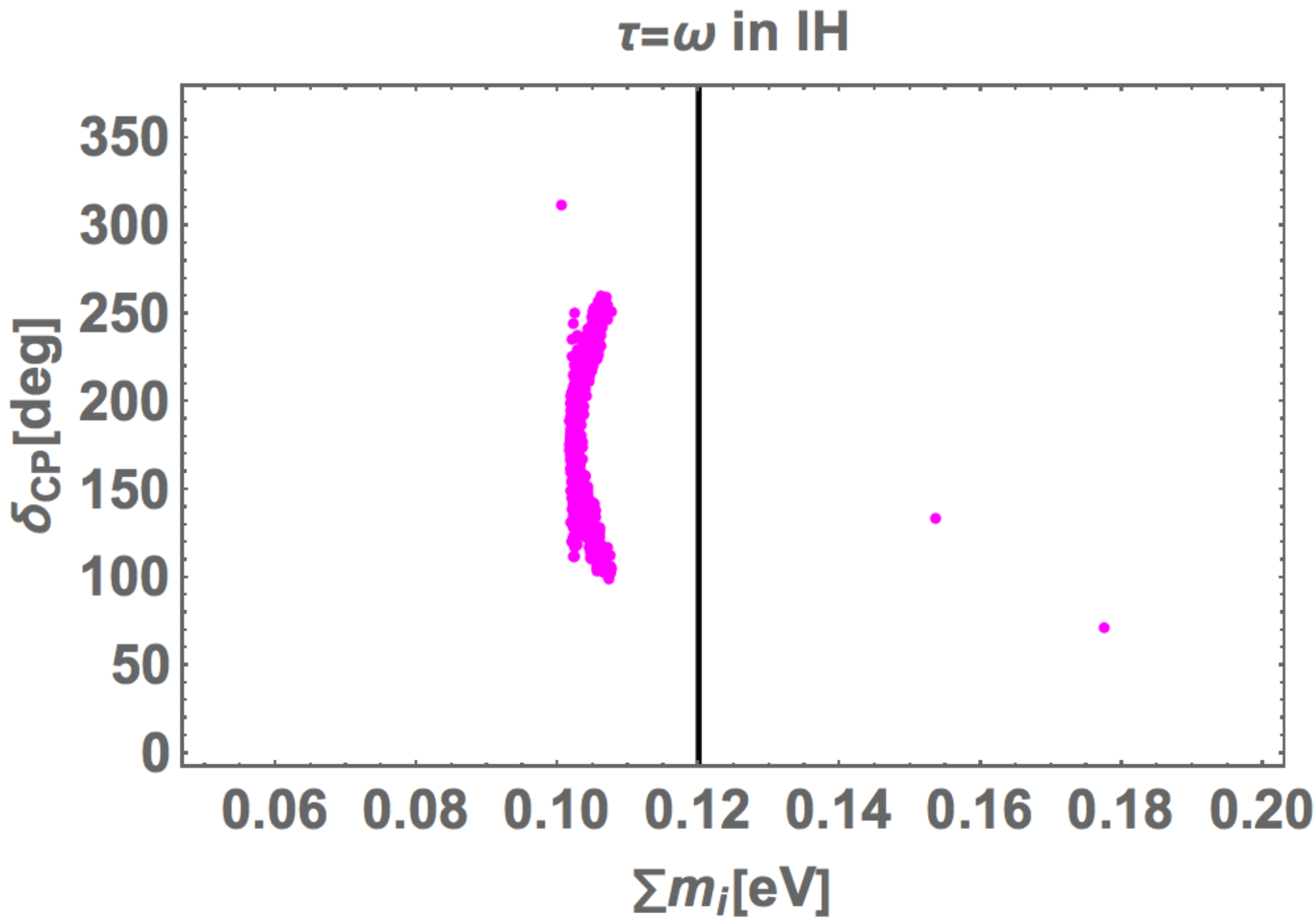}
  \end{center}
 \end{minipage}
 \caption{$|\delta \tau| < 10^{-15}$ for black, $10^{-15} \leq |\delta \tau| < 10^{-12}$ for gray, $10^{-12} \leq |\delta \tau| < 10^{-10}$ for purple, $10^{-10} \leq |\delta \tau| < 10^{-7}$ for brown, $10^{-7} \leq |\delta \tau| < 10^{-5}$ for blue green, $10^{-5} \leq |\delta \tau| < 10^{-3}$ for orange, and $10^{-3} \leq |\delta \tau| < 10^{-1}$ for magenta.}
 \label{fig.dev-tau=omega_ih}
\end{figure}
%
In Fig.~\ref{fig.dev-tau=omega_ih}, we show the several figures in terms of deviation from $\tau=\omega$ where the color legends are the same as the one in Fig.~\ref{fig.dev-tau=i_ih}.
The up-left one corresponds to the case of up-right one in Fig.~\ref{fig.chi-tau=omega_ih}.
The up-right one corresponds to the case of down-left one in Fig.~\ref{fig.chi-tau=omega_ih}.
The down-left one corresponds to the case of down-right one in Fig.~\ref{fig.chi-tau=omega_ih}.
These figures also show us that larger deviation; $10^{-3} \leq |\delta \tau| < 10^{-1}$, is requested when the neutrino oscillations are satisfied.
It is not favored by the theoretical point of view as we already discussed in Sec.~\ref{sec:moduli}.
In conclusion, in the case of $\tau=\omega$,
both the case of NH and IH would not be favored by the theoretical viewpoint.

\subsubsection{Nearby $\tau = 2i$}

\begin{figure}[H]
\begin{minipage}{0.49\hsize}
  \begin{center}
  \includegraphics[scale=0.25]{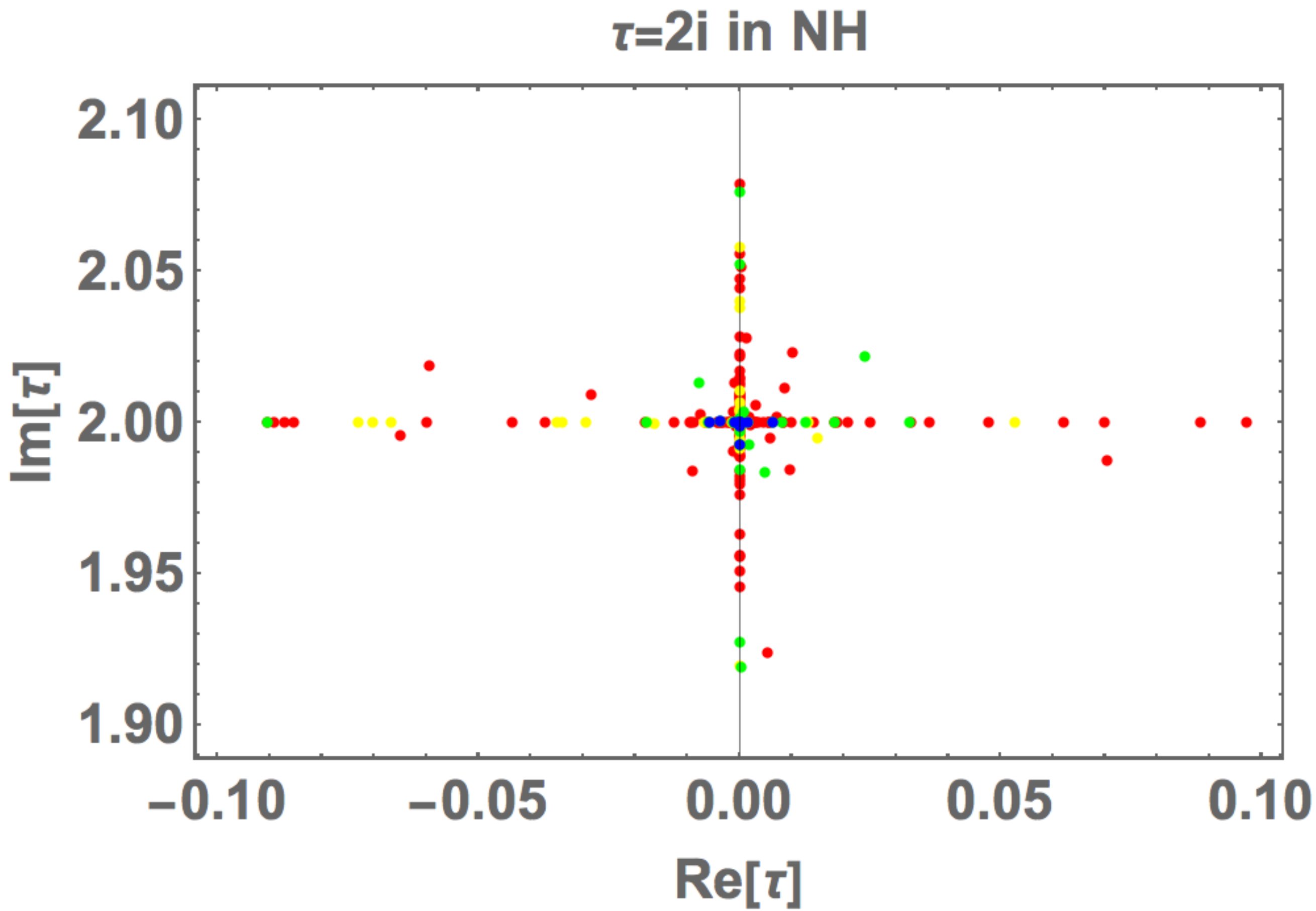}
  \end{center}
 \end{minipage}
\begin{minipage}{0.49\hsize}
  \begin{center}
  \includegraphics[scale=0.25]{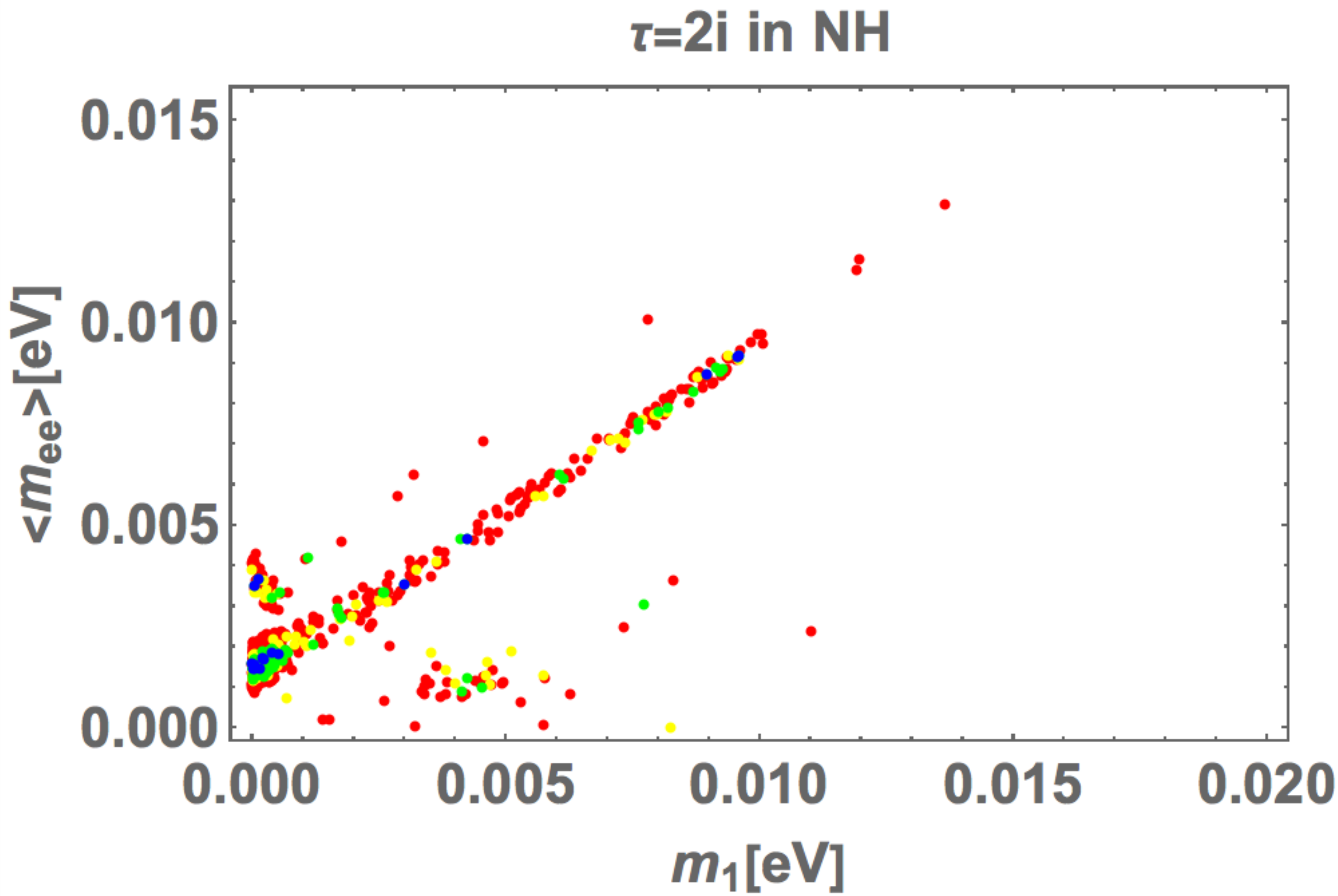}
  \end{center}
 \end{minipage}
\begin{minipage}{0.49\hsize}
  \begin{center}
  \includegraphics[scale=0.25]{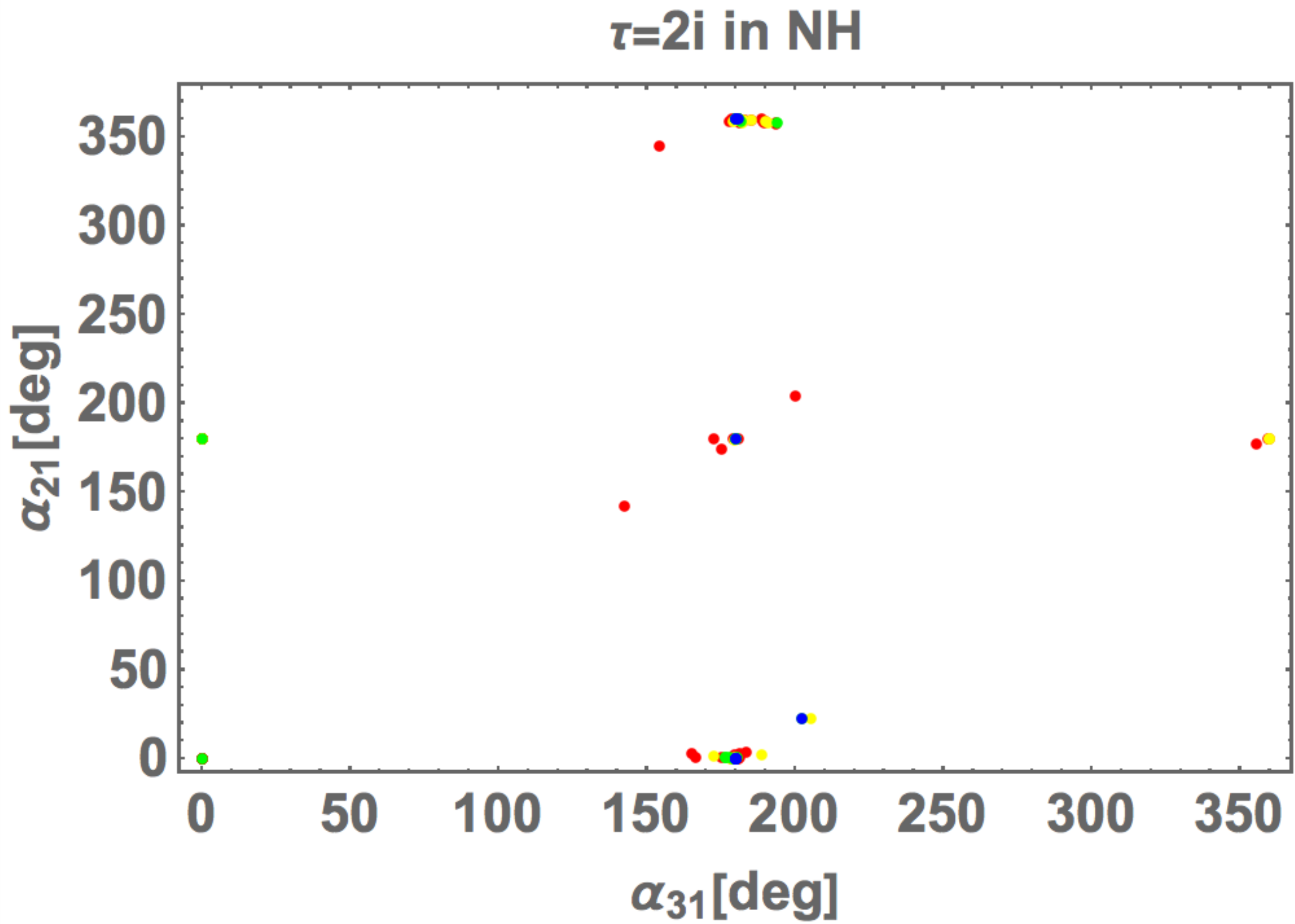}
  \end{center}
 \end{minipage}
 \begin{minipage}{0.49\hsize}
  \begin{center}
   \includegraphics[scale=0.25]{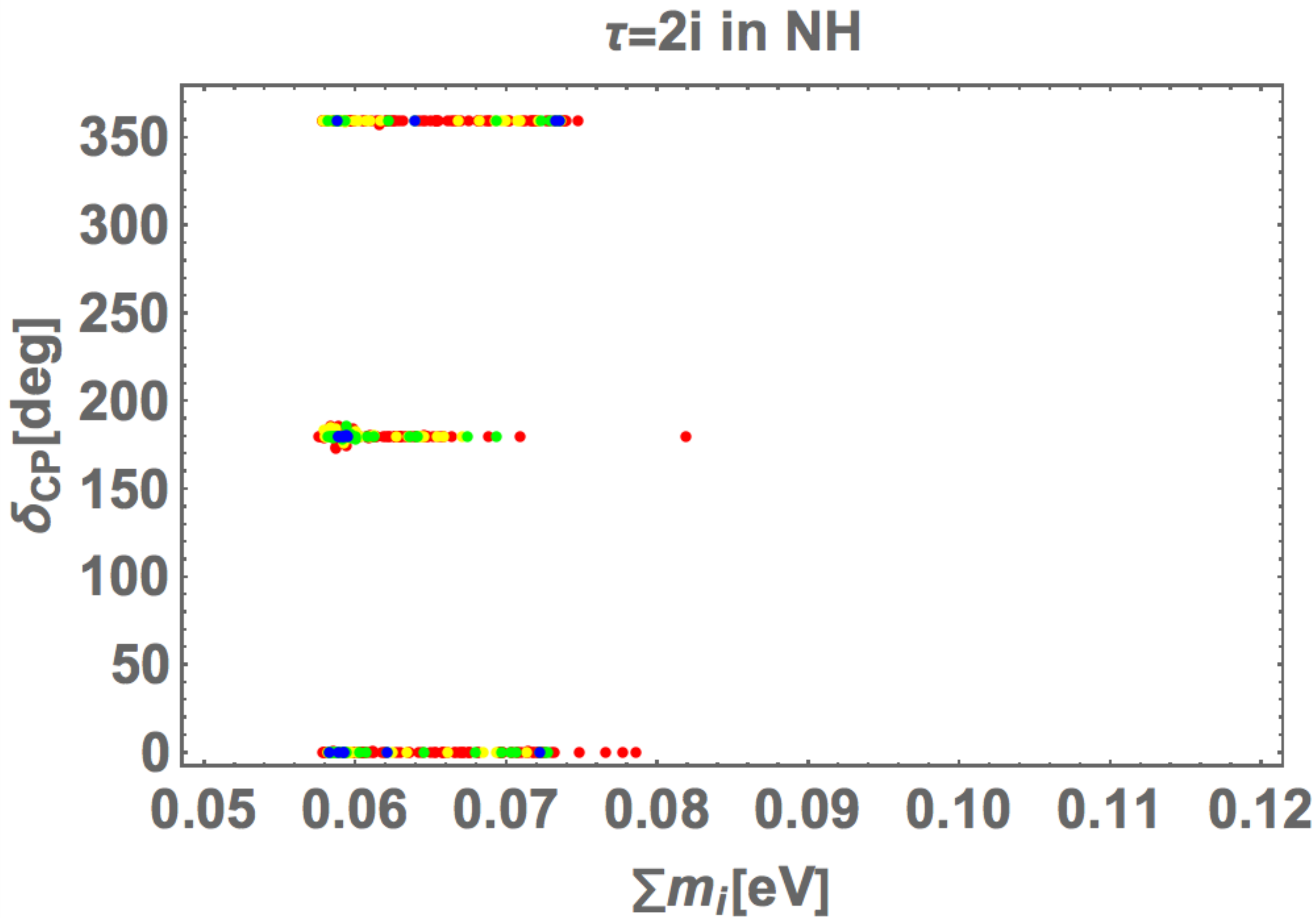}
  \end{center}
 \end{minipage}
 \caption{Each of color represents ${\rm blue} \le1\sigma$, $1\sigma< {\rm green}\le 2\sigma$, $2\sigma< {\rm yellow}\le 3\sigma$, $3\sigma<{\rm red}\le5\sigma$.}
 \label{fig.chi-tau=infty_nh}
\end{figure}
In Fig.~\ref{fig.chi-tau=infty_nh}, we show our several allowed regions on $\tau$ at nearby $\tau=2i$ in case of NH,  where the color legends are the same as the one in Fig.~\ref{fig.dev-tau=i_nh}.
The up-left one represents the allowed region of the imaginary part of $\tau$ in terms of the real part of $\tau$. The smaller $\chi$ square denoted by blue color is closest to the fixed point of $\tau=2i$,
which would be a good tendency.
The up-right one demonstrates the allowed region of neutrinoless double beta decay $\langle m_{ee}\rangle$ in terms of the lightest active neutrino mass $m_1$.
There is main linear correlation between them.
We find the allowed regions 0 eV$\le m_1\le$0.014 eV,
and 0 eV$\le \langle m_{ee}\rangle\le$0.013 eV.
The down-left one shows the allowed region of Majorana phases $\alpha_{21}$ and $\alpha_{31}$.
Both the phases allow to be $0^\circ$ or $180^\circ$.
The down-right one depicts the allowed region of Dirac phase $\delta_{\text{CP}}$ in terms of the sum of neutrino masses $\sum m_i$.
The whole allowed region of $\sum m_i$ is totally within the bound on cosmological constraint; 0.058 eV$\le\sum m_i\le$ 0.082 eV, whereas the allowed region of $\delta_{\rm CP}$ is the same as Majorana phases; $0^\circ$ or $180^\circ$.
Note that it is trivial that we find no phases since the situation is similar to the case of $\tau=i$.

\begin{figure}[H]
\begin{minipage}{0.49\hsize}
  \begin{center}
  \includegraphics[scale=0.25]{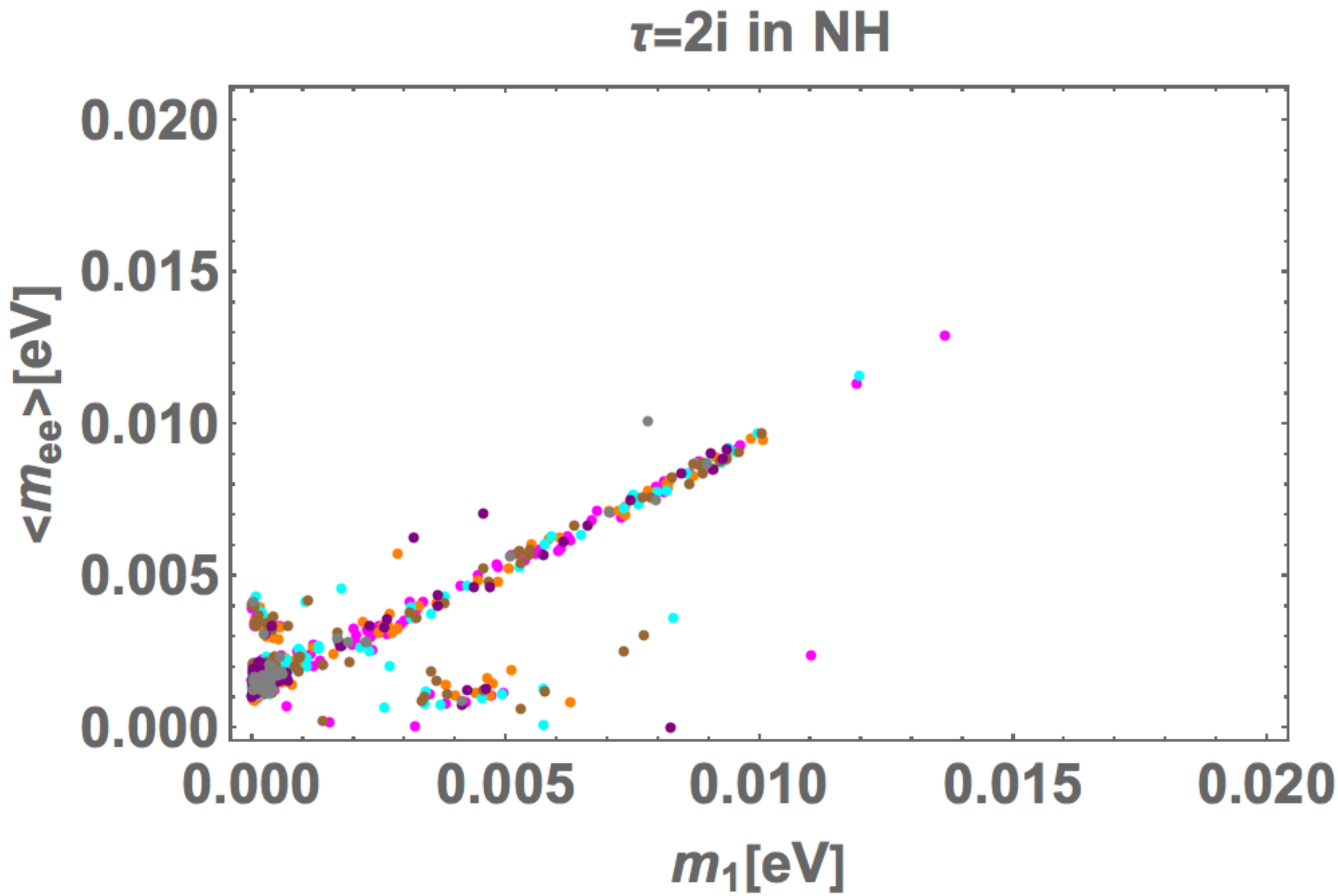}
  \end{center}
 \end{minipage}
\begin{minipage}{0.49\hsize}
  \begin{center}
  \includegraphics[scale=0.25]{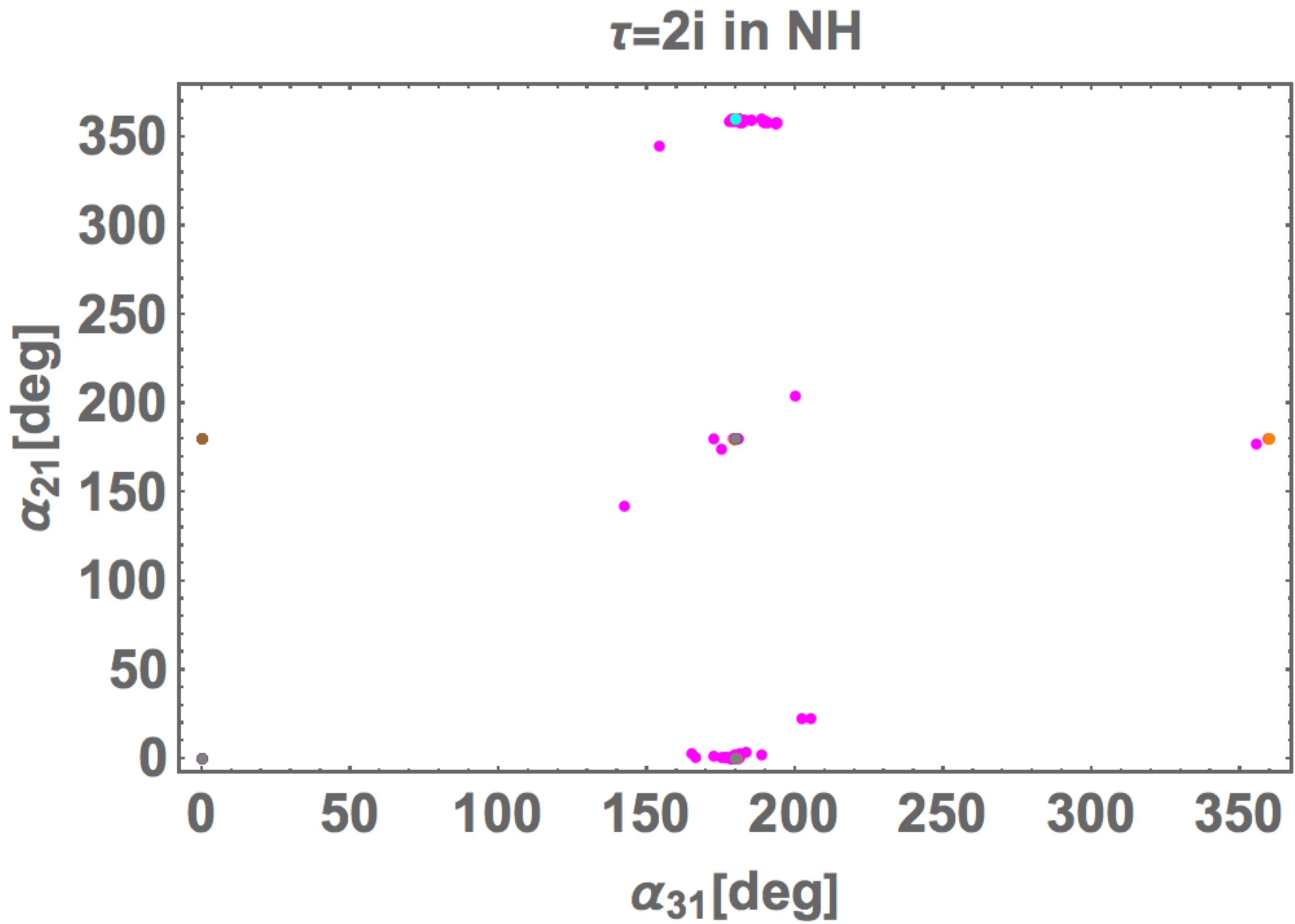}
  \end{center}
 \end{minipage}
\begin{minipage}{0.49\hsize}
  \begin{center}
  \includegraphics[scale=0.25]{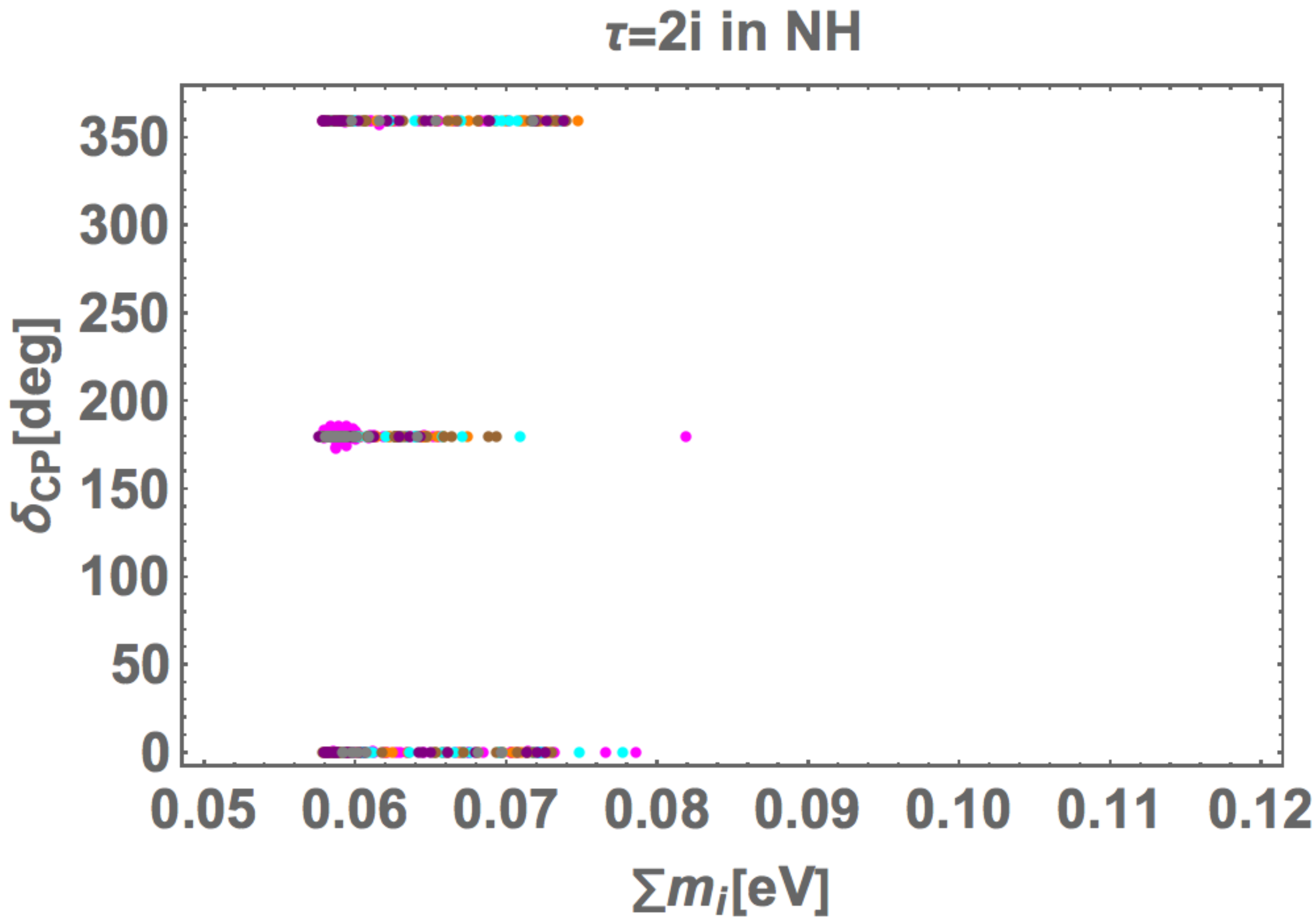}
  \end{center}
 \end{minipage}
 \caption{$|\delta \tau| < 10^{-15}$ for black, $10^{-15} \leq |\delta \tau| < 10^{-12}$ for gray, $10^{-12} \leq |\delta \tau| < 10^{-10}$ for purple, $10^{-10} \leq |\delta \tau| < 10^{-7}$ for brown, $10^{-7} \leq |\delta \tau| < 10^{-5}$ for blue green, $10^{-5} \leq |\delta \tau| < 10^{-3}$ for orange, and $10^{-3} \leq |\delta \tau| < 10^{-1}$ for magenta.}
 \label{fig.dev-tau=infty_nh}
\end{figure}
In Fig.~\ref{fig.dev-tau=infty_nh}, we show the several figures in terms of deviation from $\tau=2i$ in the same case of Fig.~\ref{fig.chi-tau=i_nh}, where the color legends are the same as the one in Fig.~\ref{fig.dev-tau=i_nh}.
The up-left one is the same as the case of up-right one in Fig.~\ref{fig.chi-tau=infty_nh}.
The up-right one is the same as the case of down-left one in Fig.~\ref{fig.chi-tau=infty_nh}.
The down-left one is the same as the case of down-right one in Fig.~\ref{fig.chi-tau=infty_nh}.
These figures suggest us that size of deviation almost run the whole ranges that are allowed by the neutrino oscillation data.

\begin{figure}[H]
\begin{minipage}{0.49\hsize}
  \begin{center}
  \includegraphics[scale=0.3]{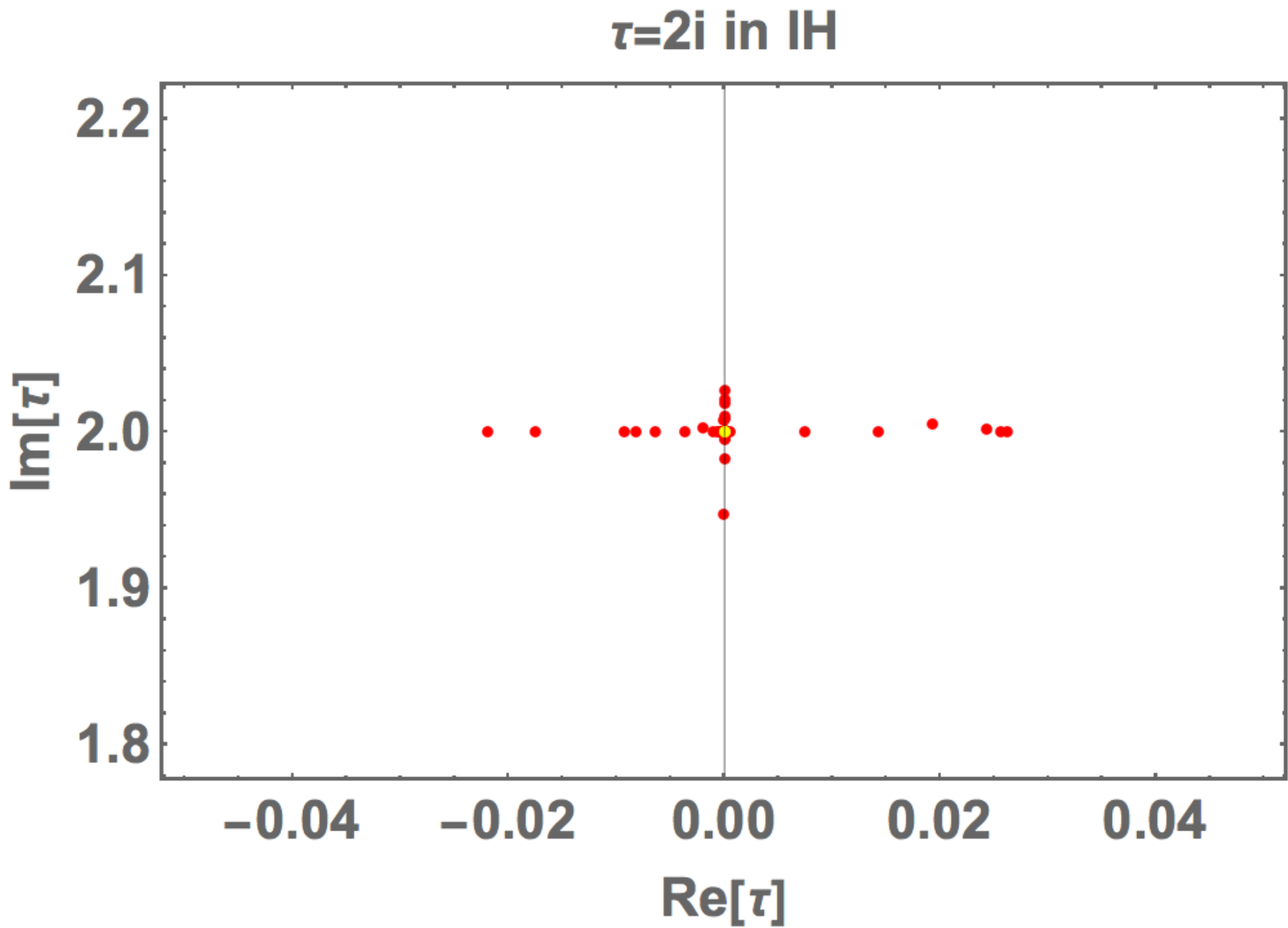}
  \end{center}
 \end{minipage}
\begin{minipage}{0.49\hsize}
  \begin{center}
  \includegraphics[scale=0.3]{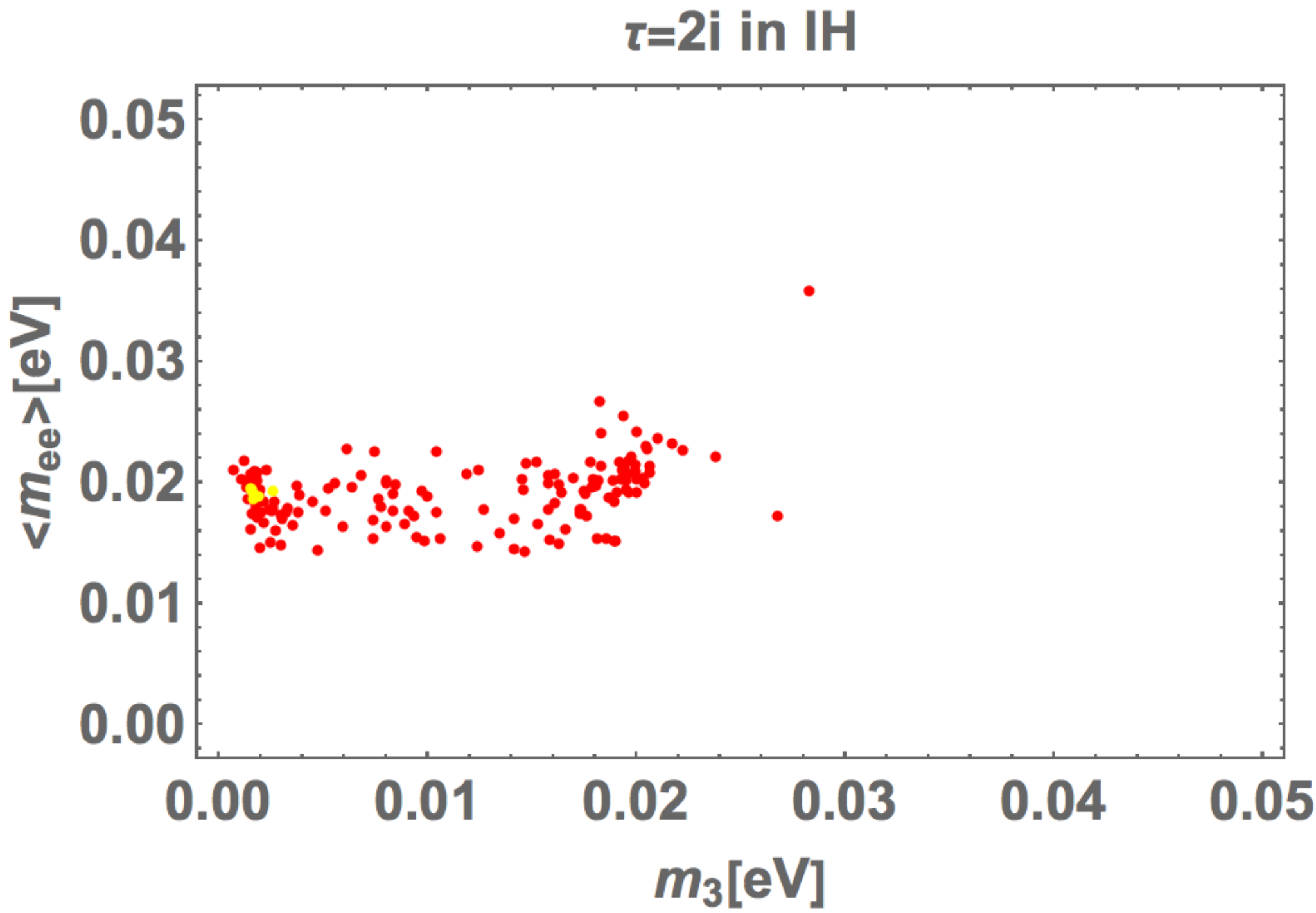}
  \end{center}
 \end{minipage}
\begin{minipage}{0.49\hsize}
  \begin{center}
  \includegraphics[scale=0.3]{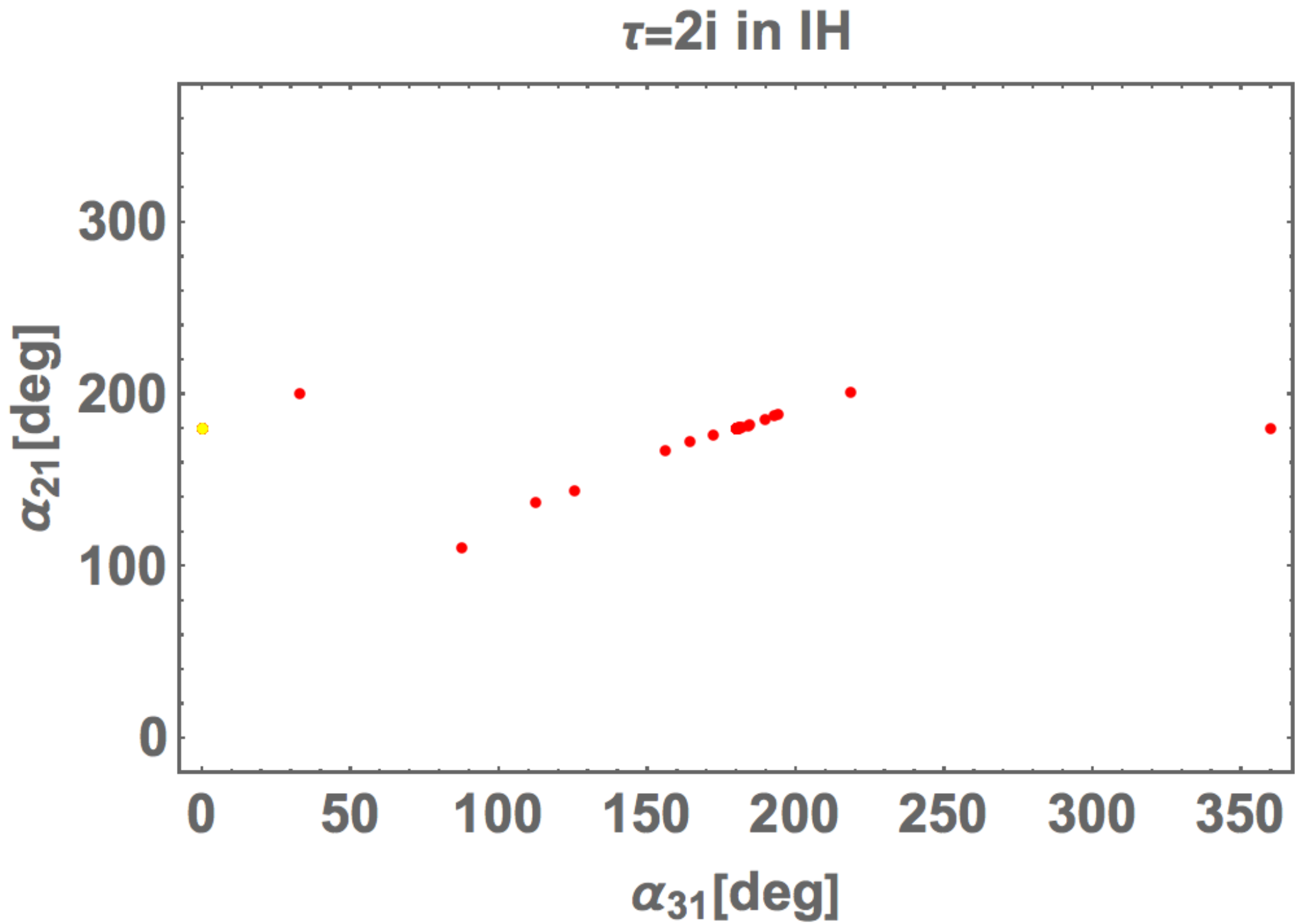}
  \end{center}
 \end{minipage}
 \begin{minipage}{0.49\hsize}
  \begin{center}
   \includegraphics[scale=0.3]{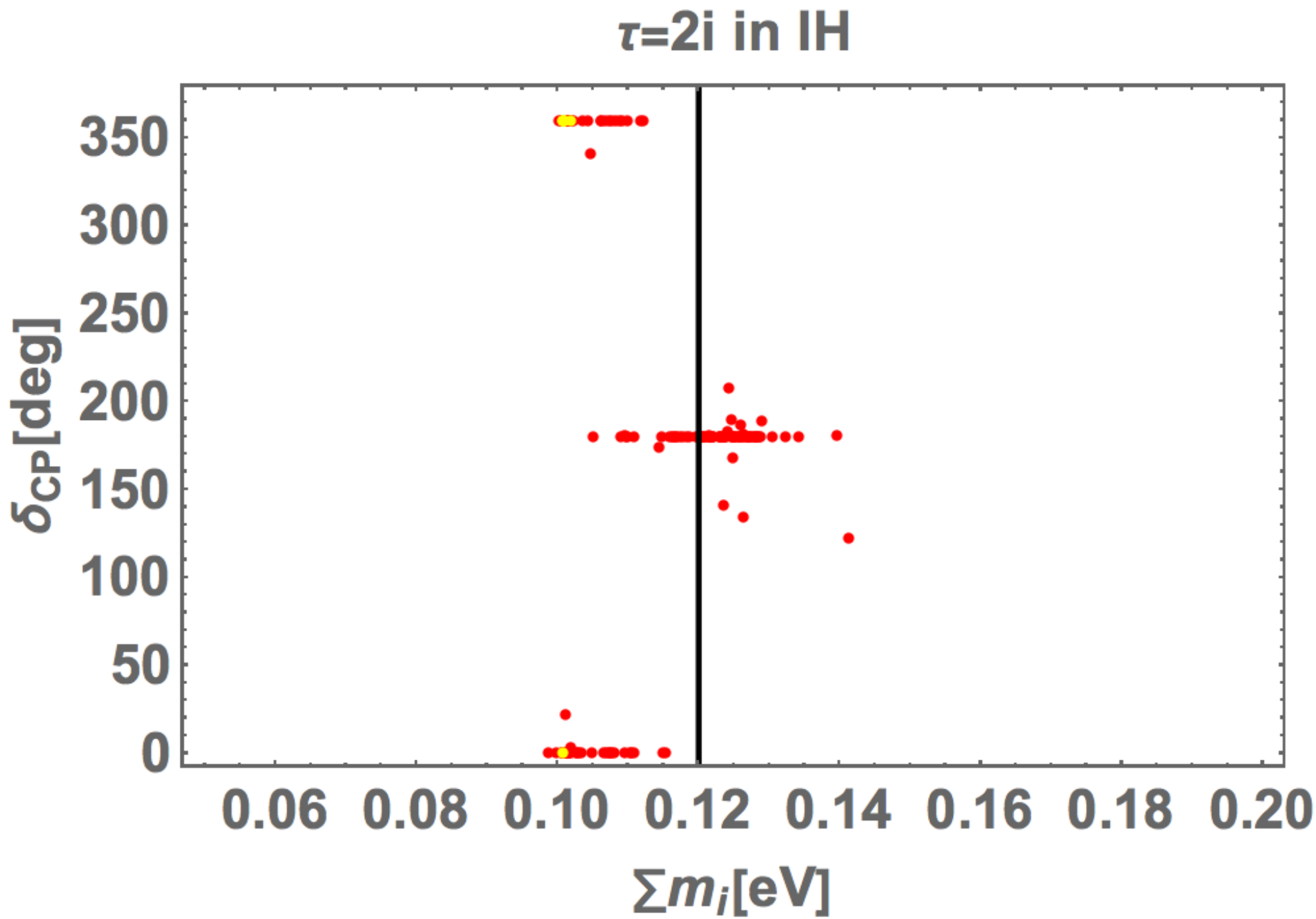}
  \end{center}
 \end{minipage}
 \caption{Each of color represents ${\rm blue} \le1\sigma$, $1\sigma< {\rm green}\le 2\sigma$, $2\sigma< {\rm yellow}\le 3\sigma$, $3\sigma<{\rm red}\le5\sigma$.}
 \label{fig.chi-tau=2i_ih}
\end{figure}
In Fig.~\ref{fig.chi-tau=2i_ih}, we show our several allowed regions on $\tau$ at nearby $\tau=2i$ in case of IH, where color legends are the same as the one of Fig.~\ref{fig.chi-tau=i_nh}. Therefore, we have found only the allowed region of $2\sigma-5\sigma$.
The up-left one represents the allowed region of the imaginary part of $\tau$ in terms of the real part of $\tau$. 
The up-right one demonstrates the allowed region of neutrinoless double beta decay $\langle m_{ee}\rangle$ in terms of the lightest active neutrino mass $m_3$.
We find the allowed regions as follows:
0 eV$\le m_3\le$0.03 eV and 0.014 eV$\le\langle m_{ee}\rangle\le$0.04 eV up to $5\sigma$, but the allowed regions are localized at nearby small masses at yellow plots.
The down-left one shows the allowed region of Majorana phases $\alpha_{21}$ and $\alpha_{31}$.
$\alpha_{21}$ is allowed by $100^\circ$ to $200^\circ$,
while $\alpha_{31}$ is wider region than $\alpha_{21}$. 
However the allowed regions are localized at nearby $\alpha_{21}=180^\circ$ and $\alpha_{31}=0^\circ$ at yellow plots.
The down-right one depicts the allowed region of Dirac phase $\delta_{\text{CP}}$ in terms of the sum of neutrino masses $\sum m_i$.
The vertical line is the upper bound on cosmological constraint. $\delta_{\rm CP}$ is allowed at the points $0^\circ$ and $180^\circ$. On the other hand, almost half the points of $\sum m_i$ would be ruled out by the cosmological bound.
Therefore, we would predict a narrow range of $0.1$eV $\le\sum m_i\le0.12$ eV that is almost the same as the one in case of $\tau=i$.

\begin{figure}[H]
\begin{minipage}{0.49\hsize}
  \begin{center}
  \includegraphics[scale=0.3]{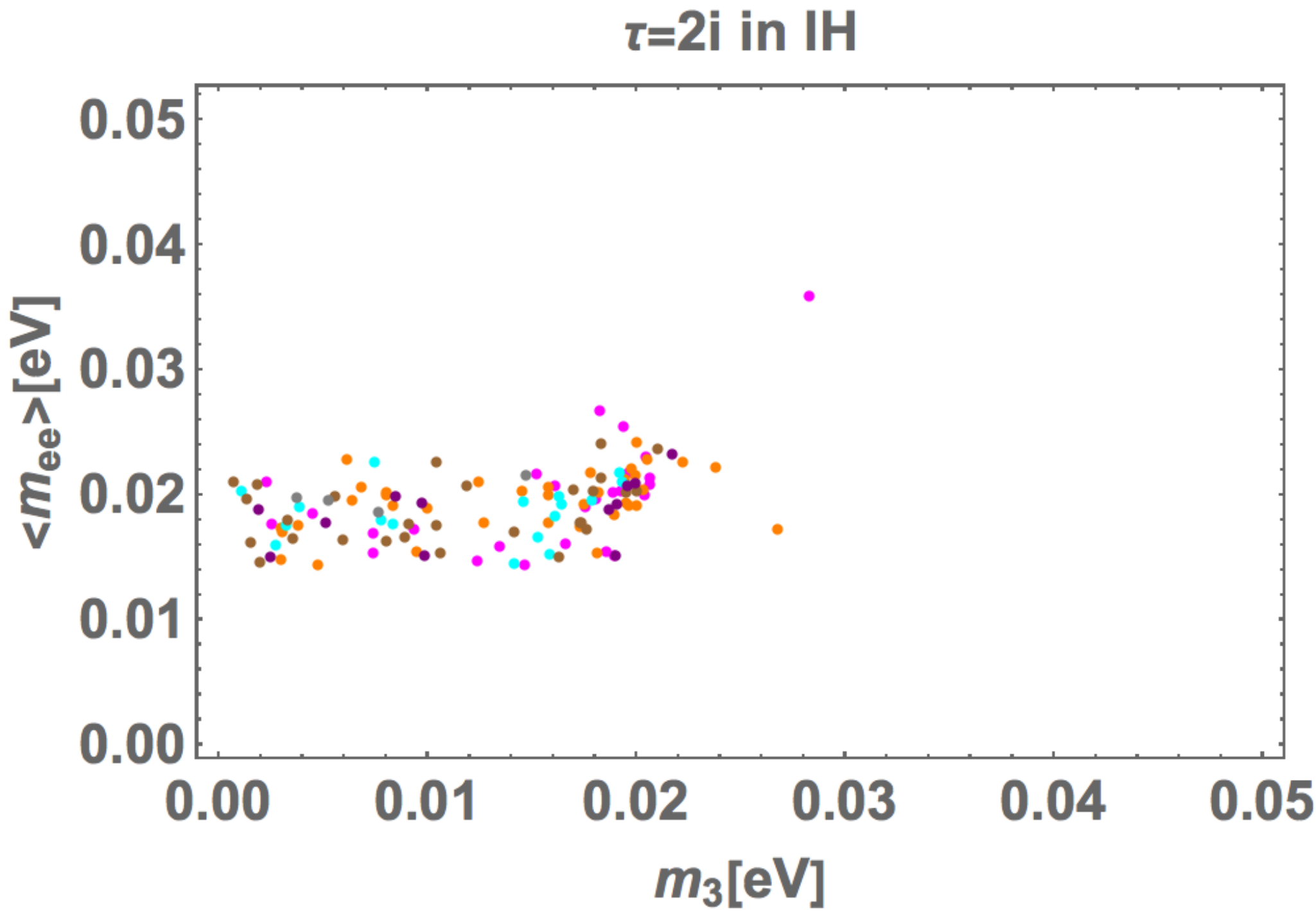}
  \end{center}
 \end{minipage}
\begin{minipage}{0.49\hsize}
  \begin{center}
  \includegraphics[scale=0.3]{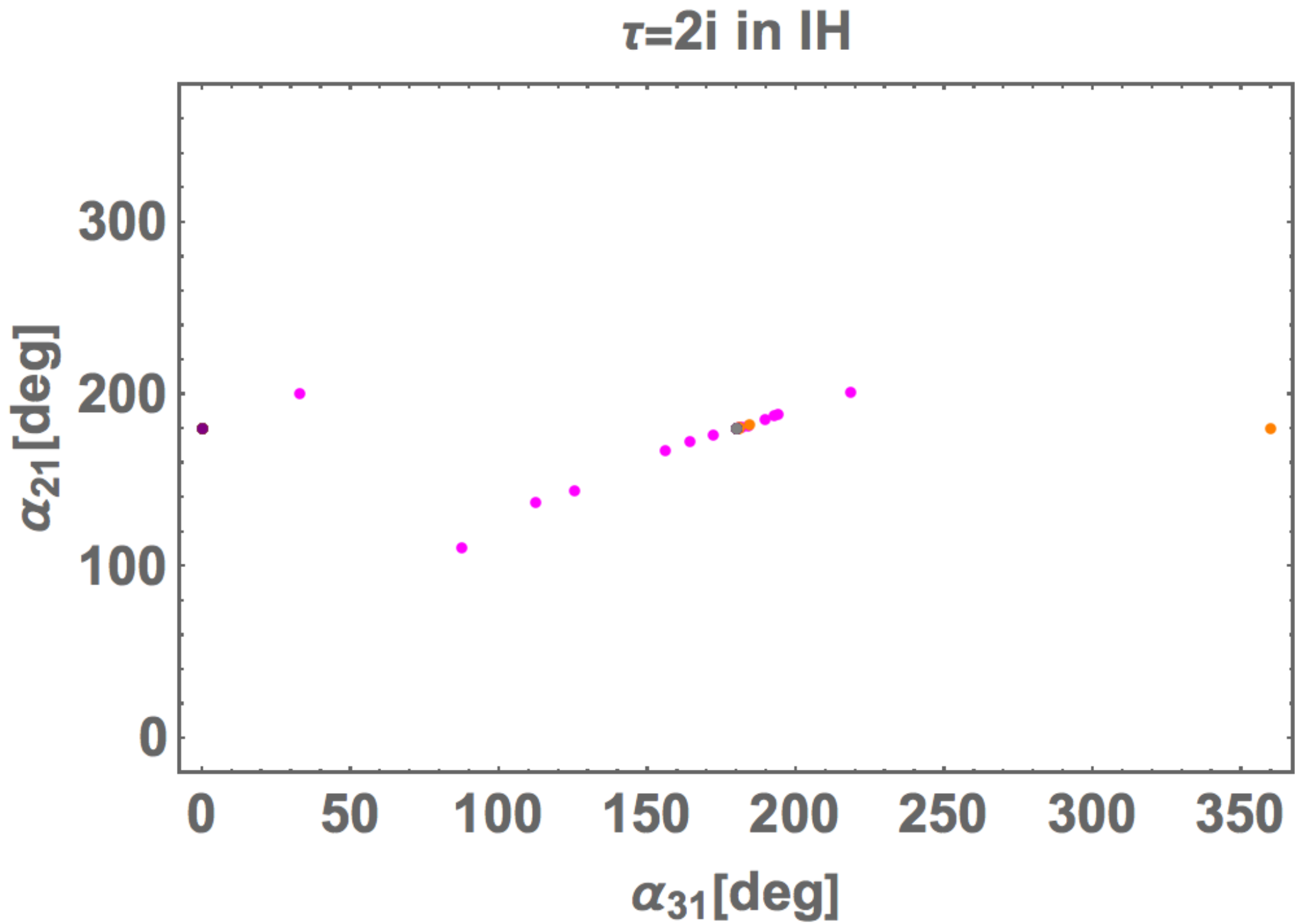}
  \end{center}
 \end{minipage}
\begin{minipage}{0.49\hsize}
  \begin{center}
  \includegraphics[scale=0.3]{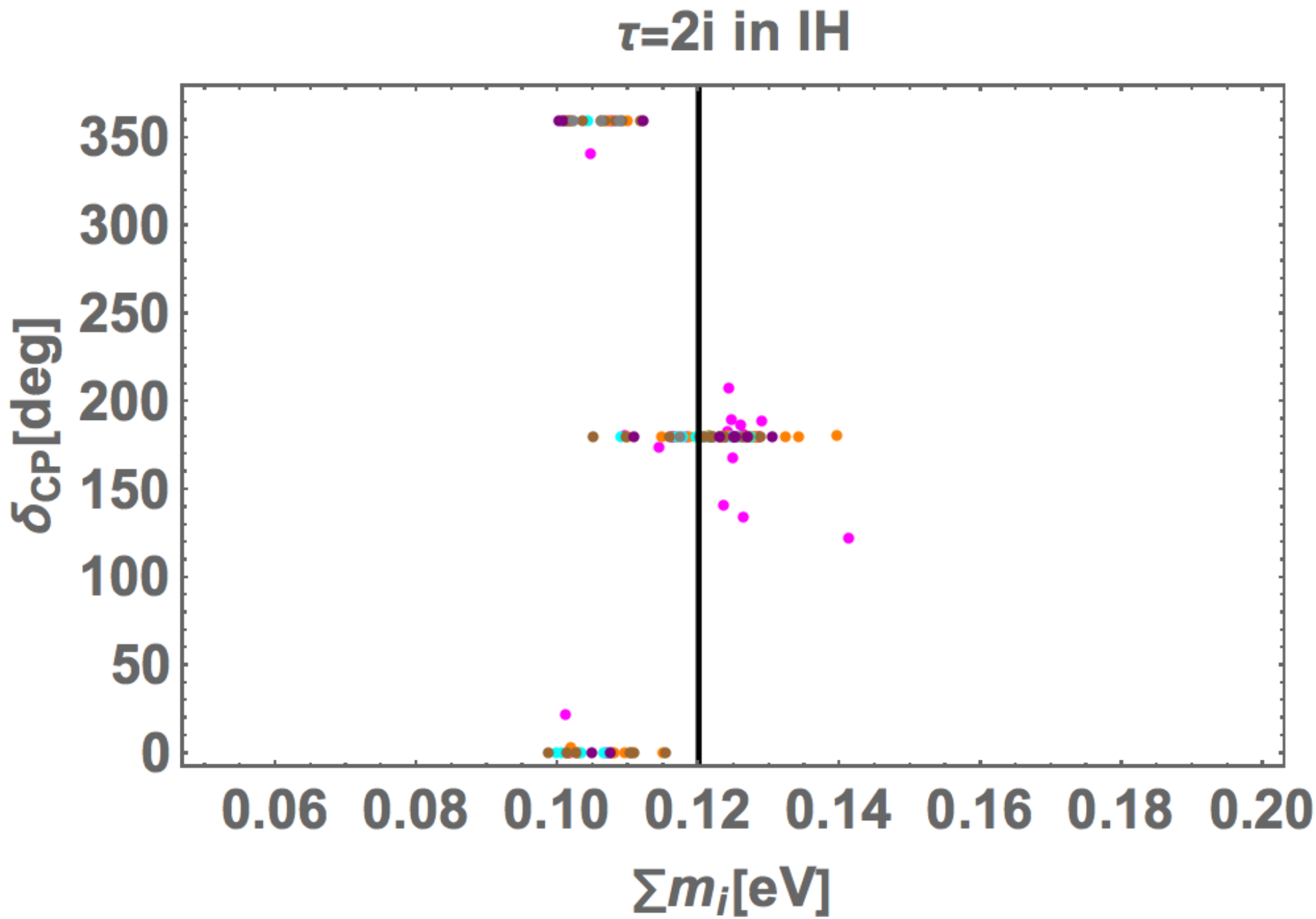}
  \end{center}
 \end{minipage}
 \caption{$|\delta \tau| < 10^{-15}$ for black, $10^{-15} \leq |\delta \tau| < 10^{-12}$ for gray, $10^{-12} \leq |\delta \tau| < 10^{-10}$ for purple, $10^{-10} \leq |\delta \tau| < 10^{-7}$ for brown, $10^{-7} \leq |\delta \tau| < 10^{-5}$ for blue green, $10^{-5} \leq |\delta \tau| < 10^{-3}$ for orange, and $10^{-3} \leq |\delta \tau| < 10^{-1}$ for magenta.}
 \label{fig.dev-tau=2i_ih}
\end{figure}
In Fig.~\ref{fig.dev-tau=2i_ih}, we show the several figures in terms of deviation from $\tau=2i$ in the same case of Fig.~\ref{fig.chi-tau=i_nh}, where the color legends are the same as the one in Fig.~\ref{fig.dev-tau=i_ih}.
The up-left one is the same as the case of up-right one in Fig.~\ref{fig.chi-tau=2i_ih}.
The up-right one is the same as the case of down-left one in Fig.~\ref{fig.chi-tau=2i_ih}.
The down-left one is the same as the case of down-right one in Fig.~\ref{fig.chi-tau=2i_ih}.
These figures suggest us that size of deviation almost run the whole ranges that are allowed by the neutrino oscillation data.

\begin{figure}[H]
\begin{minipage}{0.49\hsize}
  \begin{center}
  \includegraphics[scale=0.15]{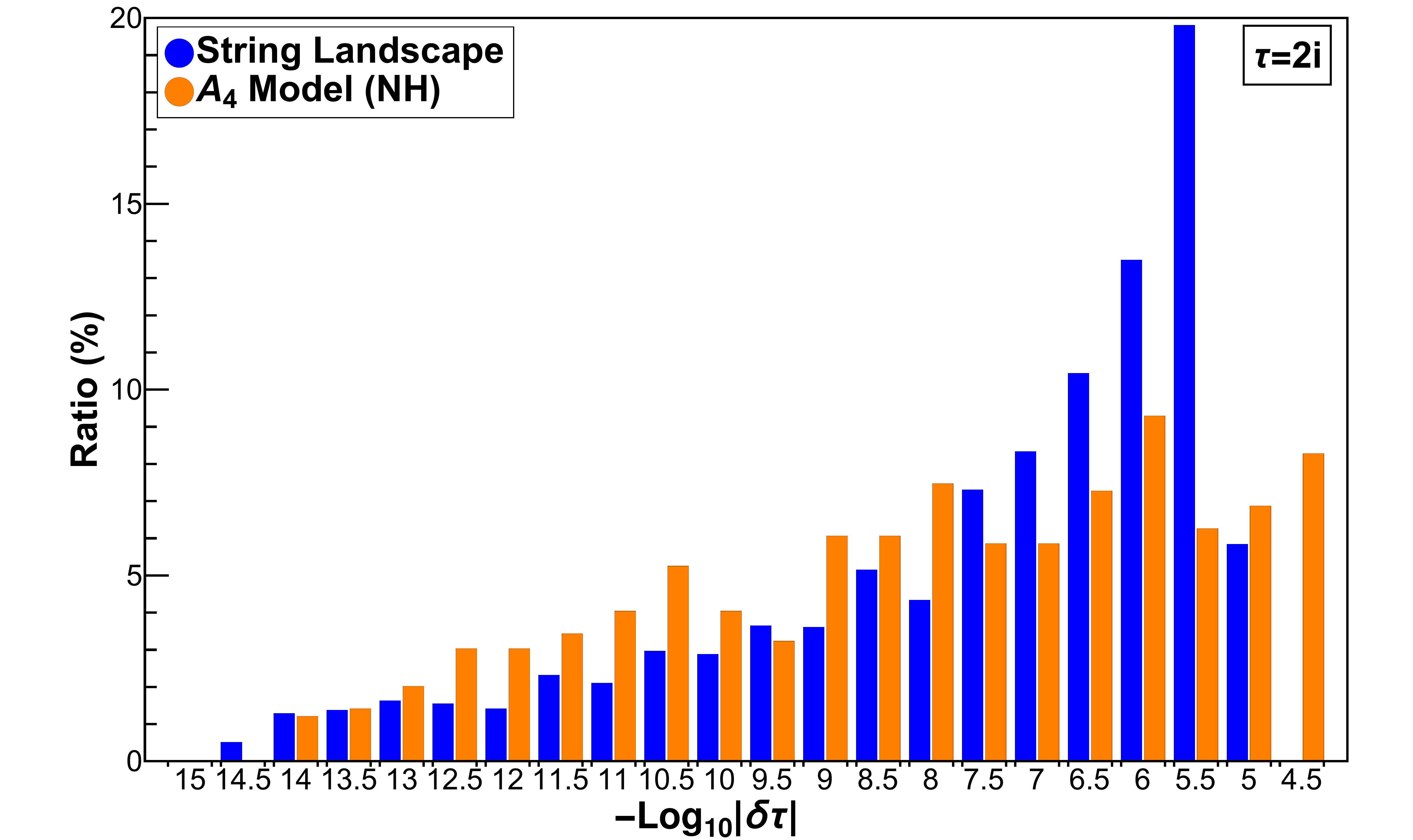}
  \end{center}
 \end{minipage}
\begin{minipage}{0.49\hsize}
  \begin{center}
  \includegraphics[scale=0.15]{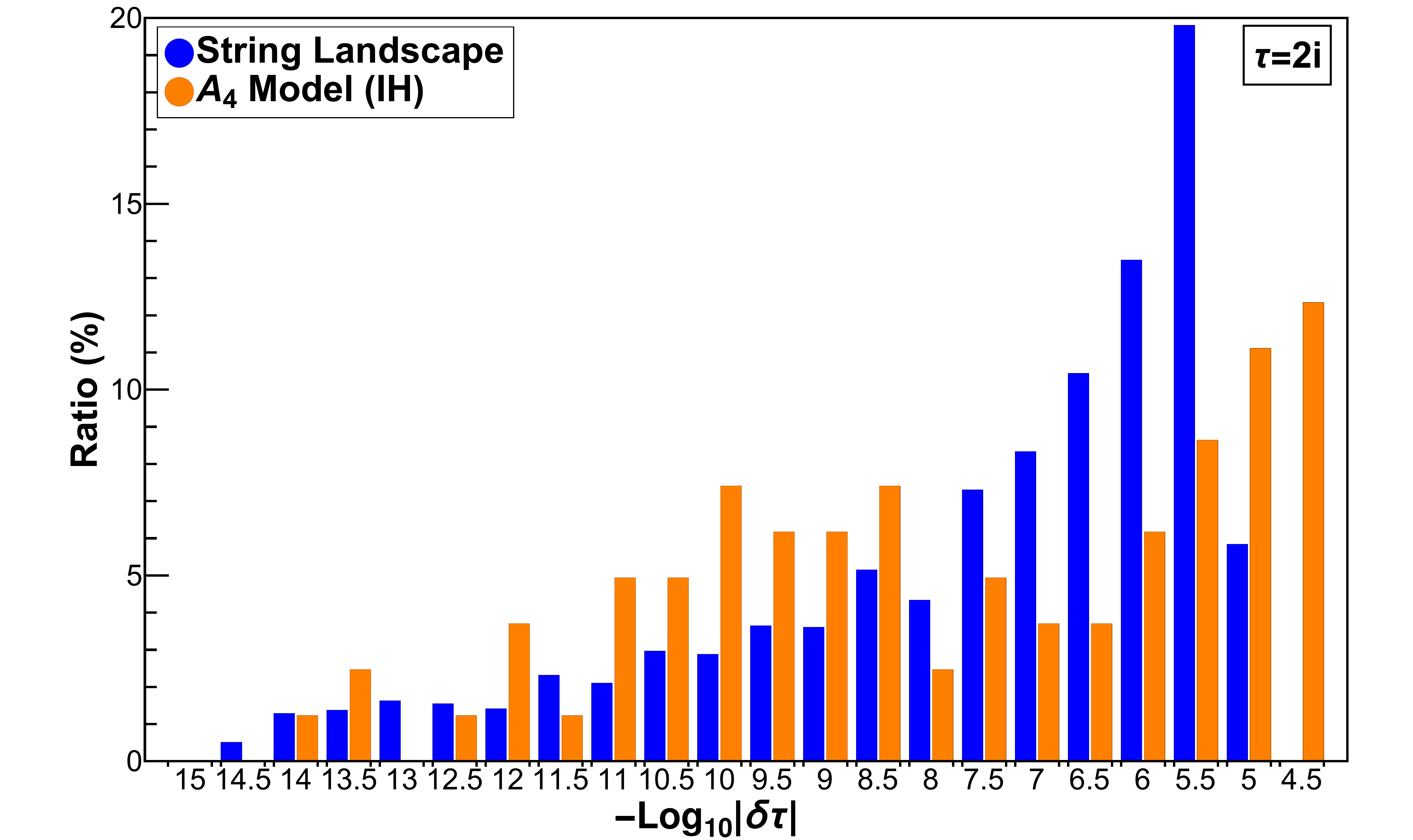}
  \end{center}
 \end{minipage}
 \caption{Ratios of the string landscape in Fig. \ref{fig:2i_up} and the $A_4$ model within $5\sigma$, where the ratios are defined as those of the number of solutions in a corresponding range of $-{\rm Log}_{10} |\delta \tau|$ to the number of whole solutions. We present the NH and the IH in the left and right panels, respectively.}
 \label{fig.ratio-tau=2i}
\end{figure}

In a similar to $\tau \simeq i$, we plot ratios of the number of solutions in a corresponding range of $-{\rm Log}_{10} |\delta \tau|$ to the number of whole solutions for both the string landscape in Fig. \ref{fig:Z2_up} and the $A_4$ model within $5\sigma$ in  Fig. \ref{fig.ratio-tau=2i}. 
It indicates both the distributions of $A_4$ model with NH and the moduli 
fields in the string landscape peak around $|\delta \tau|= {\cal O}(10^{-5})$, 
but such a signal will not be found in the IH case.

Here, we summarize our results where $\tau=\omega$ does not favor a theoretical point of view from the string landscape. Thus, we focus on $\tau=i$ and $\tau=2i$ only.
In the case of $\tau=i$ with NH, there is an intriguing tendency that the allowed region of smaller $\chi$ square is localized at smaller $\sum m_i$ that is within the cosmological bound.
Another feature is that the best fit value of Dirac CP phase $\sim195^\circ$ would be reproduced when we allow up to $5\sigma$ interval. It implies that smaller deviations $|\delta\tau|$ tend to be localized at nearby their smaller masses.
In the case of $\tau=i$ with IH, 
There are two correlations between them; one is a linear line, and another is a slightly curved one.
The solutions tend to be localized at nearby smaller mass of $m_3$ with $\langle m_{ee}\rangle=0.015,\ 0.05$ eV.
A large part of $\sum m_i$
would be ruled out by the cosmological bound.
Therefore, we would predict a narrow range of $0.1~{\rm eV}\le\sum m_i\le0.12~{\rm eV}$.
It implies that smaller deviations $|\delta\tau|$ tend to be localized at nearby their smaller masses $m_3$.
The smaller deviation would be favored in the point of view of the bound on the cosmological constraint. 
Both the distributions of $A_4$ model with NH and the moduli fields in the string landscape peak around $|\delta \tau|= {\cal O}(10^{-5})$, but such a signal is not found.
In the case of $\tau=2i$ with NH,
the smaller $\chi$ square denoted in blue is closest to the fixed point of $\tau=2i$, which would be a good tendency.
We find the allowed regions 0 eV$\le m_1\le$0.014 eV,
and 0 eV$\le \langle m_{ee}\rangle\le$0.013 eV.
The whole allowed region of $\sum m_i$ is totally within the bound on cosmological constraint; 0.058 eV$\le\sum m_i\le$ 0.082 eV.
The size of the deviation from $\tau=2i$ almost runs the whole ranges that are allowed by the neutrino oscillation data.
In the case of $\tau=2i$ with IH,
we find the allowed regions as follows:
0 eV$\le m_3\le$0.03 eV and 0.014 eV$\le\langle m_{ee}\rangle\le$0.04 eV up to $5\sigma$, but the allowed regions are localized at nearby small masses at yellow plots.
$\alpha_{21}$ is allowed by $100^\circ$ to $200^\circ$,
while $\alpha_{31}$ is wider region than $\alpha_{21}$.
Almost half the points of $\sum m_i$ would be ruled out by the cosmological bound.
Therefore, we would predict a narrow range of $0.1$eV $\le\sum m_i\le0.12$ eV that is almost the same as the one in the case of $\tau=i$.
The size of the deviation from $\tau=2i$ almost runs the whole ranges that are allowed by the neutrino oscillation data.
Both the distributions of $A_4$ model with NH and moduli fields in the string landscape peak around $|\delta\tau|={\cal O}(10^{-5})$, but such a signal is not found in the IH.

\section{Conclusions}
\label{sec:con}

The residual flavor symmetries appearing in fixed points of $PSL(2,\mathbb{Z})$ moduli space are employed in a wide variety of modular flavor models, but a small departure of the modulus from fixed points is required to realize the observed masses and mixing angles of quarks/leptons and CP-breaking effects in the bottom-up modular invariant theories. 
In this paper, we have explicitly demonstrated the breaking of residual flavor symmetry from the top-down approach.

Following Ref. \cite{Ishiguro:2020nuf}, we have studied the moduli stabilization in the context of Type IIB string theory on $T^6/(\mathbb{Z}_2\times \mathbb{Z}_2^\prime)$ orientifold. 
In Type IIB flux compactifications, it was known that $\mathbb{Z}_2$ and $\mathbb{Z}_3$ fixed points on the fundamental domain of the complex structure modulus space are statistically favored in the finite number of vacua. 
However, the volume moduli have not been stabilized yet, and a present stage of acceleration of the Universe should be realized. 
In this respect, we have incorporated non-perturbative corrections to the superpotential as well as uplifting sources to stabilize the volume moduli at the dS vacua. 
These sources naturally shift the value of $\tau$ from fixed points by a small amount. 
We find that the deviations of $\tau$ from fixed points $\langle \tau \rangle = i, w, 2i$ are statistically favored at $|\delta \tau| \simeq 10^{-5}$ and the CP symmetry $\tau\rightarrow -\bar{\tau}$ is broken in a generic choice of background fluxes. 
Since the SUSY is broken by the existence of uplifting source, the typical SUSY-breaking scale, i.e., the gravitino mass, is estimated as of ${\cal O}(10^{13})$\,GeV at the small departure $|\delta \tau| \simeq 10^{-5}$. 
In this way, the top-down approach restricts ourselves to the specific value of the modulus $\tau$ as well as the SUSY-breaking scale. 

To illustrate phenomenological implications, we analyze the concrete $A_4$ modular flavor model with an emphasis on the lepton sector. 
Under charge assignments for the lepton and Higgs sectors in Tab. \ref{tab:fields}, we have presented several predictions in the vicinity of three fixed points by a global $\chi^2$ analysis in both the normal and inverted hierarchies of neutrinos. 
It turns out that there exist many phenomenologically promising models around $\langle \tau \rangle = i$ with the normal hierarchy, whose number is compared with that of the string landscape in Fig. \ref{fig.ratio-tau=i}. 
It implies similar distributions for string and $A_4$ models with respect to $\delta \tau$. 
Furthermore, there is an intriguing tendency that allowed region of smaller $\chi$ square is localized at smaller $\sum m_i$ that is within the cosmological bound.

Before closing our paper, it is worthwhile mentioning the quasi-stable DM candidate due to the tiny deviation from the fixed points in $\tau=i,\ 2i$. In Ref.~\cite{Kobayashi:2021ajl}, especially, DM and neutrino oscillation data can simultaneously be explained at $\tau=i$ where DM is a Majorana heavy fermion with modular weight $-2$. In this set up~\footnote{In order to identify DM, we would need to assign a singlet under $A_4$ symmetry in order to avoid mixings among Majorana fermions that spoil the stability of DM. Also, we might need to construct a model with all the singlets under $A_4$ to get neutrino oscillation data. In this sense, our model has to be modified when there is DM in a theory.}, DM decays into leptons and Higgses via a Dirac term. Assuming order one free parameter and the DM mass ($m_X$) is much heavier than the leptons and Higgses, we estimate its lifetime ($\tau_X$) as follows:
\begin{align} 
\tau_X
\simeq 
1.32\times 10^{-25}\times
\left|Y^{(6)}_1\right|^{-2}
\left( \frac{1\ {\rm TeV}}{ m_X} \right) {\rm sec}.
\end{align}
When $m_X=1$ TeV, the upper limit of $\left|Y^{(6)}_1\right|$ be less than the order $10^{-21}$ in order $X$ to be a quasi-stable DM imposing
$10^{17} {\rm sec}\lesssim \tau_X$.
Here, $10^{17}$ sec is the age of the Universe.
This constraint is equivalent to  $|\delta\tau|\lesssim 5.57\times 10^{-9}$ that is within our valid parameter space.

\acknowledgments

This work was supported by JSPS KAKENHI Grant Numbers JP20K14477 (Hajime O.) and JP22J12877 (K.I). 
The work of Hiroshi O. is supported by the Junior Research Group (JRG) Program at the Asia-Pacific Center for Theoretical
Physics (APCTP) through the Science and Technology Promotion Fund and Lottery Fund of the Korean Government and was supported by the Korean Local Governments-Gyeongsangbuk-do Province and Pohang City. 
Hiroshi O. is sincerely grateful for all the KIAS members.

\appendix

\section{$A_4$ modular forms}
\label{app}

Note that the modulus-dependent modular forms are constructed by the weight 2 modular form, 
\begin{align}
Y_{\bf 3}^{(2)} = 
\begin{pmatrix}
   Y_1\\
   Y_2\\
   Y_3\\
\end{pmatrix}
,
\end{align}
with
\begin{align}
Y_1(\tau) &= \frac{i}{2\pi}\left( \frac{\eta'(\tau/3)}{\eta(\tau/3)}  +\frac{\eta'((\tau +1)/3)}{\eta((\tau+1)/3)}  
+\frac{\eta'((\tau +2)/3)}{\eta((\tau+2)/3)} - \frac{27\eta'(3\tau)}{\eta(3\tau)}  \right), \\
Y_2(\tau) &= \frac{-i}{\pi}\left( \frac{\eta'(\tau/3)}{\eta(\tau/3)}  +\omega^2\frac{\eta'((\tau +1)/3)}{\eta((\tau+1)/3)}  
+\omega \frac{\eta'((\tau +2)/3)}{\eta((\tau+2)/3)}  \right) , \label{eq:Yi} \\ 
Y_3(\tau) &= \frac{-i}{\pi}\left( \frac{\eta'(\tau/3)}{\eta(\tau/3)}  +\omega\frac{\eta'((\tau +1)/3)}{\eta((\tau+1)/3)}  
+\omega^2 \frac{\eta'((\tau +2)/3)}{\eta((\tau+2)/3)}  \right)\,, 
\end{align}
where $\eta(\tau)$ denotes the Dedekind eta-function and $\omega=e^{2\pi i /3}$. 
Recalling that the other modular forms are constructed by tensor products of $Y_{\bf 3}^{(2)}$, we list the modular forms used in our analysis:
\begin{align}
&{ Y^{\rm (4)}_{\bf 3}}(\tau)=
\begin{pmatrix}
Y_1^2-Y_2Y_3  \\
Y_3^2-Y_1Y_2 \\
Y_2^2-Y_1Y_3
\end{pmatrix}\,,
\nonumber\\
&Y_{\bf 1}^{(4)}=Y_1^2+2Y_2Y_3\,, \qquad\qquad 
Y_{\bf 1'}^{(4)}=Y_3^2+2Y_1Y_2\,. 
\nonumber\\
&{ Y^{\rm (6)}_{\bf 3}}(\tau)=Y_{\bf 1}^{(4)} { Y^{\rm (2)}_{\bf 3}}(\tau)=(Y_1^2+2Y_2Y_3)
\begin{pmatrix}
Y_1  \\
Y_2 \\
Y_3
\end{pmatrix}\,, \notag\\
&{ Y^{\rm (6)}_{\bf 3^\prime}}(\tau)=Y_{\bf 1'}^{(4)} { Y^{\rm (2)}_{\bf 3}}(\tau)=(Y_3^2+2Y_1Y_2)
\begin{pmatrix}
Y_3  \\
Y_1 \\
Y_2
\end{pmatrix}\,
.
\end{align}

\bibliography{referencesv2}{}

\providecommand{\href}[2]{#2}\begingroup\raggedright\begin{thebibliography}{10}

\bibitem{deAdelhartToorop:2011re}
R.~de~Adelhart~Toorop, F.~Feruglio and C.~Hagedorn, \emph{{Finite Modular
  Groups and Lepton Mixing}},
  \href{https://doi.org/10.1016/j.nuclphysb.2012.01.017}{\emph{Nucl. Phys. B}
  {\bfseries 858} (2012) 437}
  [\href{https://arxiv.org/abs/1112.1340}{{\ttfamily 1112.1340}}].

\bibitem{Ferrara:1989qb}
S.~Ferrara, .D.~Lust and S.~Theisen, \emph{{Target Space Modular Invariance and
  Low-Energy Couplings in Orbifold Compactifications}},
  \href{https://doi.org/10.1016/0370-2693(89)90631-X}{\emph{Phys. Lett. B}
  {\bfseries 233} (1989) 147}.

\bibitem{Lerche:1989cs}
W.~Lerche, D.~Lust and N.P.~Warner, \emph{{Duality Symmetries in $N=2$
  Landau-ginzburg Models}},
  \href{https://doi.org/10.1016/0370-2693(89)90686-2}{\emph{Phys. Lett. B}
  {\bfseries 231} (1989) 417}.

\bibitem{Lauer:1990tm}
J.~Lauer, J.~Mas and H.P.~Nilles, \emph{{Twisted sector representations of
  discrete background symmetries for two-dimensional orbifolds}},
  \href{https://doi.org/10.1016/0550-3213(91)90095-F}{\emph{Nucl. Phys. B}
  {\bfseries 351} (1991) 353}.

\bibitem{Baur:2019kwi}
A.~Baur, H.P.~Nilles, A.~Trautner and P.K.S.~Vaudrevange, \emph{{Unification of
  Flavor, CP, and Modular Symmetries}},
  \href{https://doi.org/10.1016/j.physletb.2019.03.066}{\emph{Phys. Lett. B}
  {\bfseries 795} (2019) 7} [\href{https://arxiv.org/abs/1901.03251}{{\ttfamily
  1901.03251}}].

\bibitem{Baur:2019iai}
A.~Baur, H.P.~Nilles, A.~Trautner and P.K.S.~Vaudrevange, \emph{{A String
  Theory of Flavor and $\mathscr {CP}$}},
  \href{https://doi.org/10.1016/j.nuclphysb.2019.114737}{\emph{Nucl. Phys. B}
  {\bfseries 947} (2019) 114737}
  [\href{https://arxiv.org/abs/1908.00805}{{\ttfamily 1908.00805}}].

\bibitem{Kobayashi:2018rad}
T.~Kobayashi, S.~Nagamoto, S.~Takada, S.~Tamba and T.H.~Tatsuishi,
  \emph{{Modular symmetry and non-Abelian discrete flavor symmetries in string
  compactification}},
  \href{https://doi.org/10.1103/PhysRevD.97.116002}{\emph{Phys. Rev. D}
  {\bfseries 97} (2018) 116002}
  [\href{https://arxiv.org/abs/1804.06644}{{\ttfamily 1804.06644}}].

\bibitem{Kobayashi:2018bff}
T.~Kobayashi and S.~Tamba, \emph{{Modular forms of finite modular subgroups
  from magnetized D-brane models}},
  \href{https://doi.org/10.1103/PhysRevD.99.046001}{\emph{Phys. Rev. D}
  {\bfseries 99} (2019) 046001}
  [\href{https://arxiv.org/abs/1811.11384}{{\ttfamily 1811.11384}}].

\bibitem{Ohki:2020bpo}
H.~Ohki, S.~Uemura and R.~Watanabe, \emph{{Modular flavor symmetry on a
  magnetized torus}},
  \href{https://doi.org/10.1103/PhysRevD.102.085008}{\emph{Phys. Rev. D}
  {\bfseries 102} (2020) 085008}
  [\href{https://arxiv.org/abs/2003.04174}{{\ttfamily 2003.04174}}].

\bibitem{Kikuchi:2020frp}
S.~Kikuchi, T.~Kobayashi, S.~Takada, T.H.~Tatsuishi and H.~Uchida,
  \emph{{Revisiting modular symmetry in magnetized torus and orbifold
  compactifications}},
  \href{https://doi.org/10.1103/PhysRevD.102.105010}{\emph{Phys. Rev. D}
  {\bfseries 102} (2020) 105010}
  [\href{https://arxiv.org/abs/2005.12642}{{\ttfamily 2005.12642}}].

\bibitem{Kikuchi:2020nxn}
S.~Kikuchi, T.~Kobayashi, H.~Otsuka, S.~Takada and H.~Uchida, \emph{{Modular
  symmetry by orbifolding magnetized $T^2\times T^2$: realization of double
  cover of $\Gamma_N$}},
  \href{https://doi.org/10.1007/JHEP11(2020)101}{\emph{JHEP} {\bfseries 11}
  (2020) 101} [\href{https://arxiv.org/abs/2007.06188}{{\ttfamily
  2007.06188}}].

\bibitem{Kikuchi:2021ogn}
S.~Kikuchi, T.~Kobayashi and H.~Uchida, \emph{{Modular flavor symmetries of
  three-generation modes on magnetized toroidal orbifolds}},
  \href{https://doi.org/10.1103/PhysRevD.104.065008}{\emph{Phys. Rev. D}
  {\bfseries 104} (2021) 065008}
  [\href{https://arxiv.org/abs/2101.00826}{{\ttfamily 2101.00826}}].

\bibitem{Almumin:2021fbk}
Y.~Almumin, M.-C.~Chen, V.~Knapp-P\'erez, S.~Ramos-S\'anchez, M.~Ratz and
  S.~Shukla, \emph{{Metaplectic Flavor Symmetries from Magnetized Tori}},
  \href{https://doi.org/10.1007/JHEP05(2021)078}{\emph{JHEP} {\bfseries 05}
  (2021) 078} [\href{https://arxiv.org/abs/2102.11286}{{\ttfamily
  2102.11286}}].

\bibitem{Baur:2020yjl}
A.~Baur, M.~Kade, H.P.~Nilles, S.~Ramos-Sanchez and P.K.S.~Vaudrevange,
  \emph{{Siegel modular flavor group and CP from string theory}},
  \href{https://doi.org/10.1016/j.physletb.2021.136176}{\emph{Phys. Lett. B}
  {\bfseries 816} (2021) 136176}
  [\href{https://arxiv.org/abs/2012.09586}{{\ttfamily 2012.09586}}].

\bibitem{Strominger:1990pd}
A.~Strominger, \emph{{SPECIAL GEOMETRY}},
  \href{https://doi.org/10.1007/BF02096559}{\emph{Commun. Math. Phys.}
  {\bfseries 133} (1990) 163}.

\bibitem{Candelas:1990pi}
P.~Candelas and X.~de~la Ossa, \emph{{Moduli Space of {Calabi-Yau} Manifolds}},
  \href{https://doi.org/10.1016/0550-3213(91)90122-E}{\emph{Nucl. Phys. B}
  {\bfseries 355} (1991) 455}.

\bibitem{Ishiguro:2020nuf}
K.~Ishiguro, T.~Kobayashi and H.~Otsuka, \emph{{Spontaneous CP violation and
  symplectic modular symmetry in Calabi-Yau compactifications}},
  \href{https://doi.org/10.1016/j.nuclphysb.2021.115598}{\emph{Nucl. Phys. B}
  {\bfseries 973} (2021) 115598}
  [\href{https://arxiv.org/abs/2010.10782}{{\ttfamily 2010.10782}}].

\bibitem{Ishiguro:2021ccl}
K.~Ishiguro, T.~Kobayashi and H.~Otsuka, \emph{{Symplectic modular symmetry in
  heterotic string vacua: flavor, CP, and R-symmetries}},
  \href{https://doi.org/10.1007/JHEP01(2022)020}{\emph{JHEP} {\bfseries 01}
  (2022) 020} [\href{https://arxiv.org/abs/2107.00487}{{\ttfamily
  2107.00487}}].

\bibitem{Feruglio:2017spp}
F.~Feruglio, \emph{{Are neutrino masses modular forms?}},  in \emph{{From My
  Vast Repertoire ...}: {Guido Altarelli's Legacy}}, A.~Levy, S.~Forte and
  G.~Ridolfi, eds., pp.~227--266 (2019),
  \href{https://doi.org/10.1142/9789813238053_0012}{DOI}
  [\href{https://arxiv.org/abs/1706.08749}{{\ttfamily 1706.08749}}].

\bibitem{Kobayashi:2018vbk}
T.~Kobayashi, K.~Tanaka and T.H.~Tatsuishi, \emph{{Neutrino mixing from finite
  modular groups}},
  \href{https://doi.org/10.1103/PhysRevD.98.016004}{\emph{Phys. Rev. D}
  {\bfseries 98} (2018) 016004}
  [\href{https://arxiv.org/abs/1803.10391}{{\ttfamily 1803.10391}}].

\bibitem{Penedo:2018nmg}
J.T.~Penedo and S.T.~Petcov, \emph{{Lepton Masses and Mixing from Modular $S_4$
  Symmetry}},
  \href{https://doi.org/10.1016/j.nuclphysb.2018.12.016}{\emph{Nucl. Phys. B}
  {\bfseries 939} (2019) 292}
  [\href{https://arxiv.org/abs/1806.11040}{{\ttfamily 1806.11040}}].

\bibitem{Novichkov:2018nkm}
P.P.~Novichkov, J.T.~Penedo, S.T.~Petcov and A.V.~Titov, \emph{{Modular A$_{5}$
  symmetry for flavour model building}},
  \href{https://doi.org/10.1007/JHEP04(2019)174}{\emph{JHEP} {\bfseries 04}
  (2019) 174} [\href{https://arxiv.org/abs/1812.02158}{{\ttfamily
  1812.02158}}].

\bibitem{Ding:2019xna}
G.-J.~Ding, S.F.~King and X.-G.~Liu, \emph{{Neutrino mass and mixing with $A_5$
  modular symmetry}},
  \href{https://doi.org/10.1103/PhysRevD.100.115005}{\emph{Phys. Rev. D}
  {\bfseries 100} (2019) 115005}
  [\href{https://arxiv.org/abs/1903.12588}{{\ttfamily 1903.12588}}].

\bibitem{Liu:2019khw}
X.-G.~Liu and G.-J.~Ding, \emph{{Neutrino Masses and Mixing from Double
  Covering of Finite Modular Groups}},
  \href{https://doi.org/10.1007/JHEP08(2019)134}{\emph{JHEP} {\bfseries 08}
  (2019) 134} [\href{https://arxiv.org/abs/1907.01488}{{\ttfamily
  1907.01488}}].

\bibitem{Chen:2020udk}
P.~Chen, G.-J.~Ding, J.-N.~Lu and J.W.F.~Valle, \emph{{Predictions from warped
  flavor dynamics based on the $T′$ family group}},
  \href{https://doi.org/10.1103/PhysRevD.102.095014}{\emph{Phys. Rev. D}
  {\bfseries 102} (2020) 095014}
  [\href{https://arxiv.org/abs/2003.02734}{{\ttfamily 2003.02734}}].

\bibitem{Novichkov:2020eep}
P.P.~Novichkov, J.T.~Penedo and S.T.~Petcov, \emph{{Double cover of modular
  $S_4$ for flavour model building}},
  \href{https://doi.org/10.1016/j.nuclphysb.2020.115301}{\emph{Nucl. Phys. B}
  {\bfseries 963} (2021) 115301}
  [\href{https://arxiv.org/abs/2006.03058}{{\ttfamily 2006.03058}}].

\bibitem{Liu:2020akv}
X.-G.~Liu, C.-Y.~Yao and G.-J.~Ding, \emph{{Modular invariant quark and lepton
  models in double covering of $S_4$ modular group}},
  \href{https://doi.org/10.1103/PhysRevD.103.056013}{\emph{Phys. Rev. D}
  {\bfseries 103} (2021) 056013}
  [\href{https://arxiv.org/abs/2006.10722}{{\ttfamily 2006.10722}}].

\bibitem{Wang:2020lxk}
X.~Wang, B.~Yu and S.~Zhou, \emph{{Double covering of the modular $A_5$ group
  and lepton flavor mixing in the minimal seesaw model}},
  \href{https://doi.org/10.1103/PhysRevD.103.076005}{\emph{Phys. Rev. D}
  {\bfseries 103} (2021) 076005}
  [\href{https://arxiv.org/abs/2010.10159}{{\ttfamily 2010.10159}}].

\bibitem{Yao:2020zml}
C.-Y.~Yao, X.-G.~Liu and G.-J.~Ding, \emph{{Fermion masses and mixing from the
  double cover and metaplectic cover of the $A_5$ modular group}},
  \href{https://doi.org/10.1103/PhysRevD.103.095013}{\emph{Phys. Rev. D}
  {\bfseries 103} (2021) 095013}
  [\href{https://arxiv.org/abs/2011.03501}{{\ttfamily 2011.03501}}].

\bibitem{Ding:2020msi}
G.-J.~Ding, S.F.~King, C.-C.~Li and Y.-L.~Zhou, \emph{{Modular Invariant Models
  of Leptons at Level 7}},
  \href{https://doi.org/10.1007/JHEP08(2020)164}{\emph{JHEP} {\bfseries 08}
  (2020) 164} [\href{https://arxiv.org/abs/2004.12662}{{\ttfamily
  2004.12662}}].

\bibitem{Kobayashi:2021pav}
T.~Kobayashi, H.~Otsuka, M.~Tanimoto and K.~Yamamoto, \emph{{Modular symmetry
  in the SMEFT}},
  \href{https://doi.org/10.1103/PhysRevD.105.055022}{\emph{Phys. Rev. D}
  {\bfseries 105} (2022) 055022}
  [\href{https://arxiv.org/abs/2112.00493}{{\ttfamily 2112.00493}}].

\bibitem{Kobayashi:2022jvy}
T.~Kobayashi, H.~Otsuka, M.~Tanimoto and K.~Yamamoto, \emph{{Lepton flavor
  violation, lepton $(g-2)_{\mu,\,e}$ and electron EDM in the modular
  symmetry}},  \href{https://arxiv.org/abs/2204.12325}{{\ttfamily 2204.12325}}.

\bibitem{Kobayashi:2021uam}
T.~Kobayashi and H.~Otsuka, \emph{{On stringy origin of minimal flavor
  violation}},
  \href{https://doi.org/10.1140/epjc/s10052-022-09986-4}{\emph{Eur. Phys. J. C}
  {\bfseries 82} (2022) 25} [\href{https://arxiv.org/abs/2108.02700}{{\ttfamily
  2108.02700}}].

\bibitem{Novichkov:2019sqv}
P.P.~Novichkov, J.T.~Penedo, S.T.~Petcov and A.V.~Titov, \emph{{Generalised CP
  Symmetry in Modular-Invariant Models of Flavour}},
  \href{https://doi.org/10.1007/JHEP07(2019)165}{\emph{JHEP} {\bfseries 07}
  (2019) 165} [\href{https://arxiv.org/abs/1905.11970}{{\ttfamily
  1905.11970}}].

\bibitem{Novichkov:2018ovf}
P.P.~Novichkov, J.T.~Penedo, S.T.~Petcov and A.V.~Titov, \emph{{Modular S$_{4}$
  models of lepton masses and mixing}},
  \href{https://doi.org/10.1007/JHEP04(2019)005}{\emph{JHEP} {\bfseries 04}
  (2019) 005} [\href{https://arxiv.org/abs/1811.04933}{{\ttfamily
  1811.04933}}].

\bibitem{Novichkov:2018yse}
P.P.~Novichkov, S.T.~Petcov and M.~Tanimoto, \emph{{Trimaximal Neutrino Mixing
  from Modular A4 Invariance with Residual Symmetries}},
  \href{https://doi.org/10.1016/j.physletb.2019.04.043}{\emph{Phys. Lett. B}
  {\bfseries 793} (2019) 247}
  [\href{https://arxiv.org/abs/1812.11289}{{\ttfamily 1812.11289}}].

\bibitem{Ding:2019gof}
G.-J.~Ding, S.F.~King, X.-G.~Liu and J.-N.~Lu, \emph{{Modular S$_{4}$ and
  A$_{4}$ symmetries and their fixed points: new predictive examples of lepton
  mixing}}, \href{https://doi.org/10.1007/JHEP12(2019)030}{\emph{JHEP}
  {\bfseries 12} (2019) 030}
  [\href{https://arxiv.org/abs/1910.03460}{{\ttfamily 1910.03460}}].

\bibitem{Okada:2019uoy}
H.~Okada and M.~Tanimoto, \emph{{Towards unification of quark and lepton
  flavors in $A_4$ modular invariance}},
  \href{https://doi.org/10.1140/epjc/s10052-021-08845-y}{\emph{Eur. Phys. J. C}
  {\bfseries 81} (2021) 52} [\href{https://arxiv.org/abs/1905.13421}{{\ttfamily
  1905.13421}}].

\bibitem{King:2019vhv}
S.F.~King and Y.-L.~Zhou, \emph{{Trimaximal TM$_1$ mixing with two modular
  $S_4$ groups}},
  \href{https://doi.org/10.1103/PhysRevD.101.015001}{\emph{Phys. Rev. D}
  {\bfseries 101} (2020) 015001}
  [\href{https://arxiv.org/abs/1908.02770}{{\ttfamily 1908.02770}}].

\bibitem{Okada:2020rjb}
H.~Okada and M.~Tanimoto, \emph{{Quark and lepton flavors with common modulus
  $\tau$ in $A_4$ modular symmetry}},
  \href{https://arxiv.org/abs/2005.00775}{{\ttfamily 2005.00775}}.

\bibitem{Okada:2020ukr}
H.~Okada and M.~Tanimoto, \emph{{Modular invariant flavor model of $A_4$ and
  hierarchical structures at nearby fixed points}},
  \href{https://doi.org/10.1103/PhysRevD.103.015005}{\emph{Phys. Rev. D}
  {\bfseries 103} (2021) 015005}
  [\href{https://arxiv.org/abs/2009.14242}{{\ttfamily 2009.14242}}].

\bibitem{Okada:2020brs}
H.~Okada and M.~Tanimoto, \emph{{Spontaneous CP violation by modulus $\tau$ in
  $A_4$ model of lepton flavors}},
  \href{https://doi.org/10.1007/JHEP03(2021)010}{\emph{JHEP} {\bfseries 03}
  (2021) 010} [\href{https://arxiv.org/abs/2012.01688}{{\ttfamily
  2012.01688}}].

\bibitem{Feruglio:2021dte}
F.~Feruglio, V.~Gherardi, A.~Romanino and A.~Titov, \emph{{Modular invariant
  dynamics and fermion mass hierarchies around $\tau = i$}},
  \href{https://doi.org/10.1007/JHEP05(2021)242}{\emph{JHEP} {\bfseries 05}
  (2021) 242} [\href{https://arxiv.org/abs/2101.08718}{{\ttfamily
  2101.08718}}].

\bibitem{Kobayashi:2021ajl}
T.~Kobayashi, H.~Okada and Y.~Orikasa, \emph{{Dark matter stability at fixed
  points in a modular $A_4$ symmetry}},
  \href{https://arxiv.org/abs/2111.05674}{{\ttfamily 2111.05674}}.

\bibitem{Kobayashi:2019xvz}
T.~Kobayashi, Y.~Shimizu, K.~Takagi, M.~Tanimoto and T.H.~Tatsuishi,
  \emph{{$A_4$ lepton flavor model and modulus stabilization from $S_4$ modular
  symmetry}}, \href{https://doi.org/10.1103/PhysRevD.100.115045}{\emph{Phys.
  Rev. D} {\bfseries 100} (2019) 115045}
  [\href{https://arxiv.org/abs/1909.05139}{{\ttfamily 1909.05139}}].

\bibitem{Kobayashi:2019uyt}
T.~Kobayashi, Y.~Shimizu, K.~Takagi, M.~Tanimoto, T.H.~Tatsuishi and H.~Uchida,
  \emph{{$CP$ violation in modular invariant flavor models}},
  \href{https://doi.org/10.1103/PhysRevD.101.055046}{\emph{Phys. Rev. D}
  {\bfseries 101} (2020) 055046}
  [\href{https://arxiv.org/abs/1910.11553}{{\ttfamily 1910.11553}}].

\bibitem{Kobayashi:2020uaj}
T.~Kobayashi and H.~Otsuka, \emph{{Challenge for spontaneous $CP$ violation in
  Type IIB orientifolds with fluxes}},
  \href{https://doi.org/10.1103/PhysRevD.102.026004}{\emph{Phys. Rev. D}
  {\bfseries 102} (2020) 026004}
  [\href{https://arxiv.org/abs/2004.04518}{{\ttfamily 2004.04518}}].

\bibitem{Ishiguro:2020tmo}
K.~Ishiguro, T.~Kobayashi and H.~Otsuka, \emph{{Landscape of Modular Symmetric
  Flavor Models}}, \href{https://doi.org/10.1007/JHEP03(2021)161}{\emph{JHEP}
  {\bfseries 03} (2021) 161}
  [\href{https://arxiv.org/abs/2011.09154}{{\ttfamily 2011.09154}}].

\bibitem{Novichkov:2022wvg}
P.P.~Novichkov, J.T.~Penedo and S.T.~Petcov, \emph{{Modular Flavour Symmetries
  and Modulus Stabilisation}},
  \href{https://arxiv.org/abs/2201.02020}{{\ttfamily 2201.02020}}.

\bibitem{Kobayashi:2020hoc}
T.~Kobayashi and H.~Otsuka, \emph{{Classification of discrete modular
  symmetries in Type IIB flux vacua}},
  \href{https://doi.org/10.1103/PhysRevD.101.106017}{\emph{Phys. Rev. D}
  {\bfseries 101} (2020) 106017}
  [\href{https://arxiv.org/abs/2001.07972}{{\ttfamily 2001.07972}}].

\bibitem{DeWolfe:2004ns}
O.~DeWolfe, A.~Giryavets, S.~Kachru and W.~Taylor, \emph{{Enumerating flux
  vacua with enhanced symmetries}},
  \href{https://doi.org/10.1088/1126-6708/2005/02/037}{\emph{JHEP} {\bfseries
  02} (2005) 037} [\href{https://arxiv.org/abs/hep-th/0411061}{{\ttfamily
  hep-th/0411061}}].

\bibitem{Kachru:2003aw}
S.~Kachru, R.~Kallosh, A.D.~Linde and S.P.~Trivedi, \emph{{De Sitter vacua in
  string theory}},
  \href{https://doi.org/10.1103/PhysRevD.68.046005}{\emph{Phys. Rev. D}
  {\bfseries 68} (2003) 046005}
  [\href{https://arxiv.org/abs/hep-th/0301240}{{\ttfamily hep-th/0301240}}].

\bibitem{Kikuchi:2022pkd}
S.~Kikuchi, T.~Kobayashi, K.~Nasu, H.~Otsuka, S.~Takada and H.~Uchida,
  \emph{{Modular symmetry of soft supersymmetry breaking terms}},
  \href{https://arxiv.org/abs/2203.14667}{{\ttfamily 2203.14667}}.

\bibitem{Du:2020ylx}
X.~Du and F.~Wang, \emph{{SUSY breaking constraints on modular flavor $S_{3}$
  invariant SU(5) GUT model}},
  \href{https://doi.org/10.1007/JHEP02(2021)221}{\emph{JHEP} {\bfseries 02}
  (2021) 221} [\href{https://arxiv.org/abs/2012.01397}{{\ttfamily
  2012.01397}}].

\bibitem{Kobayashi:2021jqu}
T.~Kobayashi, T.~Shimomura and M.~Tanimoto, \emph{{Soft supersymmetry breaking
  terms and lepton flavor violations in modular flavor models}},
  \href{https://doi.org/10.1016/j.physletb.2021.136452}{\emph{Phys. Lett. B}
  {\bfseries 819} (2021) 136452}
  [\href{https://arxiv.org/abs/2102.10425}{{\ttfamily 2102.10425}}].

\bibitem{Otsuka:2022rak}
H.~Otsuka and H.~Okada, \emph{{Radiative neutrino masses from modular $A_4$
  symmetry and supersymmetry breaking}},
  \href{https://arxiv.org/abs/2202.10089}{{\ttfamily 2202.10089}}.

\bibitem{Gukov:1999ya}
S.~Gukov, C.~Vafa and E.~Witten, \emph{{CFT's from Calabi-Yau four folds}},
  \href{https://doi.org/10.1016/S0550-3213(00)00373-4}{\emph{Nucl. Phys. B}
  {\bfseries 584} (2000) 69}
  [\href{https://arxiv.org/abs/hep-th/9906070}{{\ttfamily hep-th/9906070}}].

\bibitem{Betzler:2019kon}
P.~Betzler and E.~Plauschinn, \emph{{Type IIB flux vacua and tadpole
  cancellation}}, \href{https://doi.org/10.1002/prop.201900065}{\emph{Fortsch.
  Phys.} {\bfseries 67} (2019) 1900065}
  [\href{https://arxiv.org/abs/1905.08823}{{\ttfamily 1905.08823}}].

\bibitem{Candelas:1997eh}
P.~Candelas, E.~Perevalov and G.~Rajesh, \emph{{Toric geometry and enhanced
  gauge symmetry of F theory / heterotic vacua}},
  \href{https://doi.org/10.1016/S0550-3213(97)00563-4}{\emph{Nucl. Phys. B}
  {\bfseries 507} (1997) 445}
  [\href{https://arxiv.org/abs/hep-th/9704097}{{\ttfamily hep-th/9704097}}].

\bibitem{Taylor:2015xtz}
W.~Taylor and Y.-N.~Wang, \emph{{The F-theory geometry with most flux vacua}},
  \href{https://doi.org/10.1007/JHEP12(2015)164}{\emph{JHEP} {\bfseries 12}
  (2015) 164} [\href{https://arxiv.org/abs/1511.03209}{{\ttfamily
  1511.03209}}].

\bibitem{Demirtas:2019sip}
M.~Demirtas, M.~Kim, L.~Mcallister and J.~Moritz, \emph{{Vacua with Small Flux
  Superpotential}},
  \href{https://doi.org/10.1103/PhysRevLett.124.211603}{\emph{Phys. Rev. Lett.}
  {\bfseries 124} (2020) 211603}
  [\href{https://arxiv.org/abs/1912.10047}{{\ttfamily 1912.10047}}].

\bibitem{Honma:2021klo}
Y.~Honma and H.~Otsuka, \emph{{Small flux superpotential in F-theory
  compactifications}},
  \href{https://doi.org/10.1103/PhysRevD.103.126022}{\emph{Phys. Rev. D}
  {\bfseries 103} (2021) 126022}
  [\href{https://arxiv.org/abs/2103.03003}{{\ttfamily 2103.03003}}].

\bibitem{Abe:2006xi}
H.~Abe, T.~Higaki and T.~Kobayashi, \emph{{Remark on integrating out heavy
  moduli in flux compactification}},
  \href{https://doi.org/10.1103/PhysRevD.74.045012}{\emph{Phys. Rev. D}
  {\bfseries 74} (2006) 045012}
  [\href{https://arxiv.org/abs/hep-th/0606095}{{\ttfamily hep-th/0606095}}].

\bibitem{KamLAND-Zen:2016pfg}
{\scshape KamLAND-Zen} collaboration, \emph{{Search for Majorana Neutrinos near
  the Inverted Mass Hierarchy Region with KamLAND-Zen}},
  \href{https://doi.org/10.1103/PhysRevLett.117.082503}{\emph{Phys. Rev. Lett.}
  {\bfseries 117} (2016) 082503}
  [\href{https://arxiv.org/abs/1605.02889}{{\ttfamily 1605.02889}}].

\bibitem{Esteban:2020cvm}
I.~Esteban, M.C.~Gonzalez-Garcia, M.~Maltoni, T.~Schwetz and A.~Zhou,
  \emph{{The fate of hints: updated global analysis of three-flavor neutrino
  oscillations}}, \href{https://doi.org/10.1007/JHEP09(2020)178}{\emph{JHEP}
  {\bfseries 09} (2020) 178}
  [\href{https://arxiv.org/abs/2007.14792}{{\ttfamily 2007.14792}}].

\bibitem{Bjorkeroth:2015ora}
F.~Bj\"orkeroth, F.J.~de~Anda, I.~de~Medeiros~Varzielas and S.F.~King,
  \emph{{Towards a complete A$_{4} \times$ SU(5) SUSY GUT}},
  \href{https://doi.org/10.1007/JHEP06(2015)141}{\emph{JHEP} {\bfseries 06}
  (2015) 141} [\href{https://arxiv.org/abs/1503.03306}{{\ttfamily
  1503.03306}}].

\bibitem{Esteban:2018azc}
I.~Esteban, M.C.~Gonzalez-Garcia, A.~Hernandez-Cabezudo, M.~Maltoni and
  T.~Schwetz, \emph{{Global analysis of three-flavour neutrino oscillations:
  synergies and tensions in the determination of $\theta_{23}$, $\delta_{CP}$,
  and the mass ordering}},
  \href{https://doi.org/10.1007/JHEP01(2019)106}{\emph{JHEP} {\bfseries 01}
  (2019) 106} [\href{https://arxiv.org/abs/1811.05487}{{\ttfamily
  1811.05487}}].

\bibitem{Planck:2018vyg}
{\scshape Planck} collaboration, \emph{{Planck 2018 results. VI. Cosmological
  parameters}},
  \href{https://doi.org/10.1051/0004-6361/201833910}{\emph{Astron. Astrophys.}
  {\bfseries 641} (2020) A6}
  [\href{https://arxiv.org/abs/1807.06209}{{\ttfamily 1807.06209}}].

\end{thebibliography}\endgroup
\bibliographystyle{JHEP}

\end{document}